\documentclass[useAMS,usenatbib]{mn2e}
\usepackage{txfonts}
\usepackage{graphicx}
\usepackage{textcomp}
\usepackage{natbib}
\usepackage{deluxetable}

\newcommand{\um}{\textmu m }
\newcommand{\uu}{\textmu m}

\newcommand{\uuJy}{\textmu Jy}
\newcommand{\tausil}{$\rm{\tau_{9.7}}$}
\newcommand{\ssil}{$S\!_{\rm{sil}}$}
\newcommand{\lsun}{$\rm{L_{\odot}}$}

\newcommand{\msunyr}{$\rm{M_{\odot} yr^{-1}}$}
\newcommand{\lir}{$L_{\rm{IR}}$}
\newcommand{\lira}{$L^*_{\rm{IR}}$}
\newcommand{\neiii}{[Ne {\sc iii}] }
\newcommand{\neii}{[Ne {\sc ii}] }
\newcommand{\ariii}{[Ar {\sc iii}] }
\newcommand{\arii}{[Ar {\sc ii}] }
\newcommand{\oii}{[O {\sc ii}] }

\bibpunct[]{(}{)}{;}{a}{}{,}

\title[Infrared-luminous galaxies at $z$ $\sim$ 0.5--3]{Mid-infrared spectroscopy of infrared-luminous galaxies at $z$ $\sim$ 0.5--3}

\author[A. Hern\'an-Caballero et al.]{
A. Hern\'an-Caballero,$^1$\thanks{e-mail: ahc@iac.es}
I. P\'erez-Fournon,$^1$
E. Hatziminaoglou,$^2$
A. Afonso-Luis,$^1$\newauthor
M. Rowan-Robinson,$^3$
D. Rigopoulou,$^4$
D. Farrah,$^{5,6}$
C. J. Lonsdale,$^7$
T. Babbedge,$^3$\newauthor
D. Clements,$^3$
S. Serjeant,$^8$
F. Pozzi,$^9$
M. Vaccari,$^{3,10}$
F. M. Montenegro-Montes,$^{1,11}$\newauthor
I. Valtchanov,$^{3,12}$
E. Gonz\'alez-Solares,$^{13}$
S. Oliver,$^6$
D. Shupe,$^{14}$
C. Gruppioni $^{15}$\newauthor
B. Vila-Vilar\'o,$^{16}$
C. Lari,$^{11}$ and
F. La Franca$^{17}$\\
$^{1}$Instituto de Astrof\'isica de Canarias, C/ V\'ia L\'actea s/n, E-38200 La Laguna, Spain\\
$^{2}$European Southern Observatory, Karl-Schwarzschild-Str. 2, 85748 Garching bei M\"unchen, Germany\\
$^{3}$Astrophysics Group, Blackett Laboratory, Imperial College, Prince Consort Road, London SW7 2BW, UK\\
$^{4}$Department of Physics, University of Oxford, Keble Road, Oxford OX1 3RH\\
$^{5}$Department of Astronomy, Cornell University, Space Sciences Building, Ithaca, NY 14853, USA\\
$^{6}$Astronomy Centre, Department of Physics and Astronomy, University of Sussex, Falmer, Brighton BN1 9QJ, UK\\
$^{7}$Infrared Processing and Analysis Center, California Institute of Technology, Pasadena, CA 91125, USA\\
$^{8}$Centre for Astrophysics and Planetary Science, School of Physical Sciences, University of Kent, Canterbury, Kent CT2 7NR, UK\\
$^{9}$Dipartimento di Astronomia, Universit\`a di Bologna, via Ranzani 1, I-40127 Bologna, Italy\\
$^{10}$Dipartimento di Astronomia, Universit\`a di Padova, Vicolo dell'Osservatorio 5, 35122 Padua, Italy\\
$^{11}$Istituto di Radioastronomia, INAF, Via Gobetti 101, I-40129 Bologna, Italy\\
$^{12}$ESA European Space Astronomy Centre, PO Box 78, 28691 Villanueva de la Ca\~nada, Madrid, Spain\\
$^{13}$Institute of Astronomy, University of Cambridge, Madingley Road, Cambridge CB3 0HA\\
$^{14}$Spitzer Science Center, MS 220-6, Caltech, Jet Propulsion Lab, Pasadena, CA 91125, USA\\
$^{15}$INAF - Osservatorio Astronomico di Bologna, via Ranzani 1, I-40127 Bologna, Italy\\
$^{16}$University of Arizona, Steward Observatory, 933 N. Cherry Ave., Tucson, AZ 85721, USA\\
$^{17}$Dipartimento di Fisica, Universit\`a degli Studi Roma Tre, via della Vasca Navale 84, 00146 Roma, Italy
}

\begin{document}

\date{Accepted: 2009 February 18; received: 2009 February 18; in original form: 2008 December 16;}

\pagerange{\pageref{firstpage}--\pageref{lastpage}} \pubyear{2009}

\maketitle

\label{firstpage}

\begin{abstract}
\small
We present results on low-resolution mid-infrared (MIR) spectra of 70 infrared-luminous galaxies
obtained with the Infrared Spectrograph
(IRS) onboard \textit{Spitzer}. We selected sources from the European Large Area Infrared Survey
(ELAIS) with $S\!_{\rm{15}}$ $>$ 0.8 mJy and photometric or spectroscopic $z$ $>$ 1.
About half of the sample are QSOs in the optical, while the 
remaining sources are galaxies, comprising both obscured AGN and starbursts.
Redshifts were obtained from optical spectroscopy, photometric redshifts and from 
the IRS spectra. The later turn out to be reliable for 
obscured and/or star-forming sources, thus becoming an ideal complement to optical
spectroscopy for redshift estimation.

We estimate monochromatic luminosities at several restframe wavelengths, equivalent
widths and luminosities for the PAH features, and strength of the silicate 
feature in individual spectra. We also estimate integrated 8--1000 \um 
infrared luminosities via spectral energy distribution fitting to MIR 
and far-infrared (FIR) photometry from the Spitzer Wide-area 
Infrared Extragalactic survey (SWIRE) and the MIR spectrum. 
Based on these measurements, we classify the spectra using well-known infrared 
diagnostics, as well as a new one that we propose, into three types of source:
those dominated by an unobscured AGN, mostly corresponding to
optical quasars (QSOs), those dominated by an obscured AGN, and starburst-dominated sources. Starbursts 
concentrate at $z$ $\sim$ 0.6--1.0 favored by the shift of the 7.7-\um PAH band into the selection 
15-\um band, while AGN spread over the $0.5<z<3.1$ range.

Star formation rates (SFR) are estimated for individual sources from the luminosity of the PAH features.
An estimate of the average PAH luminosity in QSOs and obscured AGN is obtained from the composite spectrum
of all sources with reliable redshifts. The estimated mean SFR in the QSOs is 50--100 \msunyr, but the
implied FIR luminosity is 3--10 times lower than that obtained from stacking analysis of the FIR photometry,
suggesting destruction of the PAH carriers by energetic photons from the AGN.
The SFR estimated in obscured AGN is 2--3 times higher than in QSOs of similar MIR luminosity.
This discrepancy might not be due to luminosity effects or selection bias alone, but could
instead indicate a connection between obscuration and star formation.
However, the observed correlation between silicate absorption and
the slope of the near- to mid-infrared spectrum is compatible with the obscuration of the AGN emission 
in these sources being produced in a dust torus.

\end{abstract}
\begin{keywords}
galaxies: starburst  -- galaxies: high-redshift -- galaxies: active -- quasars: general
\end{keywords}

\section{Introduction}

Infrared-luminous Galaxies (LIRGs), those objects with infrared (IR, [8--1000\uu]) luminosities above 
$10^{11}$ \lsun\ (and the Ultraluminous Infrared Galaxies, those objects with 
$L_{IR}>10^{12}$ \lsun) are an important population for understanding the cosmological evolution 
of galaxies. At low redshifts LIRGs are relatively rare, with only a few hundred known over the 
whole sky at $z<0.2$ \citep{soifer84}. At higher redshifts they are much more common; at $z>1$ they
reach a density on the sky of several hundred per square degree, and make a substantial, perhaps 
even dominant contribution to both the Cosmic Infrared Background and the volume-averaged cosmic 
star formation rate \citep{Lagache05,Sanders96,Lonsdale06}.

Theoretical and empirical studies have shown that many LIRGs and virtually all ULIRGs at low 
redshift are interacting systems \citep{Gallimore93,Clements96, Sanders96,Borne99,Farrah01}. The 
power source behind the IR emission was initially controversial, with both star formation and AGN 
activity proposed as the sole driver, but in recent years a consensus has started to emerge, with 
most studies finding that both starbursts and AGN power low-redshift LIRGs and ULIRGs, with the star 
formation usually dominating \citep{Genzel98,farrah03,imanishi07}. At high redshifts a broadly 
similar picture is thought to exist \citep{farrah02,chapman03,takata06,berta07,bridge07}.

The ability to reliably distinguish between star formation and AGN activity in high redshift LIRGs is
essential for tracing both the cosmic star formation history and the evolution of AGN activity as a 
function of redshift. Since these galaxies are dusty, the usual optical/near-infrared diagnostics are 
not always reliable. Instead, mid-infrared (MIR) spectroscopy has proved itself the key tool for 
determining the dominant energy source in dusty galaxies 
\citep[e.g.][]{Genzel98,Rigopoulou99,Laurent00,Tran01,Weedman05}, but until recently the limited 
sensitivity of the Infrared Space Observatory (\textit{ISO}) constrained the spectroscopic study of ULIRGs to 
the local Universe. With the Infrared Spectrograph \citep[IRS,][]{Houck04} onboard \textit{Spitzer}, 
wavelength coverage and sensitivity have improved enough to allow redshift determination and 
spectroscopic classification for ULIRGs at all redshifts up to $z\sim3$ 
\citep{Houck05,Yan05,Weedman06,armus07,farrah07}.

In most cases, spectroscopic surveys of high-redshift sources with the IRS have selected sources with 
a high infrared-to-optical flux ratio \citep{Houck05,Yan05,Weedman06}. These turn out to be mostly obscured 
AGN at high redshift, with spectra dominated by silicate absorption. Starburst ULIRGs with strong PAH 
emission are rare in these samples, but prolific if the selection criteria instead include sub-mm 
detections \citep{Lutz05} or prominent stellar `bumps' in the IRAC bands \citep{Farrah08}. 

In contrast, the IRS has undertaken relatively few surveys of sources selected solely on the basis of 
MIR flux. This means that moderately to lightly absorbed sources will be under represented in IRS 
surveys, thus drawing an incomplete picture of the nature of the infrared galaxy population at high 
redshifts. In this work we present results on a sample of 70 LIRGs and ULIRGs selected in the MIR with
no constraints on the optical fluxes (other than the requirement of detection in at least 3 optical 
bands in deep imaging and photometric or spectroscopic $z$ $>$ 1. Our objectives are: (a) to classify
the sources into starburst-dominated (SB), obscured AGN-dominated (AGN2) and unobscured AGN-dominated
(AGN1), based on their MIR spectral properties, and compare this classification with that obtained 
from optical data. (b) To estimate the IR luminosity of the sources and the relative contribution of 
AGN and star formation. (c) To obtain star formation rates for the obscured and unobscured AGN. 

The paper is structured as follows: \S 2 describes the sample selection; \S 3 discusses the 
observations and data reduction; \S 4 presents the measurements and diagnostics; \S 5 compares the 
results to other high-$z$ MIR-selected samples; and \S 6 summarizes our conclusions. We adopt the 
following cosmology: $\Omega_m$ = 0.27, $\Omega_\Lambda$ = 0.73, $H_0$ = 71 km s$^{-1}$ Mpc$^{-1}$.

\section{Sample selection}

\begin{figure}
\begin{center}
\includegraphics[width=8.5cm]{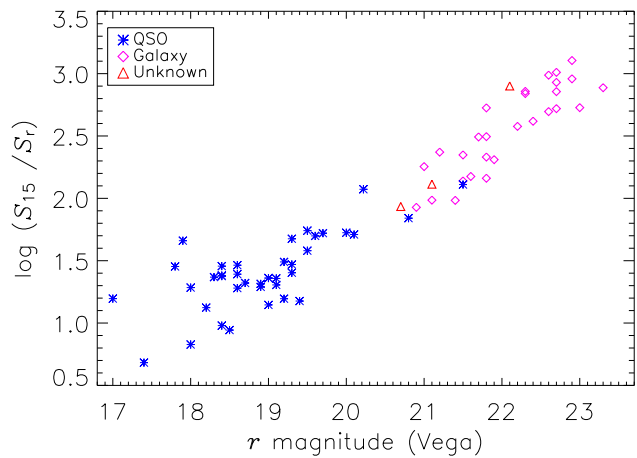}
\end{center}
\caption[IR to optical flux ratio]{Ratio of the 15-\um flux to the $r$-band flux as a function of
$r$ magnitude for the sources in the ELAIS-IRS sample. Asterisks indicate sources classified as
QSOs from their optical spectrum or SED, while diamonds represent sources classified as galaxies.
Triangles represent sources with uncertain or unknown optical class due to lack of
enough optical data.
\label{S15_vs_Sr}}
\end{figure}

The sources were selected from the European Large Area ISO Survey
\citep[ELAIS;][]{Oliver00, Rowan-Robinson04} 
final band-merged catalog \citep{Rowan-Robinson03}. 
The selection criteria required detection in the ISOCAM LW3 band \citep{Vaccari05}
with $S\!_{\rm{15}}$ $>$ 0.8 mJy, as well as detection in at least 3 optical bands
and spectroscopic or photometric redshift $z$ $\gtrsim$ 1.

The 70 objects that matched were observed with the IRS \citep{Hernan08}, 
resulting in one of the largest $z$ $\gtrsim$ 1 samples of MIR spectra of infrared-luminous 
galaxies to date. Table \ref{registro_IRS} indicates the name and coordinates of these sources.
Only the first 15 entries of each table are shown here. The complete version of the
tables is available as on-line material.

In addition to the MIR spectra, there is also optical and IR photometry available
for all sources, as well as optical spectroscopy for roughly half of them.

For sources in the northern hemisphere (ELAIS N1 and N2 fields) optical  
$U$, $g$, $r$, $i$, $Z$ photometry
comes from the Isaac Newton Telescope Wide Field Survey (WFS), with $5\sigma$
limiting magnitudes 23.4, 24.9, 24.0, 23.2 and 21.9 (Vega) respectively
\citep{Gonzalez-Solares05}.
In the southern hemisphere (ELAIS S1) optical photometry was obtained
down to $B$, $V$ $\sim$ 25, $R$ $\sim$ 24.5 \citep{Berta06}.

Eight sources (six of them in ELAIS S1) have near-infrared (NIR) photometry from 
the Two Micron All Sky Survey \citep[2MASS;][]{Skrutskie06}, 
and ten more are detected by the UKIRT Infrared Deep Sky Survey 
\citep[UKIDSS;][]{Lawrence07} in the central region of ELAIS N1.
Table \ref{fotometria-opt} summarizes the optical and NIR photometry for the ELAIS-IRS sources.

The three ELAIS fields in which the ELAIS-IRS sample was selected were
observed by Spitzer in seven IR bands from 3.6 to 160 \uu, as part
of the Spitzer Wide-Area Infrared Extragalactic Survey \citep[SWIRE;][]{Lonsdale03a,Lonsdale04}.
We performed a source match between the ELAIS-IRS catalog and the latest working catalog from
SWIRE and found that 69 out of 70 ELAIS-IRS sources are detected in at least two IRAC bands and 
in the 24-\um band from MIPS,
while 33 are detected in the 70-\um band and only 14 at 160 \um (see Table \ref{fotometria-IR}).

One source, EIRS-65, has no infrared counterpart in the SWIRE catalog and is undetected in
the IRS spectrum. Since the 15-\um detection is strong, the 15-\um source is probably real
but was associated to the wrong optical counterpart.    

Photometric redshifts were calculated using optical and NIR photometry for all sources
in the ELAIS bandmerge catalog \citep{Rowan-Robinson04}. These estimates were revised
with posteriority to the sample selection using IRAC photometry from SWIRE 
\citep{Rowan-Robinson04,Rowan-Robinson08}, leading in some cases to revised photometric $z$ $<$ 1. 
At least four and up to nine optical and NIR bands are used for photometric redshift
estimation, except in four sources whose redshifts are obtained from just three bands and are thus
very unreliable (EIRS-19,40,52,55).

Redshifts from optical spectroscopy are also available for 37 ELAIS-IRS sources, most
of them from spectroscopic follow-ups of ELAIS QSOs \citep{Afonso-Luis04,LaFranca04} or
the Sloan Digital Sky Survey \citep[SDSS;][]{York00}.

Since an optical photometric or spectroscopic redshift is required for source selection,
the sample is biased against the most obscured sources; and
the lack of constraints in the IR-to-optical flux ratio provides a large 
number of unobscured AGN 
which are absent in other high-$z$ MIR-selected samples (see \S\ref{otrasmuestras}).

39 sources are classified as QSOs by their optical spectral energy distribution (SED) or spectrum, 
while 28 are galaxies
(including starbursts and obscured AGN) and the remaining 3 are of undetermined type. 
QSOs and galaxies can be easily
distinguished by their optical-to-MIR flux ratios (Fig. \ref{S15_vs_Sr}) with little overlap
between populations.

In order to provide a good reference sample for redshift estimation and 
calibrate the MIR diagnostic tools, we have compiled a set of MIR spectra
and MIR-to-FIR photometry containing 137 well-known low-redshift sources
(the `Library'), including QSOs, Seyfert 1 and Seyfert 2 galaxies, ULIRGs and starbursts.
The MIR spectra are high- and low-resolution IRS spectra published in the literature (see Table \ref{info_biblioteca}),
while redshifts, classification and IR photometry for the sources were extracted from the NASA 
Extragalactic Database (NED).

\section{Observations and Data Reduction}

All targets were observed in the two low-resolution modules of the IRS,
Short-low (SL) and Long-low (LL),
exposing the two sub-slits on each of them (SL1, SL2, LL1 and LL2)  
and thus covering the 5--38 \um range. Two nod positions
were observed in each configuration in order to provide optimal background
subtraction. Exposure times for the LL were defined using 
ISOCAM 15-\um flux densities in order to assure an adequate S/N at this wavelength. 
Integration times for the SL were calculated from the 15-\um flux density 
assuming an M82-like SED. The observations log is shown in Table \ref{registro_IRS}.

Raw IRS data were reduced by the IRS pipeline (ver. S15.3.0) at the Spitzer Science
Center. This included ramp fitting, dark subtraction, flat-fielding and
wavelength and flux calibration.  

Since our sources are faint, additional reductions were carried out starting 
from the Basic Calibrated Data (BCD) stage.
Images with the object centred in alternate sub-slits of the same module
were subtracted (i.e: SL1-SL2, LL1-LL2) to remove sky background, and their
corresponding uncertainty and mask images were, respectively, added and
bit-wise XOR combined.\footnote{In the mask images, individual bits are set
for each pixel that meets a particular condition. In the combined mask, a
bit is set if it is set in at least one of the two original masks.}
One-dimensional spectra for both nod positions were extracted using the Spitzer
IRS Custom Extraction (SPICE) version 1.4.1. The extraction aperture was set to the
SPICE default, and the Optimal Calibration Mode, in which the signal in each pixel
is weighted with its uncertainty, was selected.

Individual spectra were flux-calibrated with SPICE using standard 
calibration files for pipeline S15.3.0, and then combined into a single spectrum
using custom IDL routines.

\begin{figure} 
\begin{center}
\includegraphics[width=8.5cm]{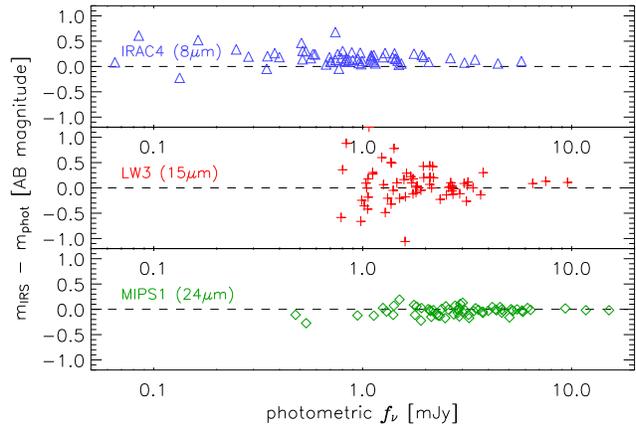}
\end{center}
\caption[Flux calibration residuals]{
Residuals obtained after subtracting the broad-band IR photometry to the synthetic photometry from 
the IRS spectrum for each of the calibration bands: 8 \um (\textit{top}), 15 \um (\textit{centre}) 
and 24 \um (\textit{bottom}).
\label{calibracion-residuos}}
\end{figure}

Comparison with
IRAC 8.0-\um and MIPS 24-\um photometry from SWIRE and ISOCAM 15-\um from ELAIS 
indicates that the relative flux
calibration uncertainty is around 14, 35 and 8 per cent at 8, 15 and 24 \uu, respectively. 
Note that
the uncertainty at 15 \um increases sharply around 1 mJy as it approximates the detection
limit of the ELAIS survey (Fig. \ref{calibracion-residuos}). The 8-\um fluxes from IRS 
data are in average 10 per cent lower than in the IRAC photometry, but we have preferred not to 
make any flux correction to the IRS spectra since the 8-\um band covers a somewhat noisy 
region of the Short-Low spectrum in the overlapping wings of the SL1 and SL2 transmission
profiles.

Wavelength calibration is performed by the IRS pipeline, and is accurate to $\sim$1/5 of the 
resolution element or $\delta\lambda$/$\lambda$ $\sim$ 0.001. 
Reduced and calibrated spectra for each of the 70 ELAIS-IRS sources are shown in
Fig. \ref{IRSobserved}.

\begin{figure*} 
\begin{center}
\includegraphics[width=15cm]{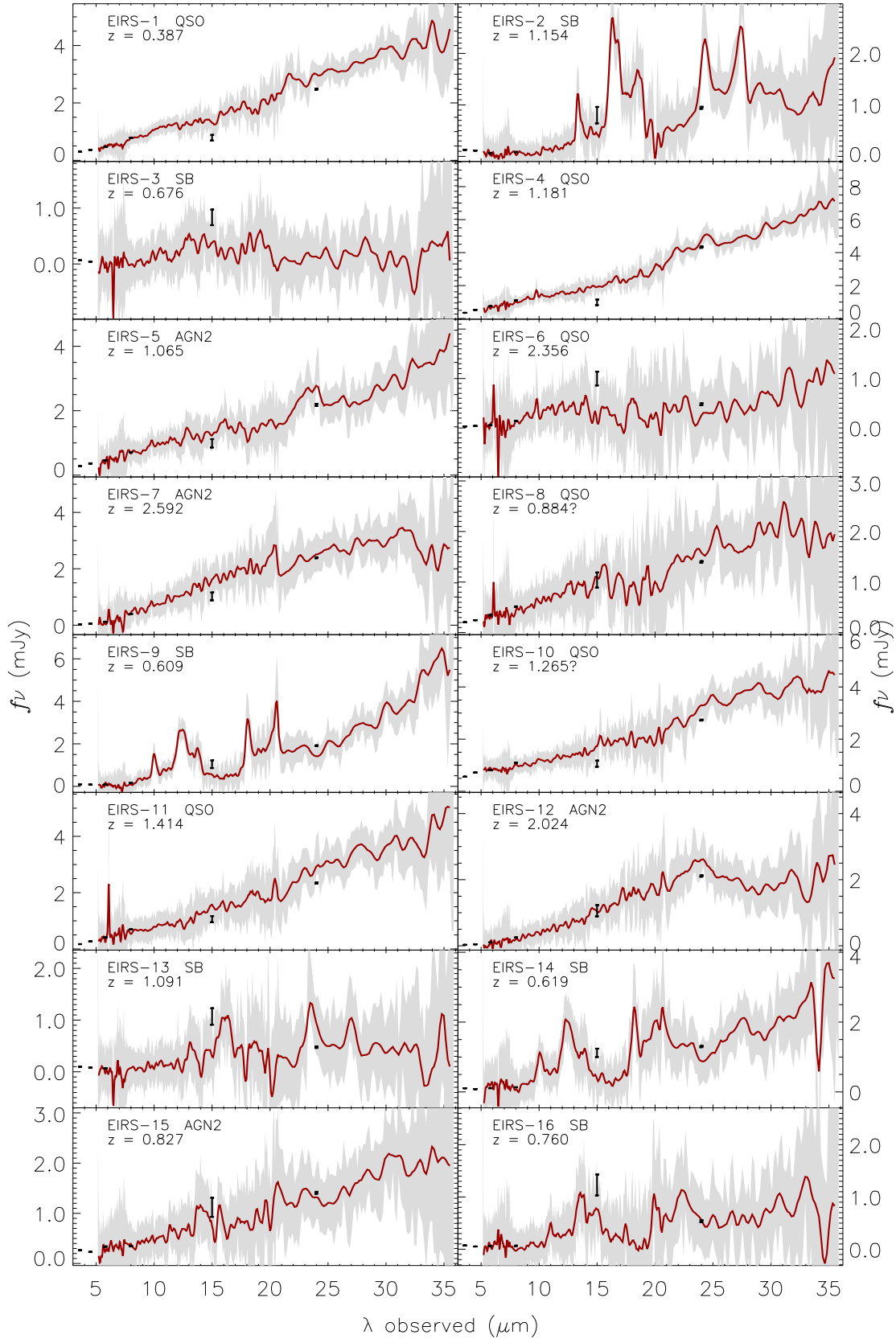}
\end{center}
\caption[Observed spectra]{Observed IRS spectra for the ELAIS-IRS sources. The shaded area represents
the $1\sigma$ uncertainty in $f_\nu$, while the solid line represents the
smoothed spectrum after a Gaussian filter with a FWHM of 3 resolution elements has been applied. 
The error bars mark the flux density measured in the IRAC (3.6, 4.5, 5.8 and 8 \uu) and MIPS (24 \uu) 
bands from SWIRE and ISOCAM (15 \uu) from ELAIS. The IR classification obtained in \S\ref{diagnostico}
is indicated as QSO (unobscured AGN-dominated), AGN2 (obscured AGN-dominated) or SB (starburst-dominated).
A question mark besides the redshift estimate indicates the measurement is unreliable.\label{IRSobserved}}
\end{figure*}

\addtocounter{figure}{-1}
\begin{figure*}
\begin{center}
\includegraphics[width=15cm]{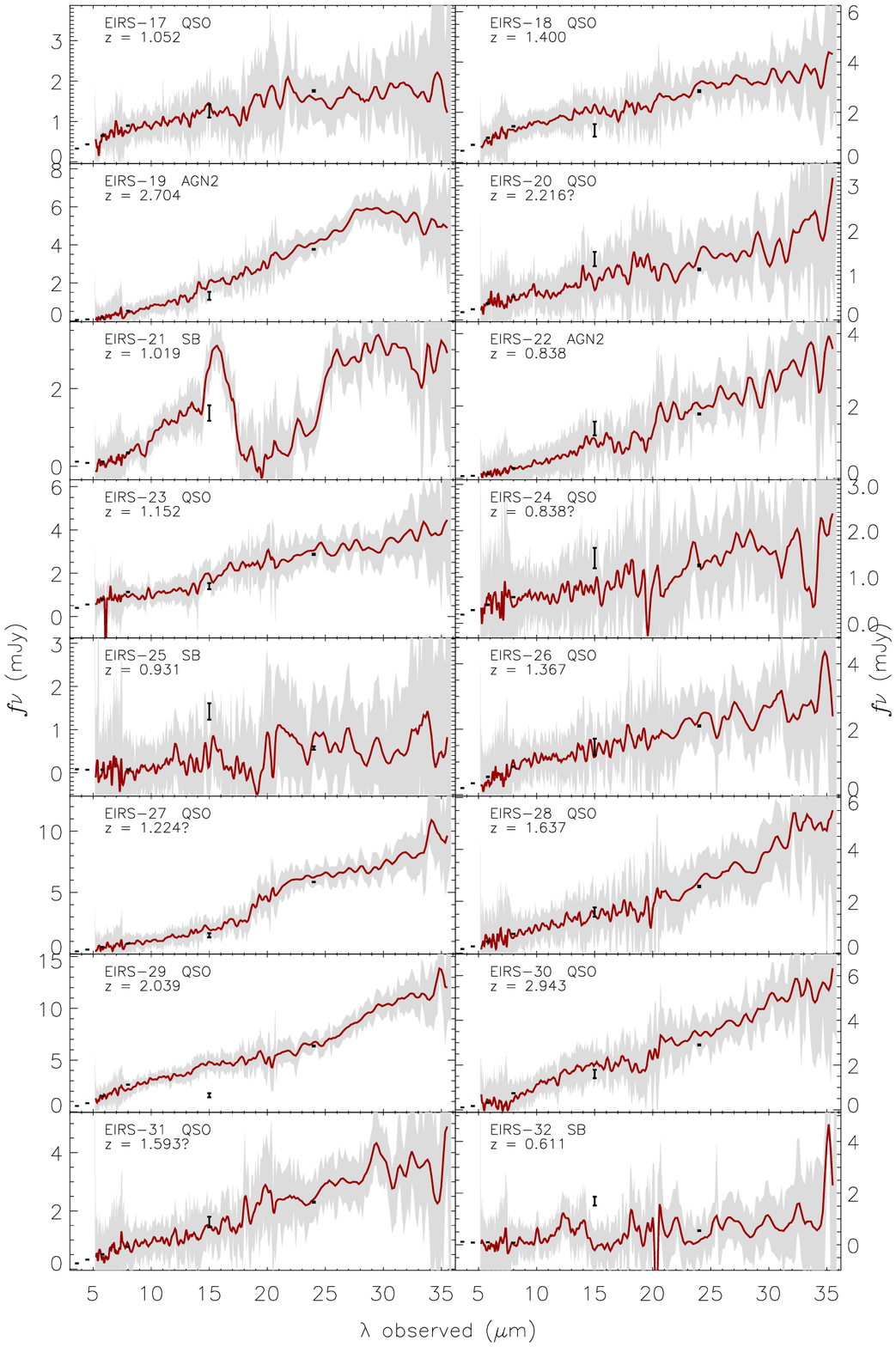}
\end{center}
\caption{Continued.}
\end{figure*}

\addtocounter{figure}{-1}
\begin{figure*}
\begin{center}
\includegraphics[width=15cm]{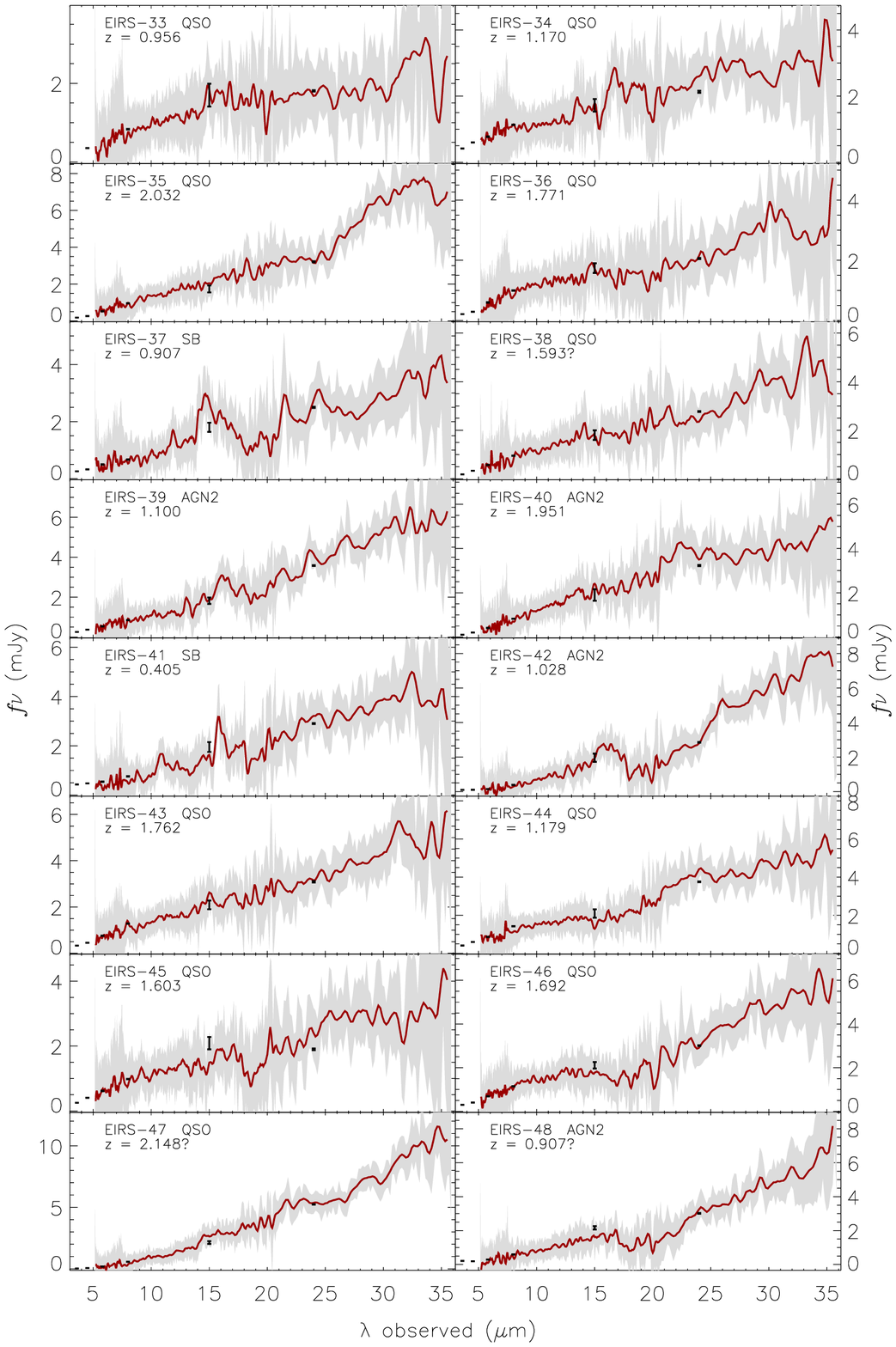}
\end{center}
\caption{Continued.}
\end{figure*}

\addtocounter{figure}{-1}
\begin{figure*}
\begin{center}
\includegraphics[width=15cm]{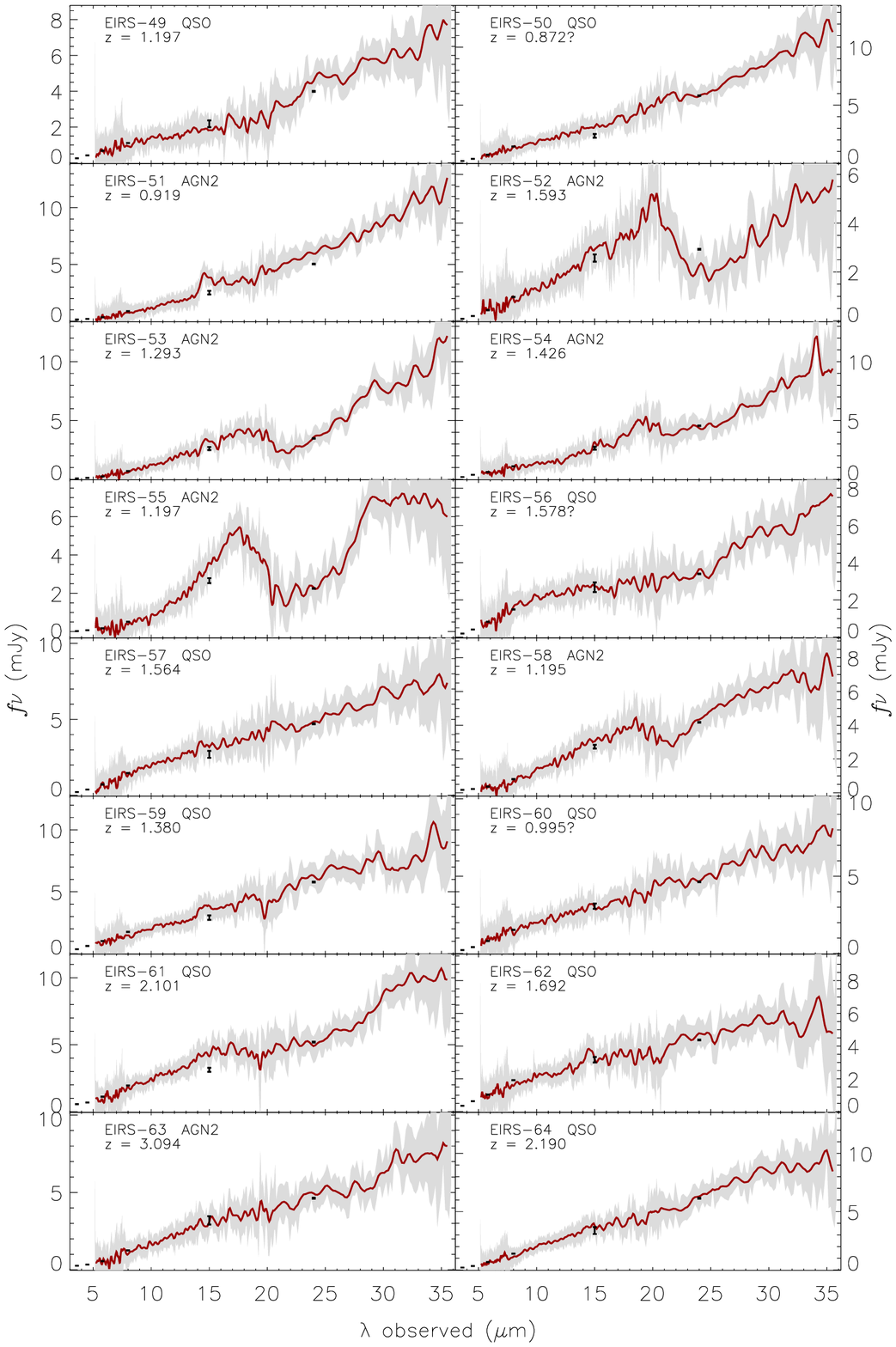}
\end{center}
\caption{Continued.}
\end{figure*}

\addtocounter{figure}{-1}
\begin{figure*}
\begin{center}
\includegraphics[width=15cm]{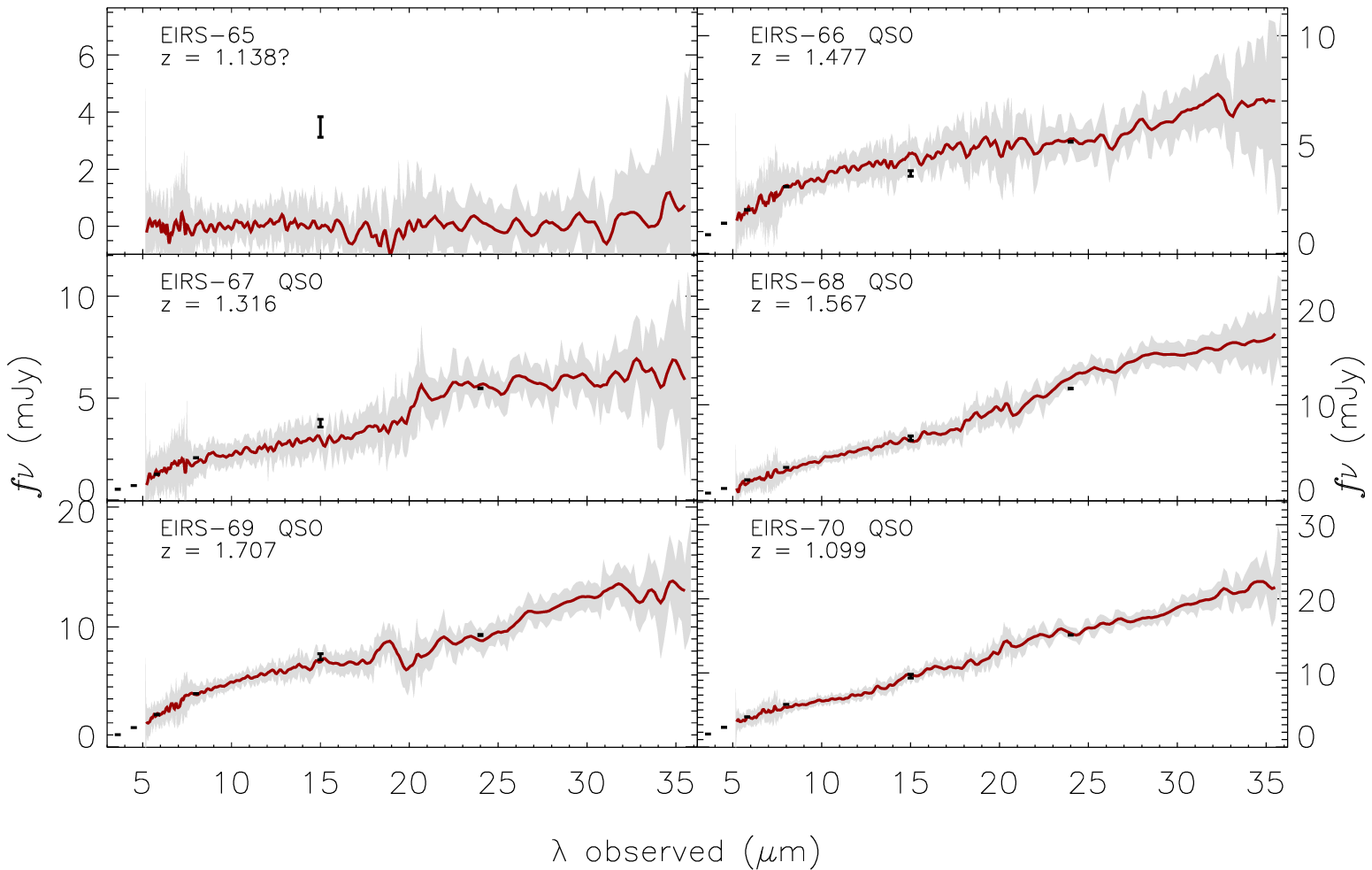}
\end{center}
\caption{Continued.}
\end{figure*}

\section{Results}

\subsection{Redshifts from the IRS spectra}

Spectroscopic redshifts ($z_{\rm{spec}}$) are available for roughly half of the sample,
all of them optical QSOs except for one obscured AGN (EIRS-7). The rest of the sources 
were selected based on photometric redshifts ($z_{\rm{phot}}$) published in the ELAIS 
bandmerged catalog \citep{Rowan-Robinson04}, 
which were later revised with the addition of IR photometry from SWIRE 
and UKIDSS after the ELAIS-IRS sample was selected \citep{Rowan-Robinson04,Rowan-Robinson08},
leading to revised $z$ $<$ 1 in some cases.

In many of the sources in the sample a redshift estimate can also be obtained from the
IRS spectrum based on the PAH features, the silicate profile and the shape of the
continuum SED ($z_{\rm{IRS}}$). To this aim, we have developed an algorithm that compares every spectrum 
with those in the Library
and determines the best-$z$ estimate by a $\chi^2$-minimization procedure
(Hern\'an-Caballero et al. in prep.).

Reliable $z_{\rm{IRS}}$ are obtained for 26 sources, while for
another 28 a $z$ estimate is obtained, but it is uncertain due to poor SED fitting or multiple 
solutions. Most sources with reliable $z_{\rm{IRS}}$ are galaxies,
because their PAH-emission and/or silicate-absorption features produce a single sharp spike
in $\chi^2$($z$), while in QSOs a smoother spectrum allows for multiple, weaker solutions.  

\begin{figure} 
\begin{center}
\includegraphics[width=8.5cm]{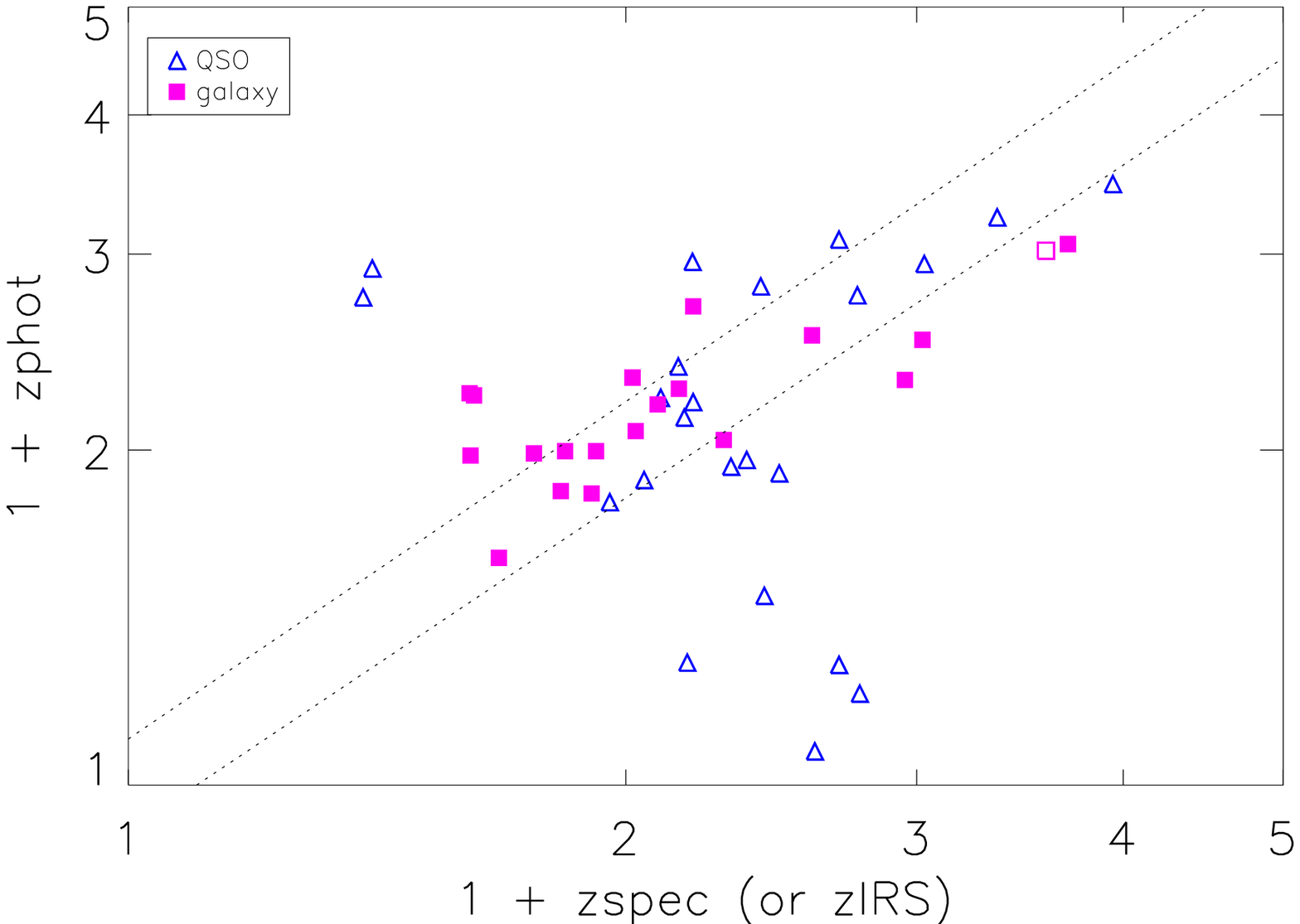}
\end{center}
\caption[zphot versus $z_{\rm{spec}}$ or $z_{\rm{IRS}}$]{Comparison of the photometric redshifts with those obtained
from optical or MIR spectroscopy for sources with reliable redshift estimations.
Triangles represent sources classified as QSOs in the optical, while squares represent galaxies. 
The open square represents EIRS-7, the only optical galaxy with redshift from optical spectroscopy. 
The dotted lines enclose photometric redshifts with errors lower than 10 per cent in $1+z$.
\label{zphot_zbest}}
\end{figure}

A comparison of optical- and MIR-spectroscopic versus photometric redshifts (Fig. \ref{zphot_zbest})
indicates that $z_{\rm{phot}}$ estimates have significant uncertainties, larger 
than those of the general ELAIS sample but similar to the uncertainty found in the $z>1$ subsample
\citep{Rowan-Robinson08}. Photometric redshifts
are significantly less accurate in optical QSOs than in galaxies, due to aliasing problems, and
may be also affected by variability issues, because the optical photometry from WFS was not 
obtained in the same epoch for all bands \citep{Afonso-Luis04}. 

Since $z_{\rm{IRS}}$ estimates are stronger in galaxies and obscured AGN, while $z_{\rm{spec}}$
is more appropriate for QSOs, the combination of both techniques allows us to
obtain reliable redshifts for 57 of 70 sources (Table \ref{origen_z}).
The remaining 13 rely on $z_{\rm{phot}}$ (EIRS-47, 60 and 65) or unreliable $z_{\rm{IRS}}$ estimates
(10 sources), and will be discarded in all redshift-sensitive analysis.

\begin{figure} 
\begin{center}
\includegraphics[width=8.5cm]{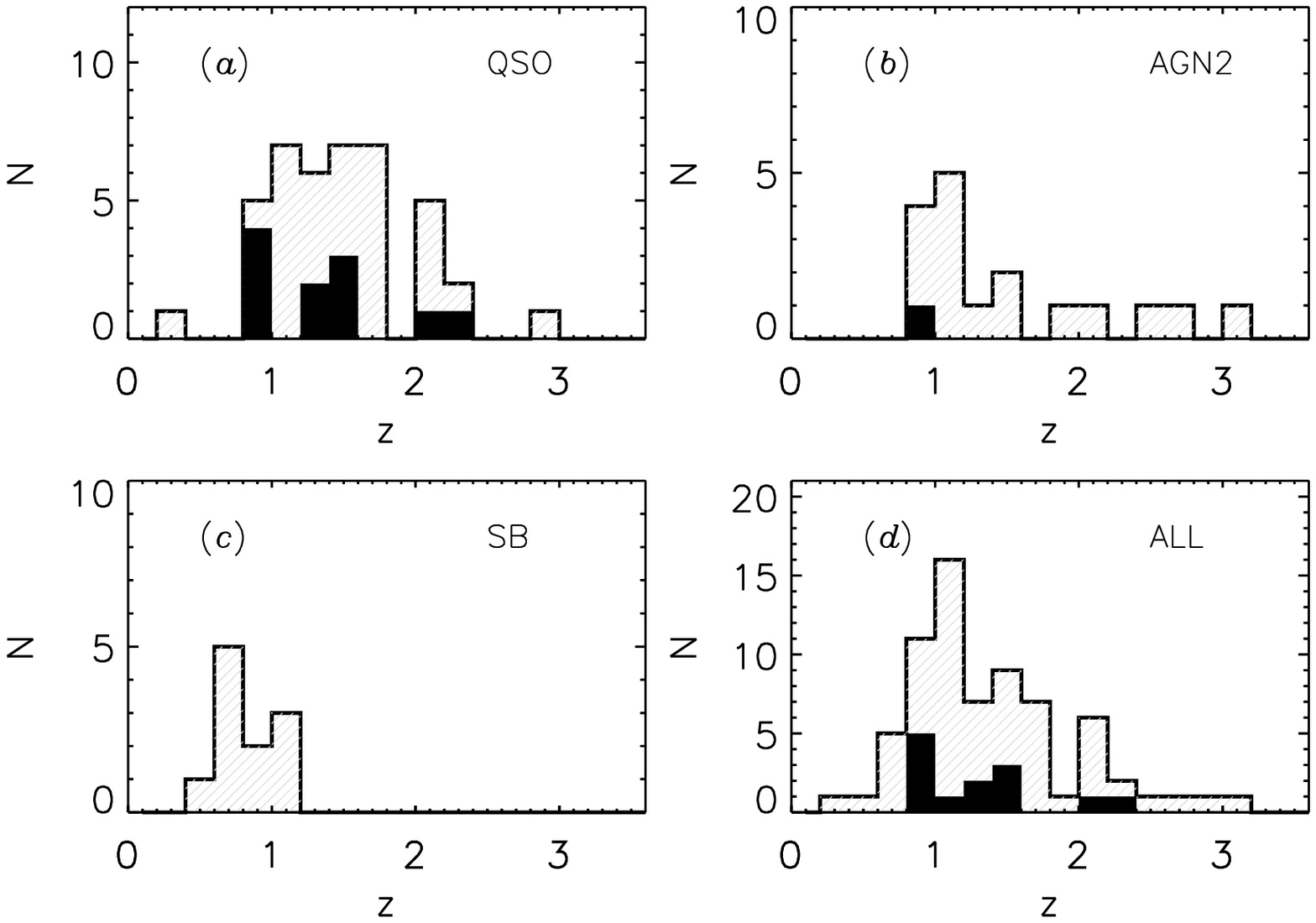}
\end{center}
\caption[Redshift distribution histograms]{
Redshift distribution of ELAIS-IRS sources as a function of IR class:
(a) unobscured AGN, (b) obscured AGN, (c) starburst galaxies, (d) all sources.
The stripped histograms represent both reliable and unreliable redshifts, while the 
solid ones include only unreliable redshifts from $z_{\rm{phot}}$ or $z_{\rm{IRS}}$.\label{zhistog}}
\end{figure}

The redshift distribution of the ELAIS-IRS sample is shown in Fig. \ref{zhistog}. Starburst galaxies
concentrate around $z$ $\sim$ 1 because the selection at 15 \um favours sources with a strong 
7.7-\um PAH feature at this redshift, while simultaneously prevents selection of higher-redshift 
sources since a typical 
starburst SED dwindles quickly for $\lambda$ $<$ 6 \uu. This behaviour is also observed
in 24-\um selected starburst galaxies, which tend to concentrate
around $z$ $\sim$ 1.7 \citep{Yan07,Farrah08}.
The redshift distribution for QSOs is smoother, with most of them in the $1<z<2$ range, but the
number of obscured AGN decreases sharply at $z$ $>$ 1.2 because of their red SED shortwards of
6 \um restframe.

\subsection{Monochromatic fluxes and luminosities\label{medidas_continuo}}

\begin{figure}[t] 
\begin{center}
\includegraphics[width=6.0cm]{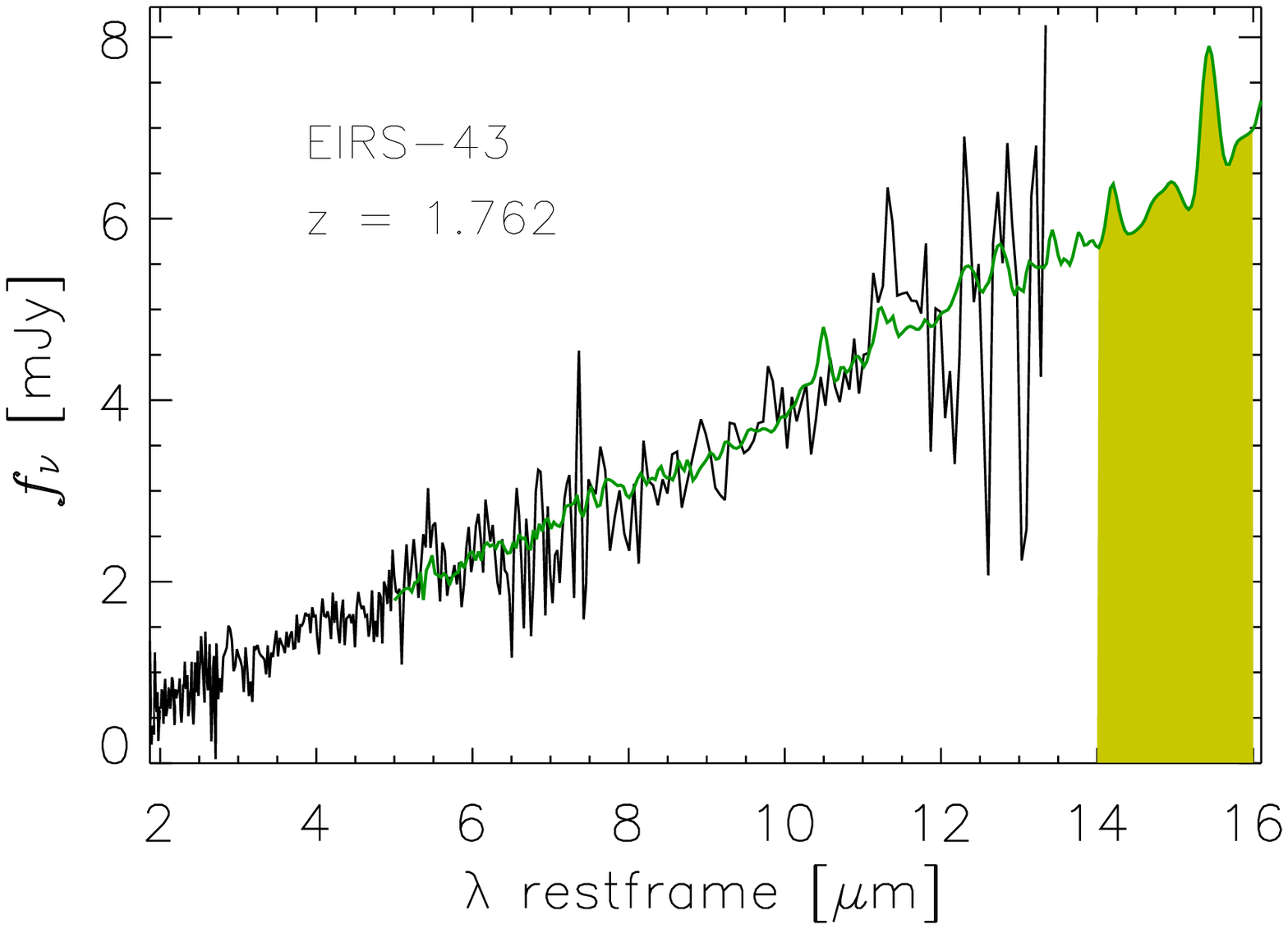}
\hfill
\includegraphics[width=6.0cm]{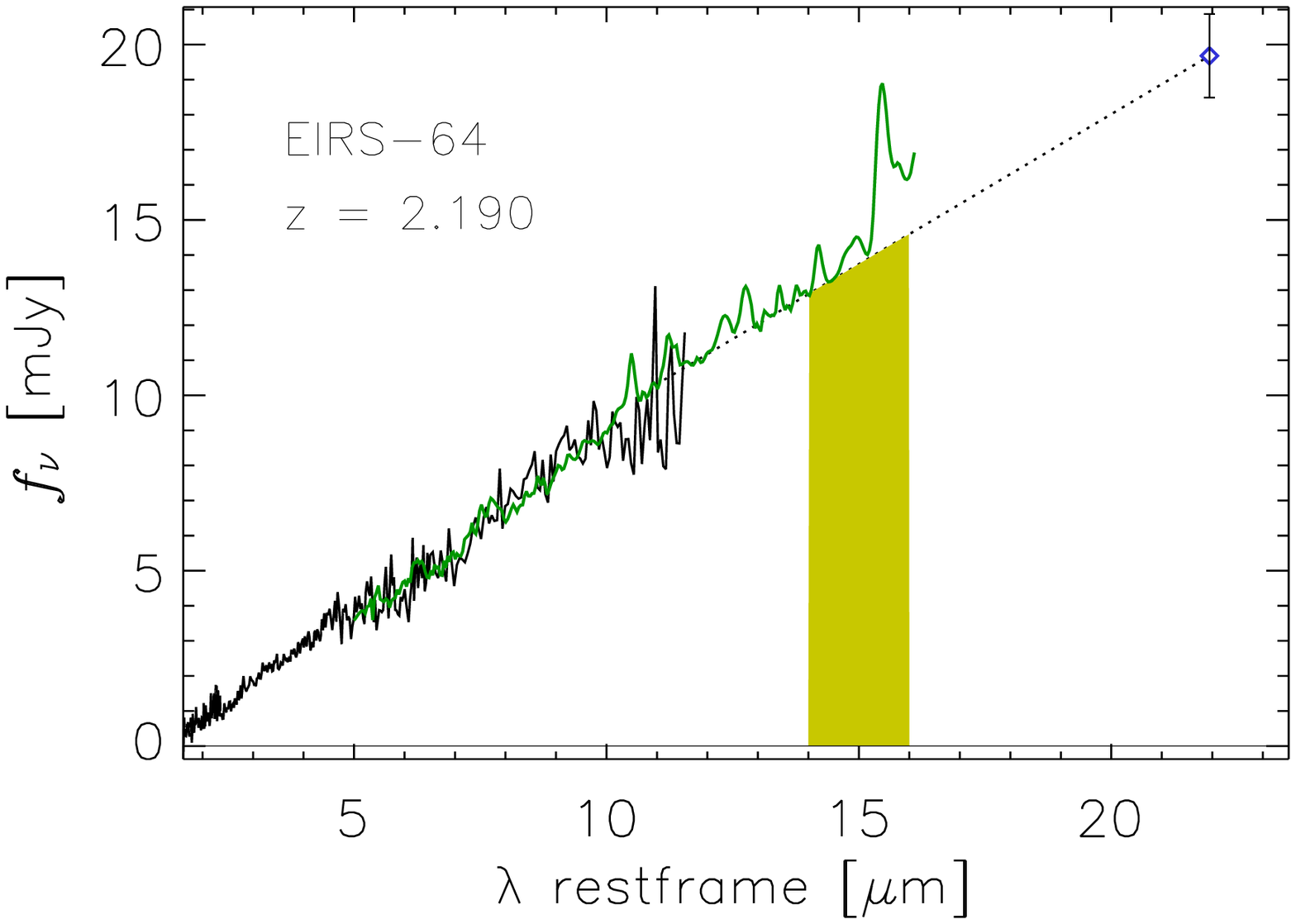}
\end{center}
\caption[15 \um flux for high redshift sources]{
Examples of the procedure followed to estimate the restframe 15-\um flux
for sources at 1.4 $<$ $z$ $<$ 2, in which the model fitted in the spectral
decomposition is extrapolated up to 16 \um restframe (\textit{top}); and $z>2$
sources, for which a linear interpolation between the end of the IRS spectrum
and the MIPS70 photometric point is used (\textit{bottom}).\label{extiende}}
\end{figure}

We calculate restframe monochromatic fluxes (f$_\nu$) and luminosities $\nu$L$_\nu$($\lambda$) at 
several wavelengths (2.2, 5.5, 7, 10 and 15 \uu).
The procedure used depends on the observed and restframe wavelength: 
for the 2.2 \um (restframe $K$ band) we interpolate the photometry in the IRAC bands (useful for sources
at 0.6 $<$ $z$ $<$ 2.65); for 5.5, 7 and 10 \um the spectrum is averaged over a narrow range
(5.3--5.8, 6.6--7.4 and 9.5--10.5 \um respectively); while for the 15-\um luminosity the procedure
is more complex: in $z<1.4$ sources it is obtained by averaging the 14--16 \um spectrum; in $1.4<z<2$ 
sources 15 \um restframe is outside the spectral range observed by IRS, but not too far, so
the 15-\um luminosity is obtained by extrapolation of the model spectrum fitted in the spectral decomposition
analysis (see \S\ref{decomposition}). In $z>2$ sources the 15-\um luminosity is interpolated between
the red end of the IRS spectrum and the MIPS70 photometric point (see Fig. \ref{extiende} for an example of the 
15-\um flux measurement in the last 2 cases). 
Table \ref{tabla_continuo} summarizes the results obtained.

\subsection{PAHs and Silicate features}

In the literature we find two main methods for measuring the flux in the aromatic bands,
which differ in the way the continuum and the PAH features are measured: a) the interpolation
method, in which the continuum underlying the aromatic bands is estimated by linear
\citep[e.g.][]{Rigopoulou99} or spline \citep[e.g.][]{Vermeij02,Spoon07} interpolation
between PAH-free regions of the spectrum, and the flux over the estimated continuum is 
attributed to the PAH feature; and b) the Lorentzian method, in which a Lorentzian or Drude profile
is assumed for the PAH features \citep{Laurent00,Smith07,Sajina07}.

We implemented the later method using a procedure in which a
double power-law for the continuum and Lorentzian profiles for the PAH features are fitted 
simultaneously. The fitting algorithm is similar to that described in \citet{Smith07}
and \citet{Sajina07} and is implemented as follows:
the continuum emission is modeled by a linear combination of two power-laws with spectral
indices $\alpha$, $\beta$ ranging from -5 to 5.
The PAH features are modeled by Lorentzian profiles $L$($b$,$\omega$,$\lambda_0$) for the five
main features (6.2, 7.7, 8.6, 11.3 and 12.7 \uu), while the rest are ignored. The amplitude of
the Lorentzian ($b$) must be non-negative, its width ($\omega$) is constrained between 0.5 and 2 times
the typical value found in the literature\footnote{$\omega$ = 0.2, 0.7, 0.4, 0.2 and 0.3 \um for the features
centred at 6.2, 7.7, 8.6, 11.3 and 12.7 \uu, respectively \citep{Li01}.} 
and the central wavelength ($\lambda_0$) can fluctuate up to 5 per cent from its expected value.

Extinction in the continuum is implemented assuming a screen model, with opacity $\tau$($\lambda$) 
following the Galactic Centre extinction law
\citep[GC;][]{Chiar06}, which has been widely used in other ULIRG samples with moderate or high silicate 
absorption \citep{Forster03,Sajina07,Polletta08}. 
Thus we fit the restframe spectra in the 5--15 \um range to the analytic expression:

 \begin{equation}
F_\lambda(\lambda)\ =\ ( A \lambda^\alpha + B \lambda^\beta ) e^{-t \tau(\lambda)} + 
  \sum_{i}^{N} L(b_i,w_i,\lambda_i)(\lambda)\label{Eq_PAHfit}  
\end{equation}

We quantify the intensity of the silicate feature using the 
`silicate strength', \ssil~ \citep[e.g.][]{Shi06,Spoon07,Maiolino07}, defined as:

\begin{equation}
$\ssil$\ =\ ln \frac{F_\lambda(\lambda_0)}{C(\lambda_0)}
\end{equation}

where $C(\lambda_0) = A \lambda^\alpha + B \lambda^\beta$  is the extinction-corrected continuum
at 9.7 \uu, and $F_\lambda(\lambda_0)$ is the observed flux (averaged in a narrow interval to
minimize the effects of the noise in the spectrum).
\ssil\ is positive for sources with silicates in emission and negative in absorption.

In two high redshift sources (EIRS-30 and EIRS-63) \ssil\ could not be
measured because 9.7 \um restframe is outside of the observed wavelength range, and in three starburst
sources (EIRS-13, 25 and 32) the mean flux in the 9.5--9.9 \um range is negative in their noisy spectra. 
For these five sources we assume \ssil\ = -\tausil, 
where \tausil\ =  $t \tau$($\lambda$=9.7\uu)
is the estimated apparent optical depth at 9.7 \um from Eq. \ref{Eq_PAHfit}.

In most sources with silicates in emission, the silicate feature is wider and centred
at longer wavelengths ($\sim$10.5 \uu) than it is found when the feature appears in absorption 
\citep[see][ for tentative explanations]{Siebenmorgen05,Netzer07}, and the extra flux in the
silicate feature tends to increase the continuum level in the fit. These effects lead to a 
systematical under estimation
of the silicate strength in sources with silicates in emission. To compensate for this,
we run a second fitting loop in sources with \tausil\ $\sim$ 0 and \ssil\ $>$ 0 in which the 
9.5--11 \um range is excluded from the fit to avoid overestimation of the continuum, and 
\ssil\ is measured at 10.5 \uu.

Table \ref{tabla_PAHfits} summarizes the results obtained from the fitting. 
Fluxes of PAH features with S/N below $2\sigma$ are considered
as non-detections and shown as upper limits.

\subsection{Infrared Luminosity}

\begin{figure} 
\begin{center}
\includegraphics[width=8.5cm]{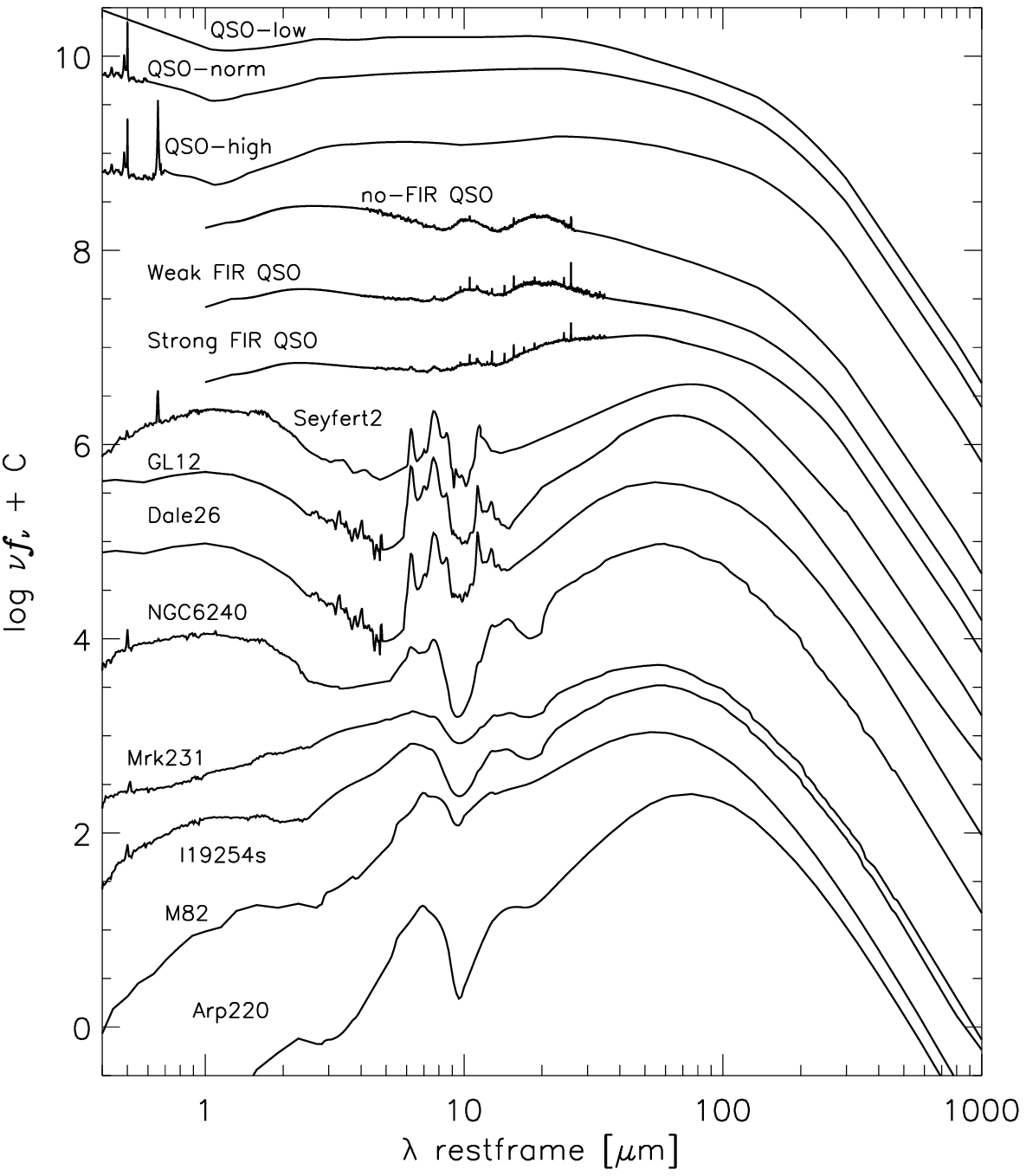}
\end{center}\caption[SED templates]{SEDs models used to estimate IR luminosity of the 
ELAIS-IRS sources. The templates Strong FIR QSO, Weak FIR QSO and no-FIR QSO from \citet{Netzer07}
have been extended longwards of 70 \um using the QSO-high template from \citet{Polletta07}. \label{SEDtemplates}}
\end{figure}

\begin{figure*}
\begin{center}
\includegraphics[width=15cm]{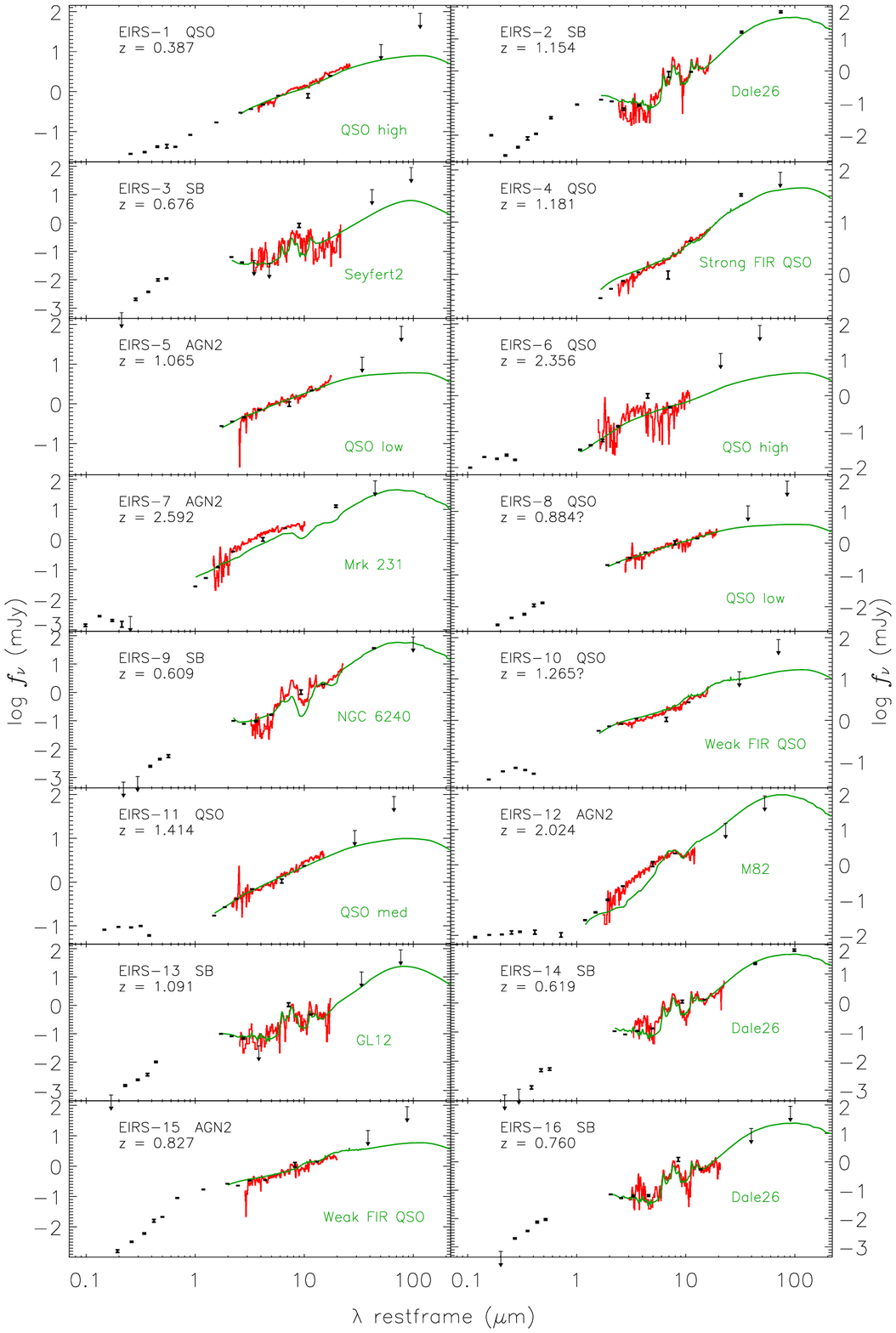}
\end{center}
\caption[Optical+IR SED]{Full optical to FIR SED of the ELAIS-IRS sources. 
The broken line represents the IRS spectrum smoothed using a boxcar filter 5 resolution
elements wide, while error bars and arrows represent photometric measurements in
the optical ($U$, $g$, $r$, $i$, $Z$), 
NIR ($J$, $K$), IRAC (3.6, 4.5, 5.8, 8.0 \uu), ISOCAM (15 \uu) and MIPS (24, 70, 160 \uu).  
The smooth continuous line represents the best fit template SED for the IR photometry.
\label{FullSED}}
\end{figure*}

\addtocounter{figure}{-1}
\begin{figure*}
\begin{center}
\includegraphics[width=15cm]{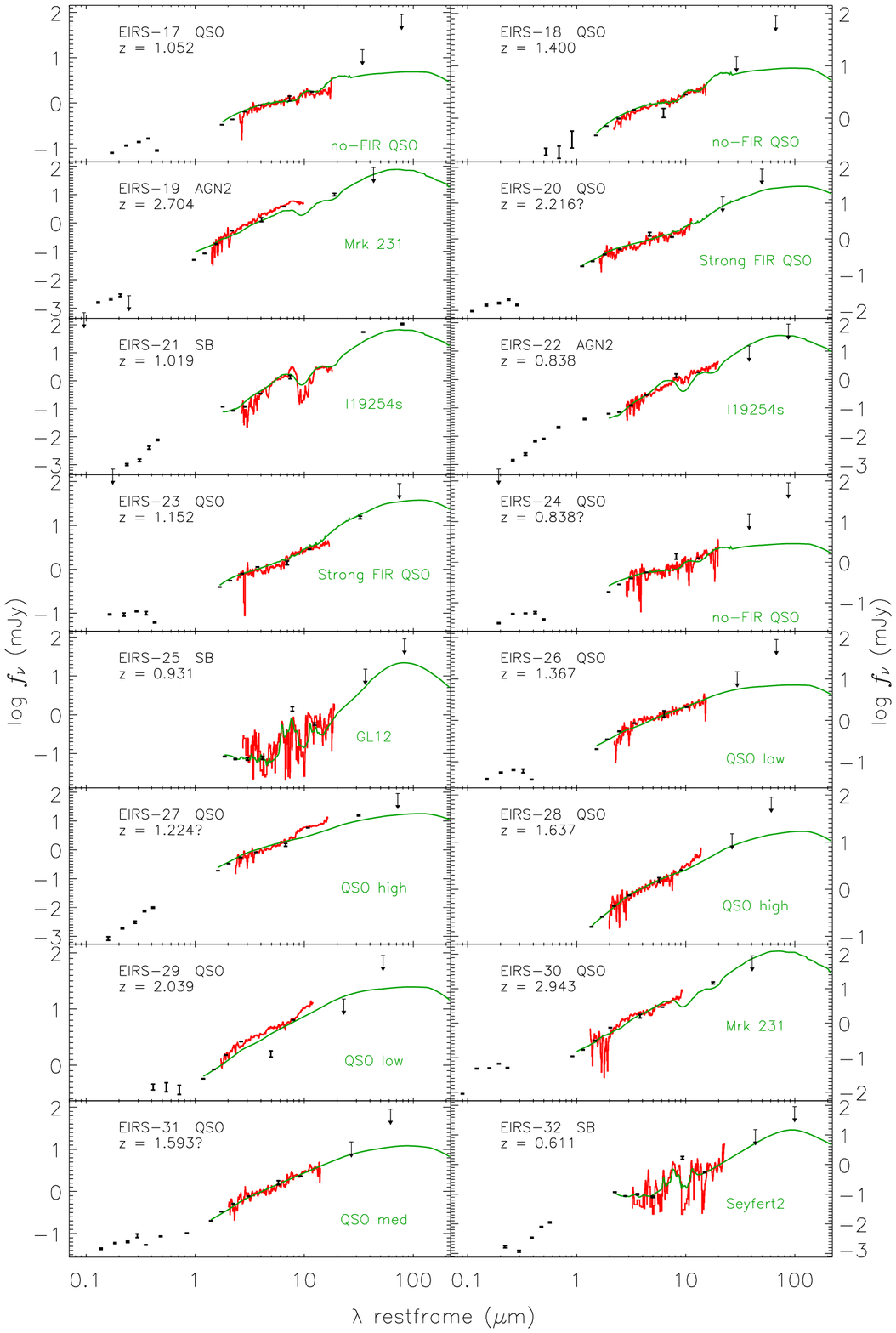}
\end{center}
\caption{Continued.}
\end{figure*}

\addtocounter{figure}{-1}
\begin{figure*}
\begin{center}
\includegraphics[width=15cm]{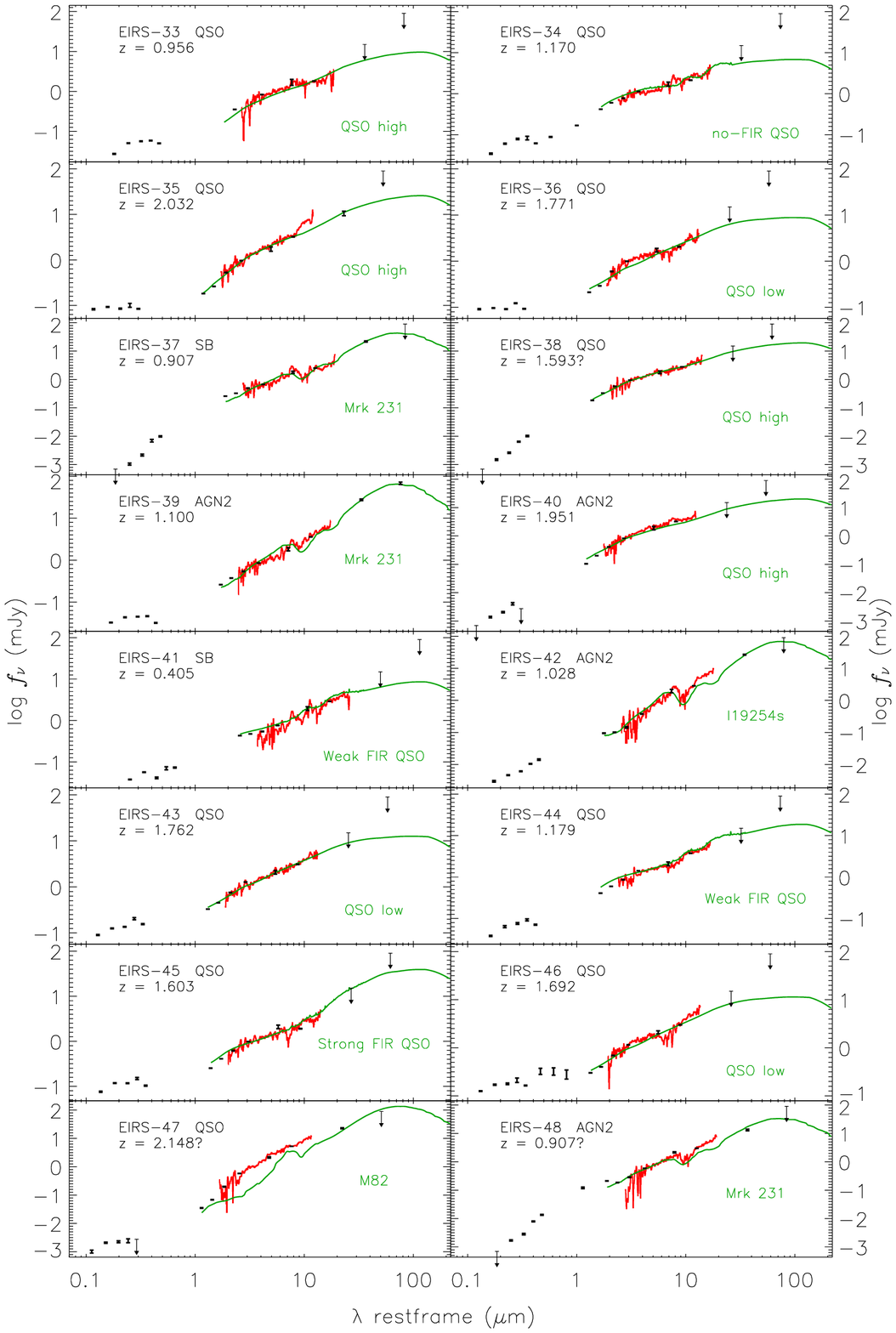}
\end{center}
\caption{Continued.}
\end{figure*}

\addtocounter{figure}{-1}
\begin{figure*}
\begin{center}
\includegraphics[width=15cm]{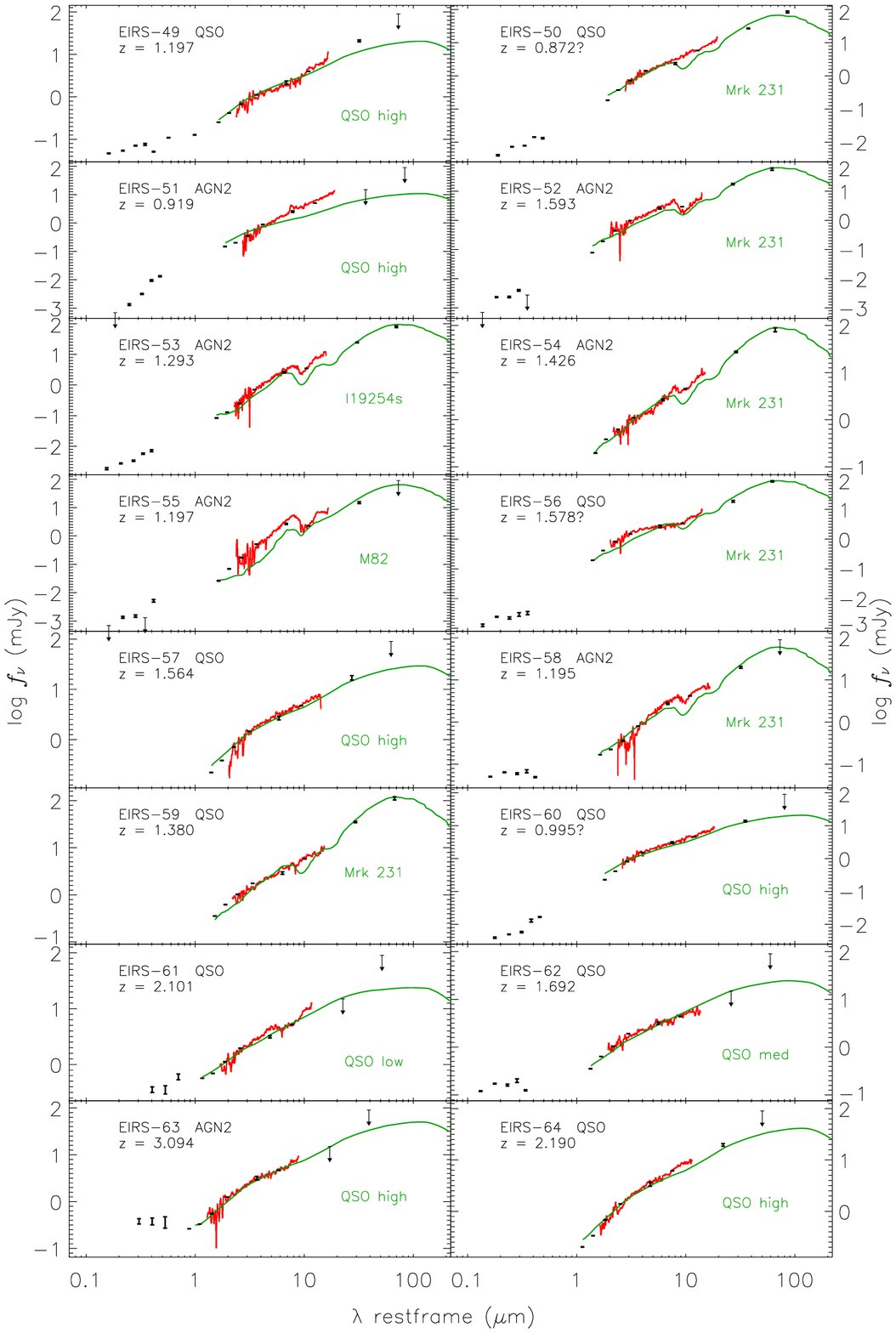}
\end{center}
\caption{Continued.}
\end{figure*}

\addtocounter{figure}{-1}
\begin{figure*}
\begin{center}
\includegraphics[width=15cm]{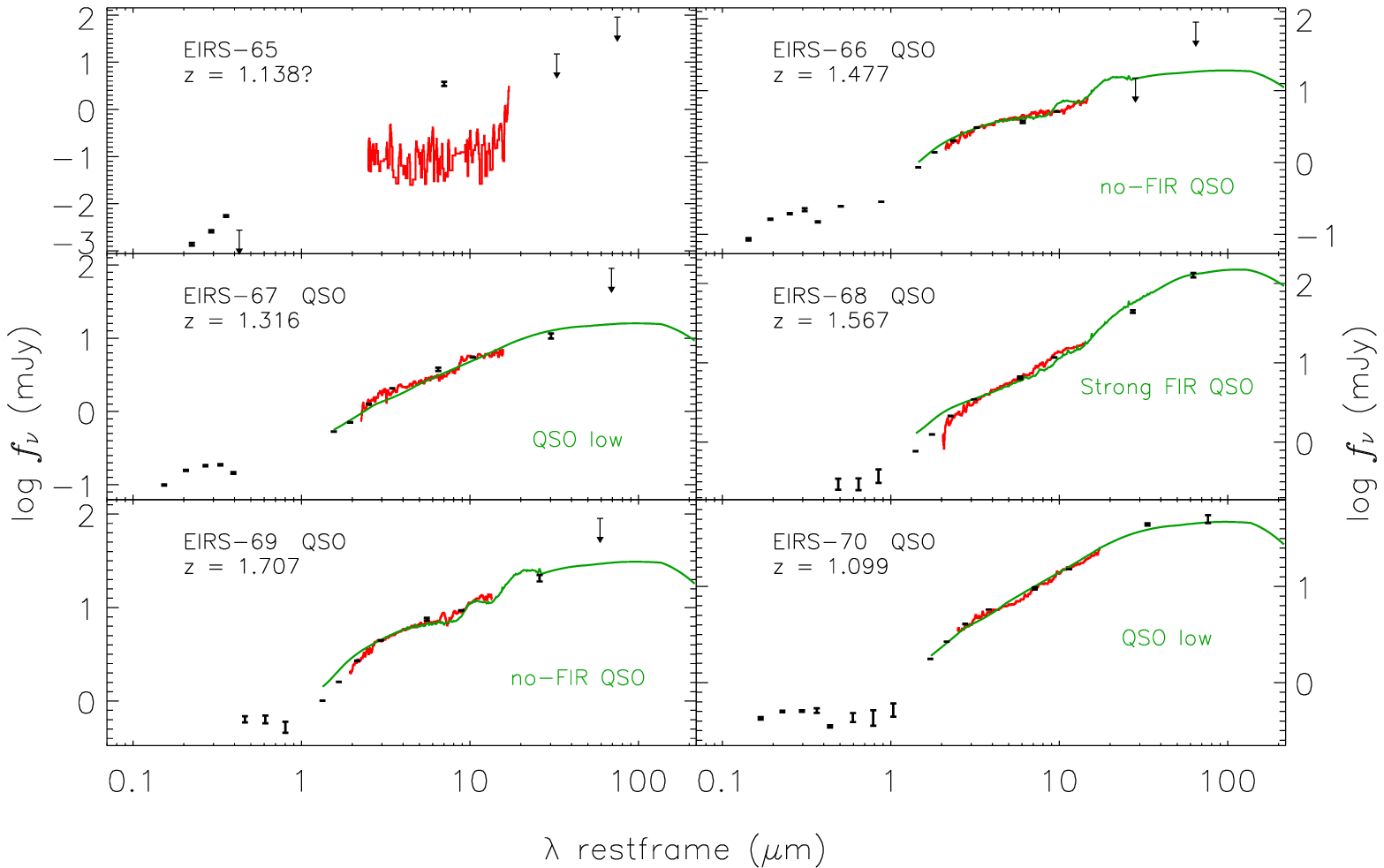}
\end{center}
\caption{Continued.}
\end{figure*}

IR luminosity (\lir) is estimated by integrating in the 8--1000 \um
range the template SED that best fits the MIR and far-infrared (FIR) photometry of each source.
Template SEDs are selected from well-known theoretical and semi-empirical models in the literature,
including four FIR models (torus, cirrus, M82 and Arp220) from \citet{Rowan-Robinson01};
one early-type, seven late-types, three starbursts, six AGN and three SB+AGN composites from   
\citet{Polletta07}; 
a set of 64 theoretical galaxy SEDs (from quiescent to active) from \citet{Dale02};  theoretical
models of normal and starburst galaxies as a function of \lir\ from \citet{Lagache03,Lagache04}; 
two empirical mean SEDs (radio-loud and radio-quiet) from a sample of 47 QSOs \citep{Elvis94};
and three mean SEDs of optically-selected QSOs with strong, weak or no detection in the FIR \citep{Netzer07}.

To reduce the number of template SEDs, the
best-fitting template for each source in the Library was obtained by fitting its mid-to-far IR photometry 
(12--850 \uu). Template SEDs that provided the best-fitting solution for less than two sources were discarded, 
and from very similar ones
\citep[e.g. Dale25 and Dale26 from][]{Dale02} only one of them was selected. This reduced the 
sample of templates to 14 models (see Fig. \ref{SEDtemplates}): M82 and Arp220 from \citet{Rowan-Robinson01}; 
GL12 from \citet{Lagache03,Lagache04}; 
Dale26 from \citet{Dale02}; NGC6240, Mrk231, I19254s, Seyfert2, QSO-norm, QSO-high
and QSO-low from \citet{Polletta07}; Strong FIR QSO, Weak FIR QSO and no-FIR QSO from \citep{Netzer07}. 
 
The best-fitting template for the ELAIS-IRS sources is determined by fitting the IRAC 3.6, 4.5, 5.8 and 8.0 \uu, 
ISOCAM 15 \uu, and MIPS 24, 70 and 160 \um photometry, 
as well as synthetic photometry in the 10--38 \um range obtained from
the IRS spectrum. The synthetic filters have Gaussian transmission profiles centred at $\lambda_c$ = 
10.4, 12.4, 14.9, 17.9, 21.5, 25.8 and 30.9 \uu, with FWHM = $\lambda_c$/6.
The photometric bands shortwards of 3.6 \um are discarded because at 
$z$ $\gtrsim$ 1 the flux in these bands is dominated by the stellar population in starbursts and obscured
AGN, or by the accretion disc in QSOs, and is poorly
correlated to the FIR emission from dust. Fitting results are shown in Fig. \ref{FullSED}.
The fits look reasonable for most sources, except for a significant under estimation of the MIR emission
in some obscured AGN with a steep slope in the IRAC bands (e.g. EIRS-12, EIRS-47, EIRS-55). These sources
fit an M82 template, probably because of the lack of a suitable obscured-AGN template.  

\begin{figure} 
\begin{center}
\includegraphics[width=8.5cm]{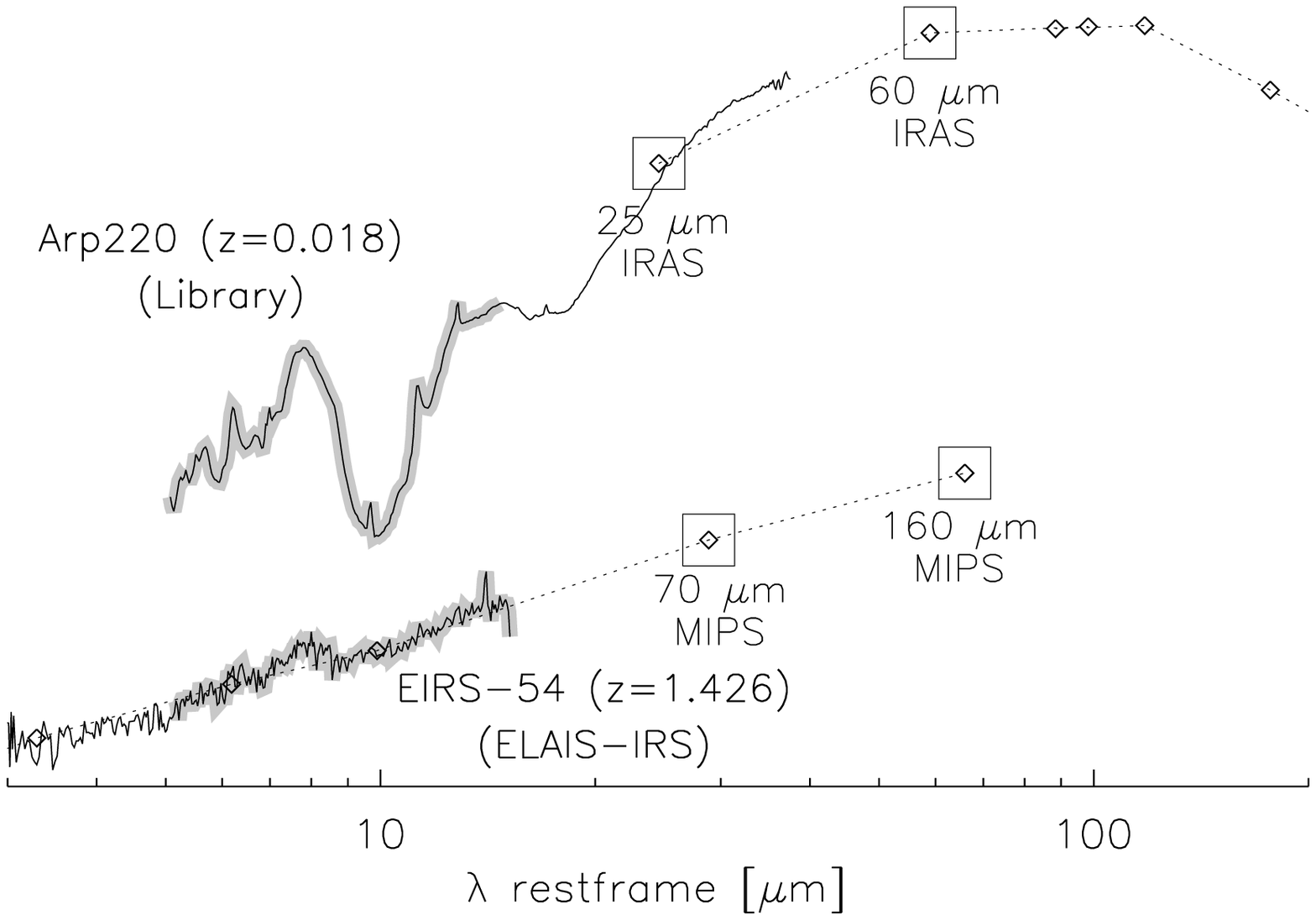}
\end{center}\caption[Correspondence between bands]{Correspondence between photometric bands used for IR luminosity 
estimations in the Library and ELAIS-IRS sources.
For ELAIS-IRS sources at $z$ $\sim$ 1.5 the MIPS70 and MIPS160 bands roughly correspond to IRAS 25 
and 60 \um photometry in local
ULIRGs. The restframe spectral range common for both samples in IRS spectroscopy (5--15 \uu) is 
highlighted.\label{correspondencia_bandas}}
\end{figure}

If we assume that the FIR SED of the ELAIS-IRS sources is not very different to that
found in the sources in the Library, the uncertainties in \lir\ for ELAIS-IRS can be estimated 
using the following procedure:\\ 
1) calculate, for the sources in the Library, the best-fitting SED
and its associated IR luminosity (\lira), using the IR photometry up to 15, 25 or 60 \um (which
roughly corresponds to 38, 70 and 160 \um for sources at $z$ $\sim$ 1.5, see Fig. 
\ref{correspondencia_bandas}).\\ 
2) compare \lira to the `true' IR luminosity (\lir) obtained by fitting all the FIR and sub-mm
photometry available.

The mean value of the logarithmic error in \lir, \mbox{$\Delta$log(\lir) = $\mid$log(\lir) - log(\lira)$\mid$,}
is around 0.05 when \lira is estimated using all the available photometry up to 60 \um (equivalent to detection
up to 160 \um in a $z\sim1.5$ ELAIS-IRS source), 0.15  when using the photometry up to 25 \um
(equivalent to 70 \um detected ELAIS-IRS sources), and 0.23 if only photometry up to 15 \um is used (equivalent to ELAIS-IRS
sources undetected at 70 and 160 \uu). 

In addition to these errors, uncertainty in the MIPS70 and MIPS160 photometry (up to 
50 per cent in sources near the detection limit) and the bias introduced by the use of a discrete set of
templates must also be taken into account. Considering these extra error sources, a factor 2--3 of uncertainty 
seems reasonable for most 
sources, significantly lower than the factor 5--10 estimated for extrapolations using only the MIPS24 photometry
\citep{Dale05}.

\begin{figure} 
\begin{center}
\includegraphics[width=8.5cm]{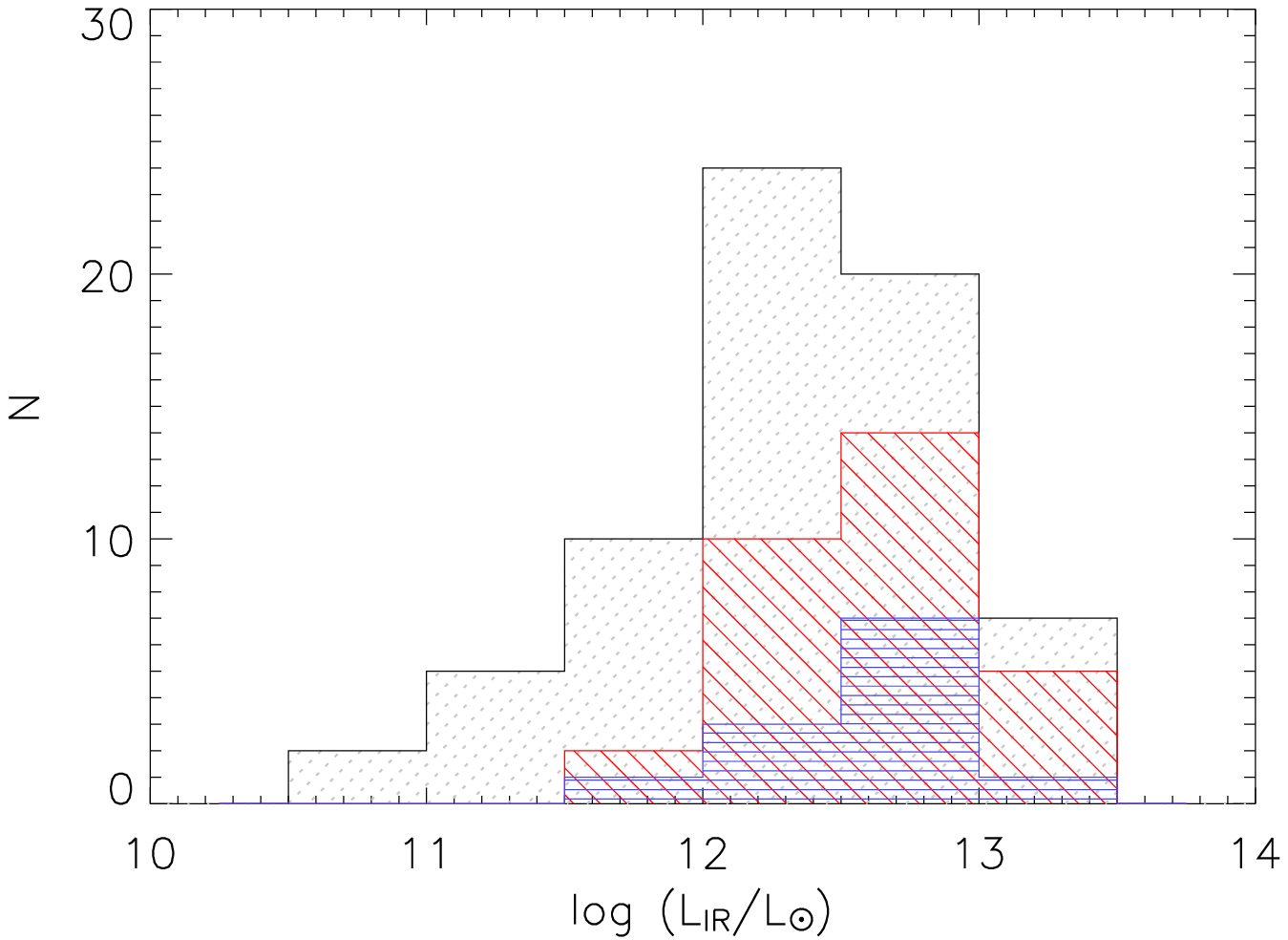}
\end{center}\caption[IR luminosities]{
Histogram of \lir for the ELAIS-IRS sources as calculated from SED fitting.
The dot-filled histogram represents the whole sample, while those of the subsamples with MIPS70
and MIPS160 detection are drawn in diagonal and horizontal lines, respectively.\label{histograma_LumIR}}
\end{figure}

Most ELAIS-IRS sources have estimated IR luminosities in the ULIRG range, with some extreme cases in the Hyper-luminous
range (\lir $>$ 10$^{13}$ \lsun) and 1/4 of the sample in that of LIRGs. 
Fig. \ref{histograma_LumIR} shows a histogram of the 
luminosity distribution of the sample. It is noteworthy that the sources detected at 70 and 160 \um
concentrate in the higher luminosity tail of the distribution. This is an expected outcome given the relatively
shallow depth of the MIPS70 and MIPS160 observations, and as a consequence, the most accurate \lir\ 
estimations are obtained in the most luminous sources. Table \ref{LumIR_SFR} indicates the best-fitting SED and
estimated IR luminosity for the ELAIS-IRS sources.

\subsection{Star Formation Rates}

For sources dominated by star formation, the most robust measurement of the star formation rate is obtained
from the integrated 8--1000 \um IR luminosity (\lir). For starburst galaxies, \citet{Kennicutt98} finds the relation:
\begin{equation}
SFR\ [\rm{M_{\odot} yr^{-1}}]\ =\ 1.72 \times 10^{-10}\ \textit{L}_{\rm{IR}}\ [\rm{L_\odot}]\\
\label{SFR_kennicutt}
\end{equation} 

Unfortunately, \lir\ has big uncertainties for individual sources, and in sources with AGN activity
there is not a straightforward method to isolate the starburst and AGN contributions to \lir. 

Many recent papers find a strong correlation between the luminosity of the PAH bands ($L_{\rm{PAH}}$) and \lir\
in starburst-dominated sources \citep[e.g.][]{Rigopoulou99,Peeters04,Brandl06}.
The ratio $L_{\rm{PAH}}$/\lir\ depends on the properties of the interstellar medium, and thus it varies with the
luminosity of the sources: in normal spiral galaxies, with quiescent star formation, the mean ratio is highest
\citep{Smith07}, in starburst galaxies it is somewhat lower, and in local ULIRGs with massive star formation,
it is lowest \citep{Rigopoulou99,Lu03,Netzer07}. 
Nevertheless, the dependency of $L_{\rm{PAH}}$/\lir\ with \lir\ is small relative to the dispersion between
sources of the same luminosity: \citet{Rigopoulou99} find for the 7.7-\um PAH a mean value of 
8.1 $\times$ 10$^{-3}$ in starburst galaxies and 5.5 $\times$ 10$^{-3}$ in starburst-dominated ULIRGs.
\citet{Brandl06} find that the 6.2-\um PAH luminosity relates to \lir\ trough a power-law of index
1.13 in a sample spanning 2 orders of magnitude in \lir.

Some authors use the 6.2-\um PAH because its measurement suffers little contamination from 
other MIR features \citep{Peeters04,Brandl06} while others prefer the 7.7-\um PAH because it is the 
strongest PAH band \citep{Rigopoulou99,Lutz03}, or even use the combined flux of two or more features 
\citep[e.g.][]{Lu03,Farrah08}. 

The estimated luminosity in the PAH features depends strongly on the measurement procedure: \citet{Smith07}
find discrepancies of a factor 1.7 at 6.2 \um and 3.5 at 7.7 \um depending on whether the continuum is
calculated by spectral decomposition into Lorentzian+continuum or it is interpolated between adjacent
regions.
This suggests that the discrepancies in the value of $L_{\rm{PAH}}$/\lir\ found by different
authors (see Table \ref{tabla_LPAH_LIR}) are mainly due to systematics.
In addition, the integrated infrared luminosity is not the same throughout the literature:
some authors define \lir\ in the range 1--1000 \um \citep[e.g.][]{Farrah08}, while others use only the
FIR (42--122 \uu, \citet{Lu03}; 40--500 \uu, \citet{Peeters04}). 

Because of these issues, we obtain our own estimate of the average $L_{\rm{PAH}}$/\lir\ in the
starburst-dominated ELAIS-IRS sources before using the PAH luminosity to estimate SFR in the
whole sample.
The mean values of $L_{\rm{PAH}}$/\lir\ found in the 7 starburst-dominated sources with no
sign of significant AGN emission (EIRS-21, 25, 37 and 41 excluded) are:
$L_{\rm{62}}$/\lir\ = 0.012 $\pm$ 0.007,
$L_{\rm{77}}$/\lir = 0.038 $\pm$ 0.016 and 
$L_{\rm{113}}$/\lir = 0.011 $\pm$ 0.007.

The ratio found for the 7.7-\um PAH is similar to that obtained by \citet{Lutz03} in a sample of
starburst galaxies and \citet{Smith07} in normal and starburst galaxies. It agrees with that found
by \citet{Rigopoulou99} in starburst galaxies and ULIRGs if we take into account the factor 3.5 
decrement calculated by \citet{Smith07} for fluxes measured by interpolation of the continuum using splines. 
For the 6.2-\um feature, our ratio is in agreement with those of \citet{Smith07} and \citet{Farrah08}
but it is roughly double the ratio found by \citet{Spoon04} in a sample of normal and starburst galaxies
even after correction for the differences in the continuum estimation procedure.

\begin{figure} 
\begin{center}
\includegraphics[width=8.5cm]{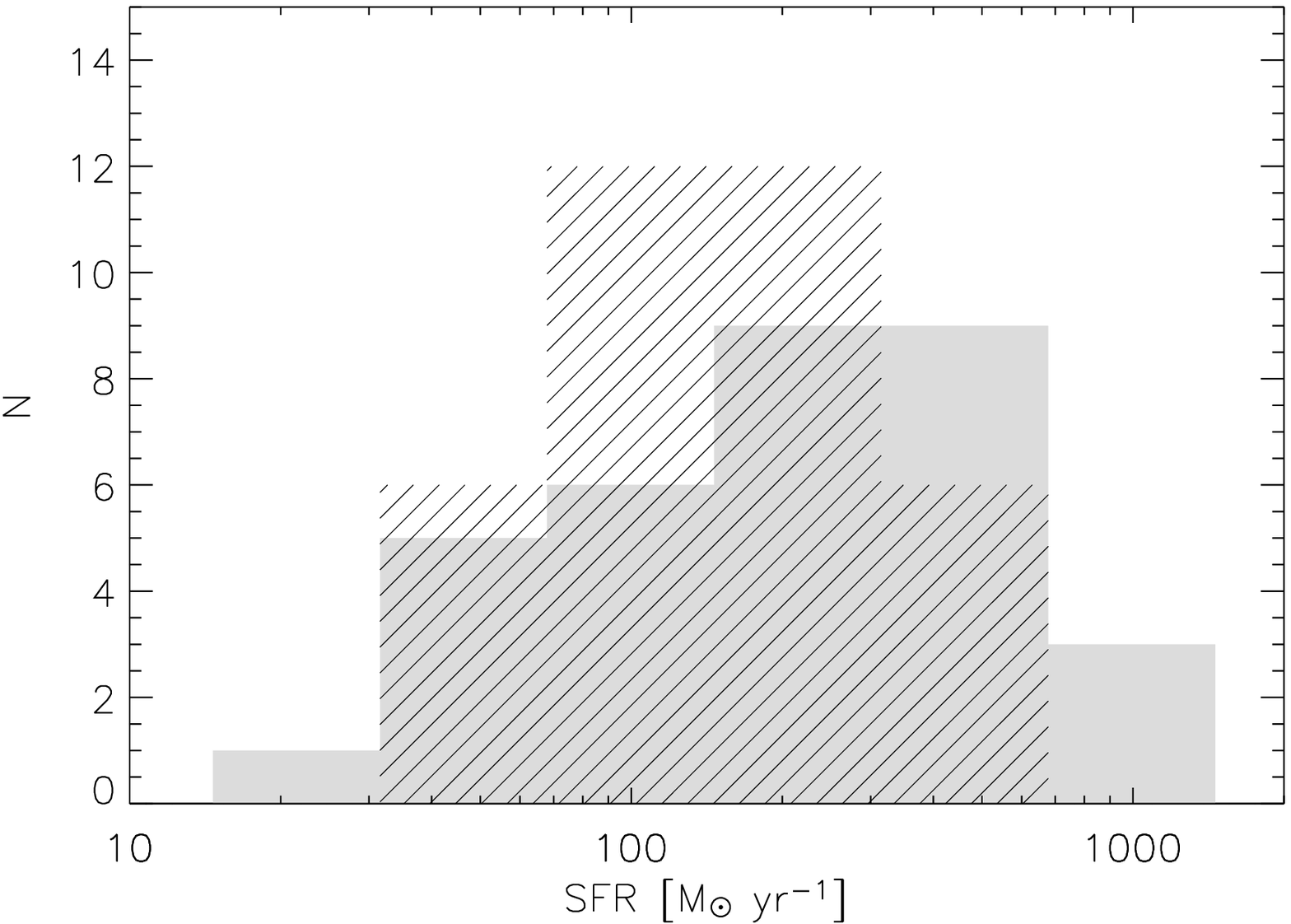}
\end{center}
\caption[SFR Histogram]{Histogram of star formation rates as estimated from the luminosity
of the 6.2, 7.7 and 11.3 \um PAH features in individual ELAIS-IRS sources. The solid histogram
represent SFR estimates in sources where at least two PAH features are detected,
while the stripped histogram indicates upper limits for sources with detection in just one or none
of the PAH bands.\label{histograma_SFR}}
\end{figure}

By applying the calibration in Eq. \ref{SFR_kennicutt}, and assuming that the properties of the interstellar
medium in which the PAH emission originates are not significantly altered by the AGN activity, the SFR in
the whole sample can be calculated as:
\begin{eqnarray}
SFR\ [\rm{M_{\odot} yr^{-1}}] & = & 1.40 \times 10^{-8}\ L_{\rm{62}}\ [\rm{L_\odot}]\label{SFR_eq1}\\
SFR\ [\rm{M_{\odot} yr^{-1}}] & = & 4.56 \times 10^{-9}\ L_{\rm{77}}\ [\rm{L_\odot}]\label{SFR_eq2}\\
SFR\ [\rm{M_{\odot} yr^{-1}}] & = & 1.52 \times 10^{-9}\ L_{\rm{113}}\ [\rm{L_\odot}]\label{SFR_eq3}
\end{eqnarray} 

For sources in which two or three PAH features are detected with S/N $>$ 2 
we estimate SFR from their combined luminosity, while if only one or no PAH features
are detected, we average estimates from individual upper limits.

Table \ref{LumIR_SFR} shows the estimated SFR for ELAIS-IRS sources. They range from 
$\sim$10 to $\sim$1000 \msunyr. The stated uncertainties account for
the uncertainty in the PAH flux only; if we consider the uncertainty in the IR luminosity and the dispersion
in $L_{\rm{PAH}}$/\lir\ of the starburst-dominated
sources, a more realistic estimate would be a factor 2 uncertainty in SFR.

Figure \ref{histograma_SFR} shows de distribution of SFR in the ELAIS-IRS sample. Most estimates are in the range
150--600 \msunyr, but if we consider upper limits for sources with undetected PAHs, which account
for more than half of the AGN population, many of them have probably SFR $<$ 100 \msunyr.

\section{Source Classification\label{clasificacion}} 

\subsection{Spectral Decomposition\label{decomposition}}

A powerful tool for diagnostics is decomposition 
into several spectral components, which can provide considerable insight to the physics of the 
sources \citep[e.g.][]{Tran01}. 

MIR emission from active galaxies arises mostly from HII regions, Photodissociation Regions (PDRs) and
AGN \citep{Laurent00}, so we have used a simple model comprising the 
superposition of three spectral templates (AGN, PDR and HII) obscured by a screen of dust. 

The spectral decomposition is performed by fitting the 5--15 \um restframe range 
of the spectrum to a parametrized $F_{\lambda}(\lambda)$ of the form: 

\begin{equation}
F_{\lambda}(\lambda) = e^{-b \tau(\lambda)} \Big{(}a_{1} f_{\rm{AGN}}(\lambda) + 
a_{2} f_{\rm{HII}}(\lambda) + a_{3} f_{\rm{PDR}}(\lambda) \Big{)}
\end{equation}

\noindent where the free parameters ($b$, $a_1$, $a_2$, $a_3$) are calculated using a 
Levenberg--Marquardt $\chi^2$-minimization algorithm, and must have non-negative values.

To keep the number of free parameters low, a single optical depth value is applied to the 
three spectral components, but for most sources the solution is almost identical when two
optical depth values (one for the AGN and other for the PDR and HII components) are used,
with an increased $\chi^2$ value due to the extra free parameter. 

$\tau$($\lambda$) is obtained from the GC extinction law, while
$f_{\rm{AGN}}$ is represented by the IRS spectrum of the Seyfert 1 galaxy NGC 3515 \citep{Buchanan06},
$f_{\rm{PDR}}$ by the ISOCAM spectrum of a PDR in the reflection nebula NGC 7023 \citep{Cesarsky96a} and
$f_{\rm{HII}}$ by the ISOCAM spectrum of M17 in the vicinity of OB stars \citep{Cesarsky96b}.

\begin{figure} 
\begin{center}
\includegraphics[width=8.5cm]{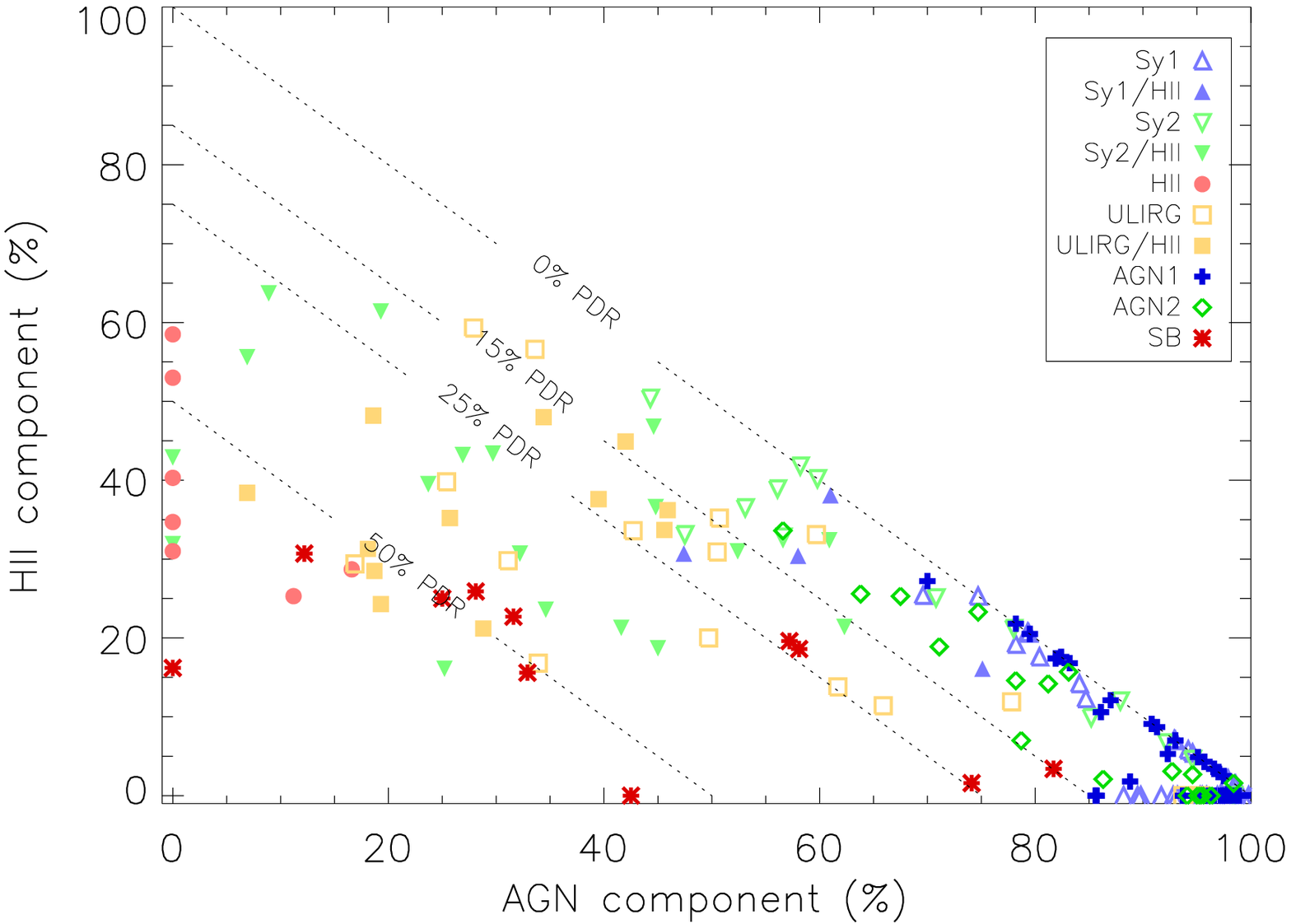}
\end{center}
\caption[Spectral Decomposition Diagram]{
Relative contribution to the integrated 5--15 \um restframe luminosity of the AGN and HII spectral 
components as derived from the spectral decomposition for sources in the Library and ELAIS-IRS
samples. Sources from the Library are plotted with different symbols according to their
optical-spectroscopic classification from NED: QSOs and Seyfert 1 galaxies (triangles), 
obscured quasars and Seyfert 2 galaxies (inverted triangles), 
starburst galaxies (circles), and ULIRGs (squares). Solid symbols indicate sources with a
strong starburst component, while open symbols represent mostly pure-AGN sources.
Symbols for the ELAIS-IRS sources indicate their MIR classification: unobscured AGN (plus signs), 
obscured AGN (diamonds) or starbursts (asterisks).
Dotted lines indicate the relative contribution of the PDR spectral component to the 5--15 \um
restframe luminosity.\label{descomposicion_espectral}}
\end{figure}

We quantify the contribution of each spectral component to the MIR spectrum as the ratio of its
integrated luminosity to total luminosity in the 5--15 \um restframe range, $r_{\rm{AGN}}$,
$r_{\rm{HII}}$ and $r_{\rm{PDR}}$ for the AGN, HII and PDR components.
Figure \ref{descomposicion_espectral} shows the distribution of spectral components for the 
Library and the ELAIS-IRS sample.

\subsection{Diagnostic Diagrams\label{diagnostico}}

A number of MIR diagnostic diagrams have been proposed to determine the nature of infrared galaxies
\citep[e.g.][]{Genzel98,Lutz98,Rigopoulou99,Laurent00,Peeters04,Spoon07}. 
They rely mainly on the strength of the PAH features to infer the dominant energy source 
(AGN or starburst), and on the
slope of the continuum and sign of the silicate strength (emission or absorption) to 
distinguish obscured and unobscured sources.
 
\begin{figure*}
\begin{center}
\includegraphics[width=8.7cm]{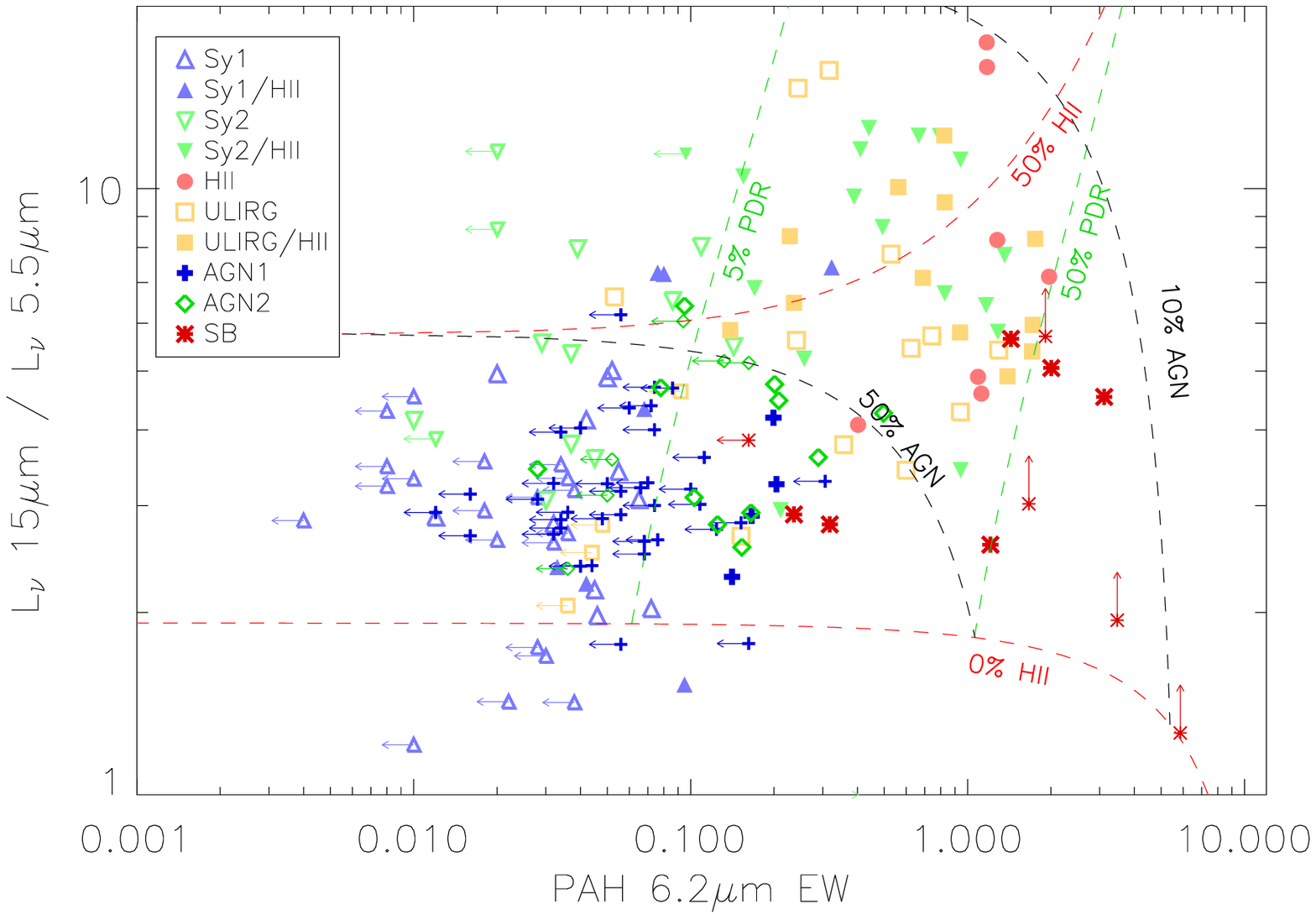}\hfill
\includegraphics[width=8.7cm]{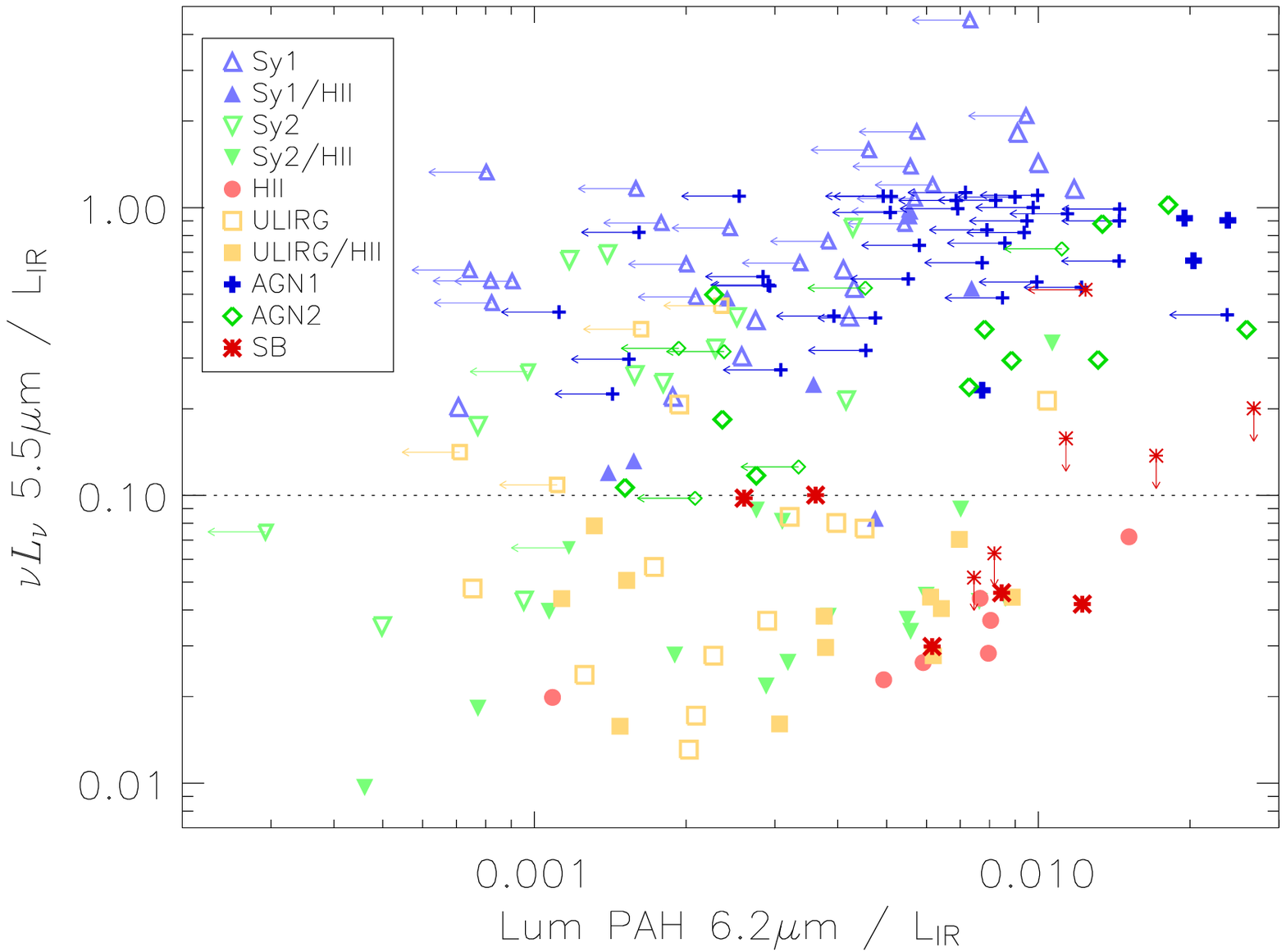}\\
\vspace{0.2cm}
\includegraphics[width=8.7cm]{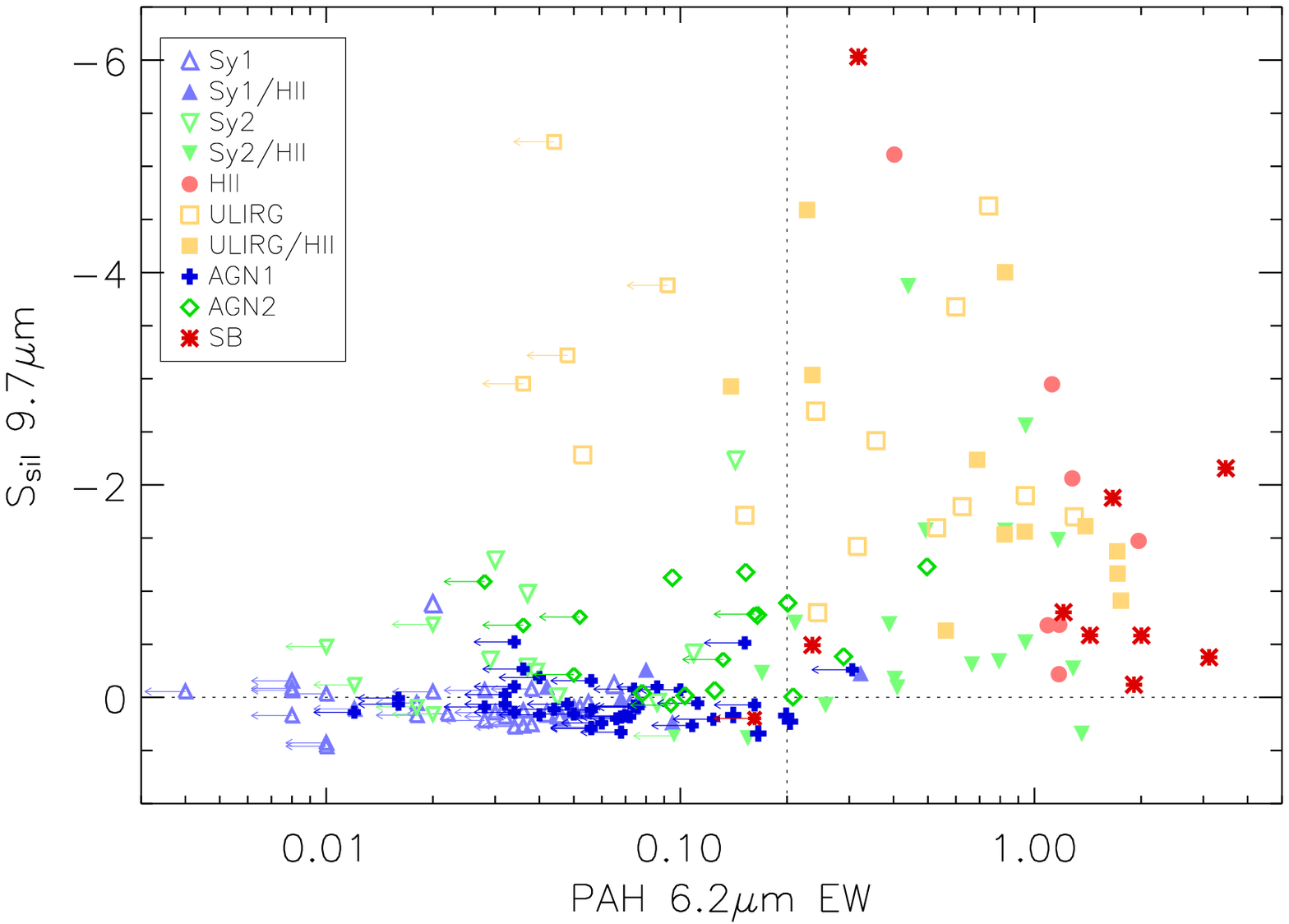}\hfill
\includegraphics[width=8.7cm]{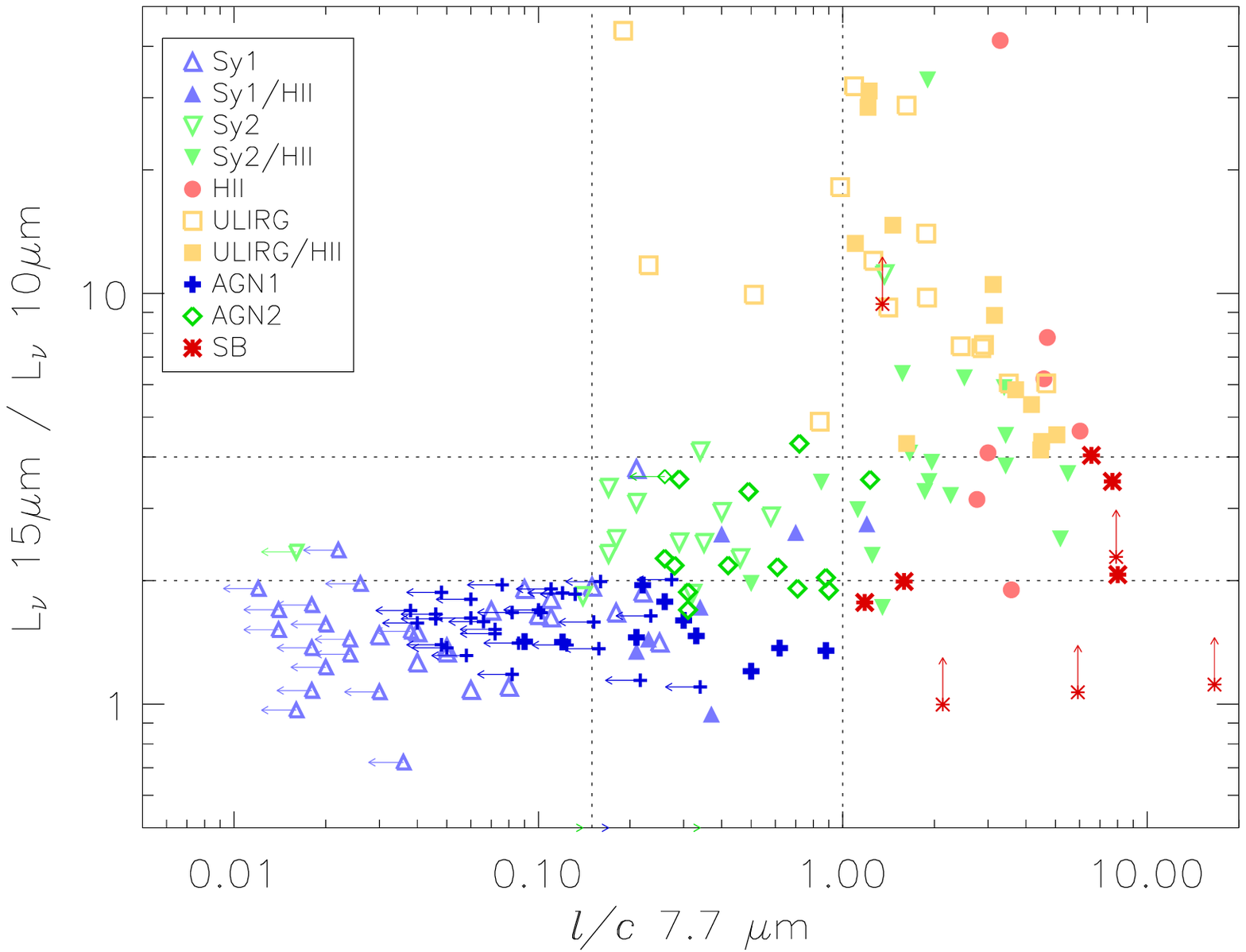}
\caption[Diagnostic Diagrams]{MIR diagnostic diagrams applied to sources in the Library and ELAIS-IRS,
adapted from \citet{Laurent00} (\textit{top left}), \citet{Peeters04} (\textit{top right}) and 
\citet{Spoon07} (\textit{bottom left}). A new diagnostic diagram based on the 7.7-\um PAH line-to-continuum
ratio and 15 \um to 10 \um luminosity ratio is also shown (\textit{bottom right}). Symbols are the same as
in Fig. \ref{descomposicion_espectral}. Dotted lines represent limiting values between populations
for parameters used as diagnostic criteria (see Table \ref{tabla_criterios}). Dashed lines in the Laurent
diagram represent mixing lines between the AGN (NGC 3515), HII (M17) and PDR (NGC 7023) templates used in 
the spectral decomposition discussed in \S\ref{decomposition}.\label{diagnosticos}}
\end{center}
\end{figure*}

There is a strong correlation between the optical and MIR spectrum of AGN: type 2 AGN usually show a red MIR spectrum, 
and the silicate feature in absorption, while most type 1 AGN show a bluer MIR spectrum, with stronger NIR emission,
and the silicate feature absent or in emission. Nevertheless, the match between 
populations is not perfect, and many mismatches (most notably, type 1 AGN with silicates 
in absorption and type 2 with silicates in emission) have been 
reported \citep[e.g.][]{Buchanan06,Brand07,Hao07}.

A silicate emission feature in type 2 AGN can be reproduced by a low-Si, low-tau smooth
torus model \citep{Fritz06}, or by silicate emission being produced in an extended region,
such as a dusty Narrow Line Region \citep{Sturm05}.
Silicate absorption in type 1 AGN can be observed if the Broad Line Region is partially obscured,
maybe by dust in the galaxy. A very remarkable case is Mrk 231 \citep{Lonsdale03b,Weedman05},
which shows a broad emission-line spectrum reddened with $A_V$ $\sim$ 2 in the
optical \citep{Boksenberg77,Lipari94}, and strong silicate absorption in the MIR \citep{Spoon02}.

Since making assumptions on the Seyfert type of AGN 
based on their IR properties alone would be misleading, we will refrain from doing that and 
will instead classify the AGN as obscured or unobscured in the IR.

Using some well known diagnostics along with some others from our own we 
separate the MIR spectra of the ELAIS-IRS sources into three main categories: starburst-dominated (SB), 
unobscured AGN-dominated (AGN1) and obscured AGN-dominated (AGN2). Every diagnostic criterion is 
calibrated using the known nature of sources 
in the low-redshift Library, including Seyfert type for AGN, and a final MIR classification
is obtained as the most repeated result among all criteria. 

Measurements on the ELAIS-IRS and Library spectra were obtained using exactly the same procedures, to
prevent distortions in the diagnostic diagrams due to systematics. Limits between populations
in the diagnostic diagrams are determined based solely on the Library spectra, and are 
fine-tuned to minimize the number of misclassifications within the Library. 

Figure \ref{diagnosticos} shows an adaptation of three well known diagnostic diagrams
by \citet{Laurent00}, \citet{Peeters04} and \citet{Spoon07}, as well as a new one that we propose.

In the diagnostic diagram of \citet{Laurent00} the 6.2-\um PAH equivalent width ($EW_{\rm{62}}$) easily separates
starburst-dominated sources from the AGN-dominated ones, 
while the ratio between 15 and 5.5 \um restframe spectral luminosities ($L_{\rm{15}}$/$L_{\rm{55}}$) 
distinguishes sources dominated by hot dust from an unobscured AGN (low $L_{\rm{15}}$/$L_{\rm{55}}$)
or cold dust by an obscured AGN or starburst (high $L_{\rm{15}}$/$L_{\rm{55}}$).
The value of $EW_{\rm{62}}$ that best separates starburst-dominated and AGN-dominated sources from the
Library is $\sim$ 0.2, while in $L_{\rm{15}}$/$L_{\rm{55}}$ we find too much overlap between populations
and prefer to drop it as a diagnostics criterion.

In the diagram from \citet{Peeters04} the ratio between $L_{\rm{55}}$ and \lir\ is related to the relative amount of
hot and cold dust in the galaxy, being higher for unobscured AGN and lower for starbursts and
deeply embedded sources. The 6.2-\um PAH luminosity ($L_{\rm{62}}$) to \lir\ ratio depends on the 
relative contribution of unobscured star formation to the IR luminosity, and is higher for unobscured
starbursts or AGN+SB composite sources, while lower for the dust enshrouded ones. Most ELAIS-IRS sources
have too high upper limits on this ratio to be meaningful, so we also drop this parameter as diagnostic
criterion. 

The diagram from \citet{Spoon07} is based on $EW_{\rm{62}}$ and the strength of the silicate
feature. Galaxies populate their diagram along two branches: one horizontal branch 
from AGN-dominated to starburst-dominated spectra, and one diagonal branch from highly obscured nuclei 
to starburst-dominated spectra. They hypothesize that the separation into two branches 
reflects two different configurations 
of the obscuring dust: clumpy in the horizontal and non-clumpy (homogeneous shell) in the diagonal.
In our adaptation of the Spoon diagram (Fig. \ref{diagnosticos}, bottom left) it is
noteworthy the lack of ELAIS-IRS sources in the high-extinction, weak-PAH region of the diagram, where
many local ULIRGs are located. 
The silicate strength is used to separate obscured and unobscured AGN, but most 
sources have a small absolute value of \ssil, and its uncertainty can lead to a wrong 
classification in some cases.

We propose a new diagnostic diagram, inspired by the one from Spoon but optimized for 
low S/N sources (Fig. \ref{diagnosticos}, bottom right). In our modified Spoon diagram, $EW_{\rm{62}}$ is replaced 
by the line (peak) to continuum ratio of the 
7.7-\um PAH feature ($l$/$c_{\rm{77}}$). This parameter has been extensively used to distinguish starbursts from
AGN \citep[e.g.][]{Lutz98,Rigopoulou99,Weedman06}, and works better in low S/N sources because the
7.7-\um feature is stronger than the one at 6.2 \uu. The silicate strength
is replaced by a more straight-forward and model-independent measurement: the ratio between $L_{\rm{15}}$ and
the spectral luminosity averaged in a narrow band centred at 10 \um restframe ($L_{\rm{10}}$). 
This ratio separates type 1 and 2 AGN from the Library more efficiently 
than \ssil\ does, because it combines the two main distinctive factors: absorption/emission in the silicate band
and slope of the continuum. In addition, it is less sensitive than \ssil\ to noise in the spectrum.
Its main drawback is that 15 \um restframe is unobservable by IRS for sources at $z$ $>$ 1.5, but it can
be interpolated using the MIPS70 photometry as we showed in \S\ref{medidas_continuo}. 
Type 1 and 2 AGN from the Library are better separated in this diagram
than in the original from Spoon in both axes, and only the type 1 AGN with a strong starburst contribution overlap significantly
with the type 2 ones. This overlap is produced by a higher slope of the continuum and depleted silicate feature 
in type 1 AGN with a starburst component.

Finally, the spectral decomposition diagram in Fig. \ref{descomposicion_espectral} can also be used for
diagnostics, since starburst-dominated sources locate near the left border, because
of strong PDR and/or HII component, while AGN-dominated sources concentrate in the lower-right corner.

From the five diagnostic diagrams described, we extract six criteria to classify 
infrared galaxies (Table \ref{tabla_criterios}).

Criteria C1 to C4 distinguish AGN from starbursts based on the strength of the PAH features and MIR-to-FIR
luminosity ratios, while C5 and C6 rely on the silicate feature to separate obscured and unobscured AGN.

The final class assigned to each source is the most repeated one among the six criteria.
This combination of multiple criteria allows for a more robust classification than that obtained using
a single diagnostic diagram, since it is less likely to be altered by statistical uncertainties. 
This is most significant in the lower S/N spectra, and also in obscured AGN,
which some criteria often classify as unobscured AGN or starbursts because 
typical values for obscured AGN lay in a narrow range between those of starbursts and unobscured AGN.
  
For sources in the Library, the IR classification is consistent with the optical class for 
more than 90 per cent of them (see Table \ref{resumen_diagnostico_biblioteca}).
An analysis of the results from individual criteria shows that mismatches are caused by two main factors:
a) strong star formation in type 1 AGN, which leads to classification as obscured AGN via high values of
$l$/$c_{\rm{77}}$ and $L_{\rm{15}}$/$L_{\rm{10}}$. b) silicate emission in type 2 AGN, 
which confuses the two criteria based on the silicate feature.

In the ELAIS-IRS sample we expect a somewhat higher mismatch rate because the lower S/N of the
spectra increases the likelyhood of incorrect classification by some of the individual criteria.
In all diagnostic diagrams the ELAIS-IRS sources
are plotted with symbols representing their final IR class, so that mismatches between individual 
criteria and the final classification can be easily recognized.

Based on the criteria presented above, 41 sources are classified as unobscured AGN (AGN1), and 17 are 
obscured AGN (AGN2). The remaining 11 sources are classified as starbursts (SB), including
a few sources with strong continuum (EIRS-21, EIRS-25, EIRS-37), which could host an obscured AGN,
and an AGN1+SB composite (EIRS-41). 
Table \ref{tabla_diagnostico} shows the classification from individual criteria as well as the 
final IR class for the whole ELAIS-IRS sample. 

A comparison of the optical versus IR classification for ELAIS-IRS sources 
(Table \ref{resumen_diagnostico_ELAIS-IRS}) shows 12 mismatches:
seven are optical galaxies with unreliable redshifts classified in the IR as unobscured
AGN because of weak or absent PAH features or presence of silicate emission.
Some of them (EIRS-8, EIRS-38) present a rather flat slope in the IRAC bands, indicating low extinction,
while others (EIRS-47, EIRS-50, EIRS-56, EIRS-60) show a steep slope and FIR emission, suggesting a high
level of extinction. The other galaxy, EIRS-27, shows a very strong silicate emission feature,
similar to that found in the Seyfert 2 galaxy IRAS F01475-0740 \citep{Buchanan06,Hao07}.
The remaining five sources are optical QSOs with reliable spectroscopic redshifts, which are classified as 
obscured AGN in the IR due to intense PAH emission and a small but negative value of \ssil. In these
sources, the dust in star-forming regions may have depleted the silicate emission feature in the AGN spectrum
enough to make it appear in absorption. 
In three of these sources (EIRS-5, EIRS-39 and EIRS-54)
the criterion in $L_{\rm{15}}$/$L_{\rm{10}}$ predicts AGN1 while \ssil\ $<$ 0. 
A fourth source, EIRS-63, is at $z$ = 3.094 and the silicate 
feature moves out of the range observed by IRS, so the IR diagnostic is inconclusive. Finally, EIRS-58 has a 
deep silicate feature and is classified as
AGN2 by both $L_{\rm{15}}$/$L_{\rm{10}}$ and \ssil, but its optical SED is blue and shows a power-law spectrum
in the optical and UV up to the Lyman break.

The best fitting IR SED and MIR classification are consistent with each other for most sources. All sources 
classified as unobscured AGN fit best to one of the seven QSO SEDs, except EIRS-47, which depends on a 
photometric redshift and is obscured in the optical.  
Obscured AGN also select mainly QSO templates (most significantly Mrk 231, a reddened type-1 QSO with 
strong silicate absorption), but 1/3 of them fit best M82 or I19254s (an obscured AGN + starburst composite).
Starburst sources favour pure-starburst (GL12 and Dale26) or composite (Seyfert 2, I19254s and NGC 6240) templates,
except for two sources with strong MIR continuum (EIRS-41 and EIRS-37) which fit best QSO templates.

\subsection{Averaged spectra} 
    
We have used the spectral classification in \S\ref{diagnostico} to calculate average spectra of each
population, which have better S/N than the spectra of individual sources.
In the average spectrum we only include sources with reliable redshift estimations, to 
prevent as much as possible the dilution of fine details by the dispersion in $z$.

To calculate the averaged spectrum, individual spectra are shifted to the 
rest-frame, normalized at 7 \um and interpolated in a common array of wavelengths uniformly 
distributed in log($\lambda$). For each value of $\lambda$ we discard the highest and lowest value of
$F_\lambda$ and the remaining ones are combined using a weighted average.

\begin{figure} 
\begin{center}
\includegraphics[width=8.5cm]{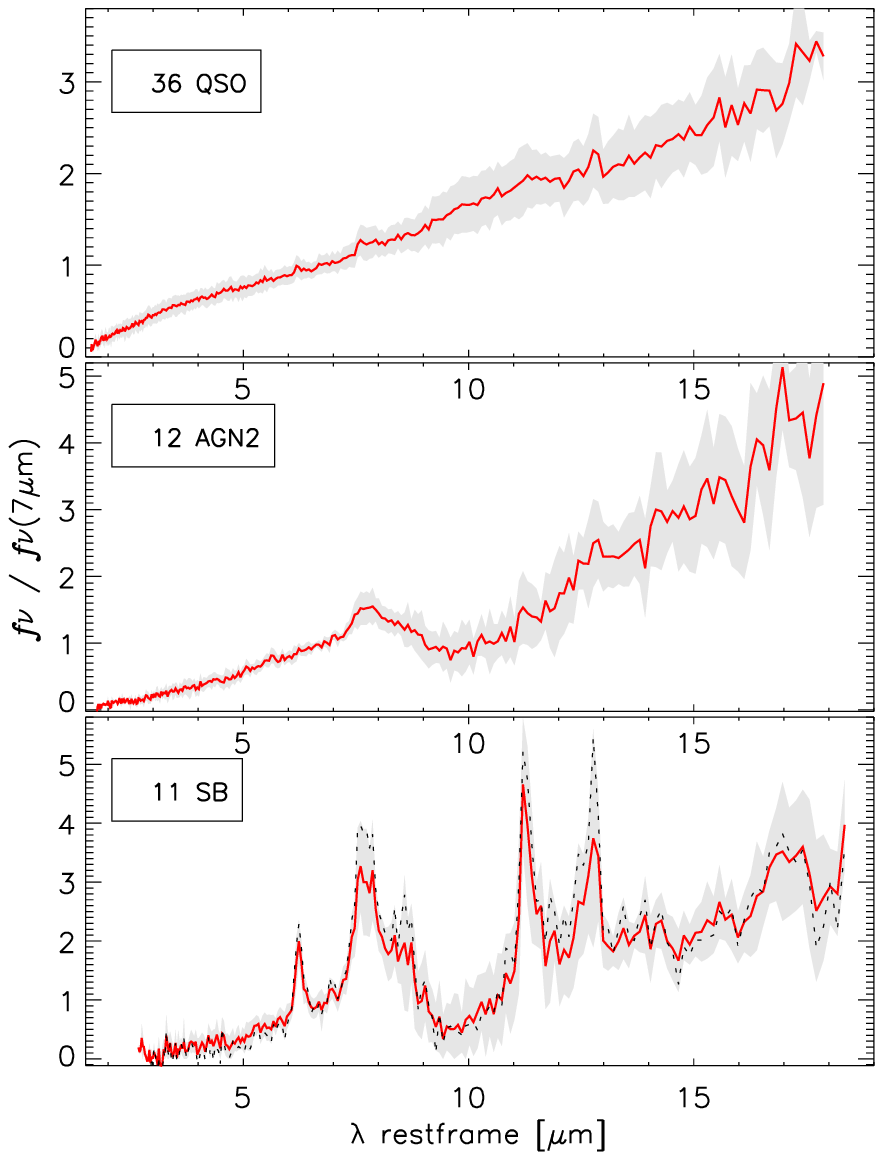}
\end{center}
\caption[Averaged spectra]{Composite spectrum of ELAIS-IRS sources separated based on their
IR classification: unobscured AGN (\textit{top}), obscured AGN (\textit{centre}) and starbursts 
(\textit{bottom}). The continuous line represents the average 
spectrum, while the shaded area represents the $1\sigma$ dispersion of individual spectra.
The dashed line in the bottom panel represents the average starburst spectrum after removal
of sources with a strong continuum component (EIRS-21, 25, 37 and 41).\label{espectros_promedio}}
\end{figure}

Figure \ref{espectros_promedio} shows the averaged spectra for starbursts, optical QSOs and 
obscured AGN (optical galaxies classified as AGN in the IR) in the sample.

\begin{figure}
\begin{center}
\includegraphics[width=8.5cm]{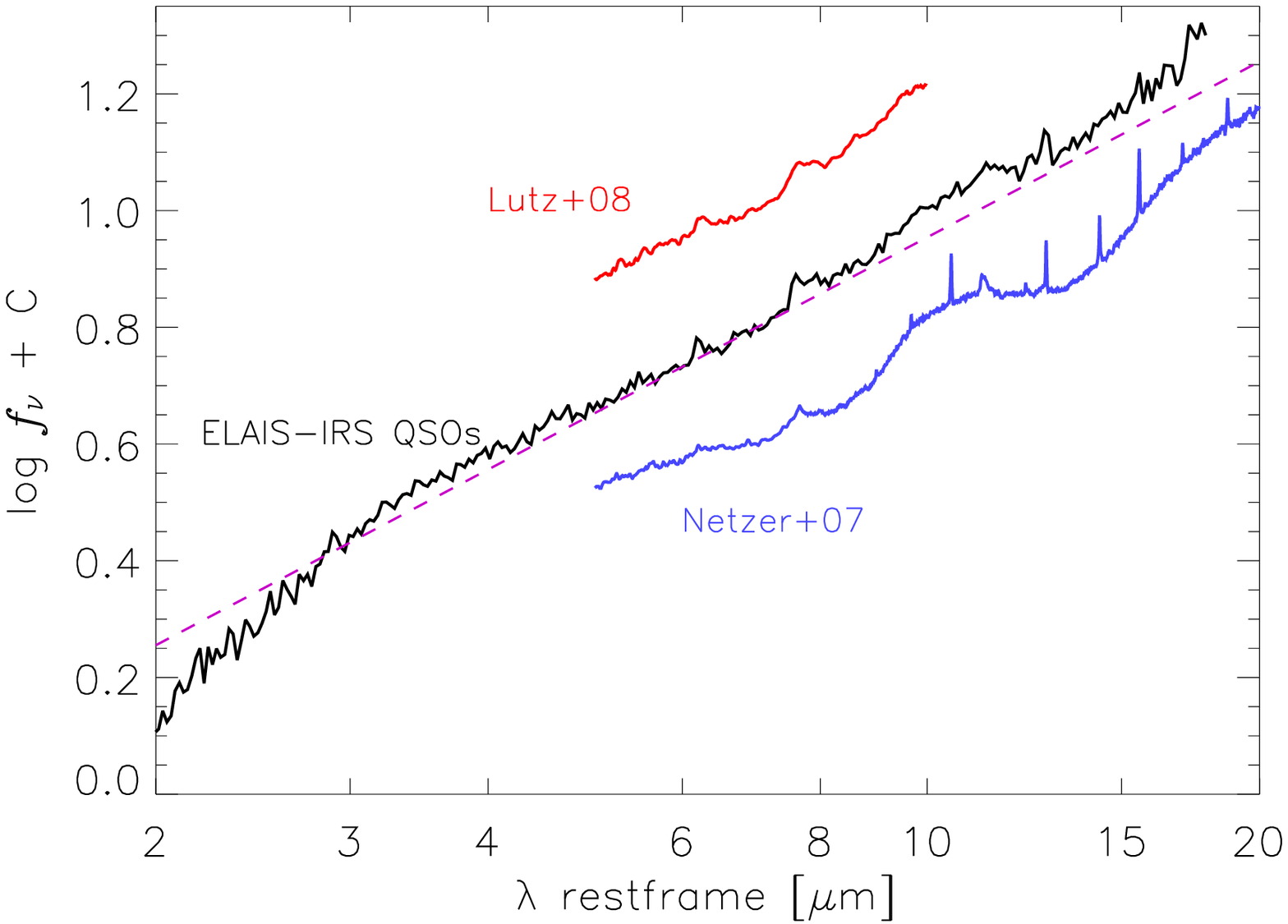}
\end{center}
\caption[Averaged spectra]{Composite spectrum of the ELAIS-IRS QSOs compared to 
the composite spectrum of a local sample of PG QSOs from \citet{Netzer07} (Netzer+07) and
a sample of 12 type 1 mm-selected QSOs at $z$ $\sim$ 2 from \citet{Lutz08} (Lutz+08).
A power law $f_\nu$ $\propto$ $\nu^{-\alpha}$ with index $\alpha = 1$ is shown for reference 
(dashed line).\label{espectro_QSOs}}
\end{figure}

The quasar spectrum can be approximated by a power-law of index $\alpha$ = -1
in the range 4--16 \um (Fig. \ref{espectro_QSOs}), which is slightly steeper than the index 
found in the averaged spectrum
of a sample of local QSOs \citep{Netzer07}. In addition, the silicate feature appears in
emission in the averaged spectrum, but with a very low profile. These findings suggest the
ELAIS-IRS quasars could be more obscured than average, nearby optically selected QSOs.
See \S\ref{geometry} for a discussion on the connection between the silicate feature and the 
MIR spectral index.

The slope of the spectrum increases shortwards of 3 \um towards the 1 \um minimum of the SED,
as has also been observed in other optically selected quasar samples 
\citep[e.g.][]{Hatziminaoglou05}. This is interpreted within the torus model as thermal emission 
from graphite grains at temperatures up to $\sim$1500 K.

In the averaged spectrum there are tentative detections of some ionic lines of neon,
 \neii 12.81 \um and \neiii 15.56 \uu, but since they have low/medium 
excitation potentials, their emission can also be excited by star formation.
The PAH features are weak but detected, with $EW_{\rm{62}}$ $\sim$ 0.015 and $EW_{\rm{77}}$ $\sim$ 0.05.
In \S\ref{SFR_in_AGN} we estimate mean PAH luminosities and star formation rates for these sources.

\begin{figure} 
\begin{center}
\includegraphics[width=8.5cm]{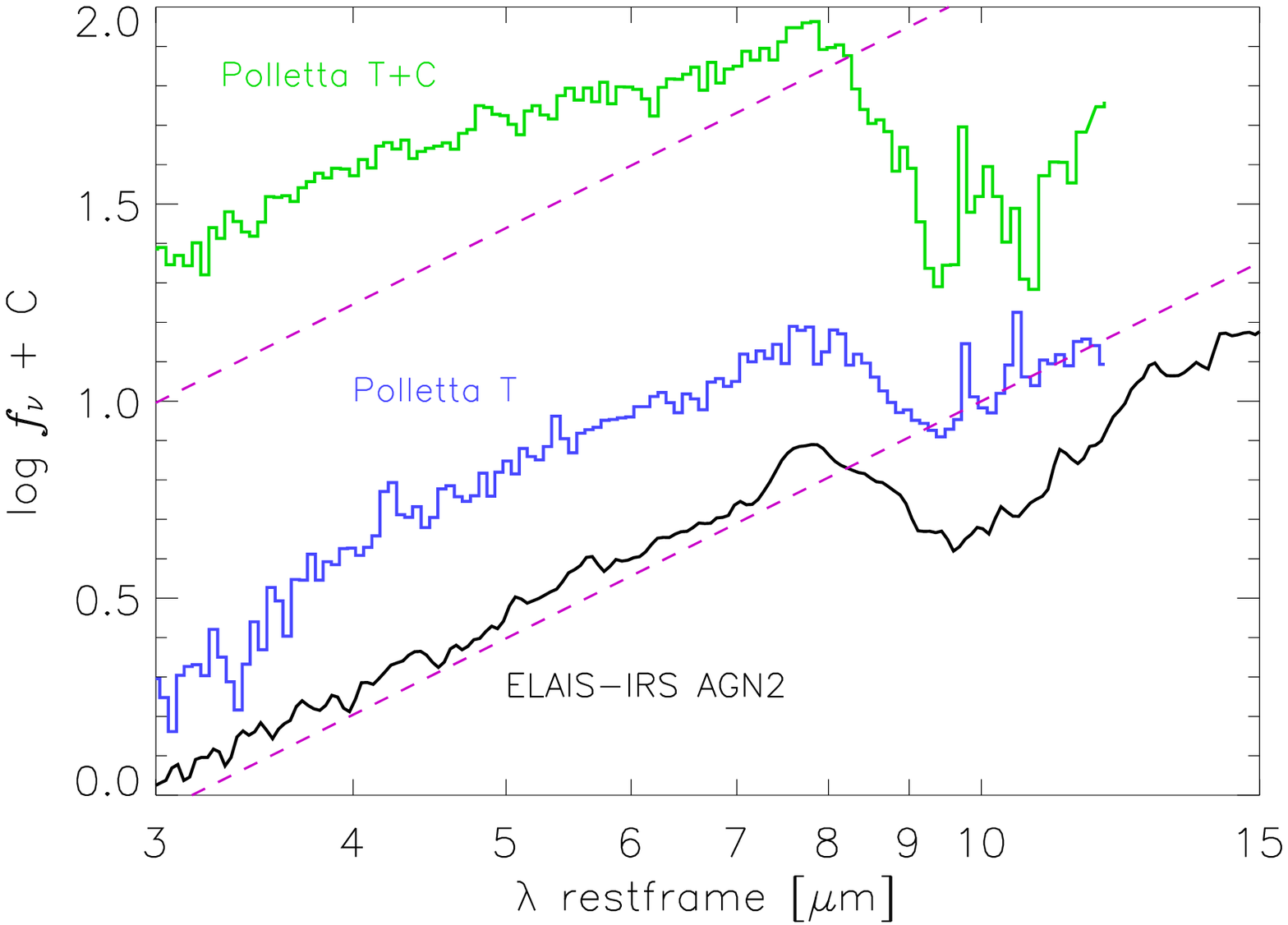}
\end{center}
\caption[Averaged spectra AGN2]{Composite spectrum for the ELAIS-IRS AGN2 sources compared to
the composite spectrum of X-ray selected, higher-luminosity QSO2s from \citet{Polletta08} in which the
NIR-to-MIR spectrum is best fitted by a torus model (Polletta T) or torus + cold absorber (Polletta T+C).
A power-law $f_\nu$ $\propto$ $\nu^{-\alpha}$ with index $\alpha = 2$ is shown for reference (dashed lines).
\label{AGN2_highz}}
\end{figure}

The averaged spectrum of the obscured AGN shows a steep slope at $\lambda$ $<$ 8 \um ($\alpha$ = -2) 
but it flattens longwards of the silicate trough at $\sim$10 \um (Fig. \ref{AGN2_highz}). The depth of the silicate
feature is relatively small (\tausil\ $\sim$ 0.7), as has been observed in many Seyfert 2 and
QSO2s \citep[e.g.][]{Sturm05,Weedman06,Deo07,Hao07}.  
However, the requirement of optical 
spectroscopic or photometric redshift for selection may have reduced the average silicate absorption by
biasing the sample against the most obscured sources. 

Emission lines for ionic species are very weak or absent, with the significant exception of \neii
12.81 \uu. The 7.7-\um PAH feature seems rather strong, but it could be magnified by a peak in the
underlying continuum emission at 8 \um caused by the onset of the silicate absorption feature. The other PAH
features are weak, with only the one at 11.3 \um clearly detected. 
The 6.2 \um feature is missing even if it is detected in many individual spectra. This may be 
caused by the cancellation effect produced by strong 6.0 \um absorption bands by water ice or 
hydrocarbons in some sources \citep{Sajina07}. 

In the averaged spectrum of the starburst-dominated galaxies we find, in addition to the usual PAH
bands at 6.2, 7.7, 8.6, 11.3 and 12.7 \uu, a few additional, fainter bands, centred at 5.2, 5.8,
14.2 and 17.0 \uu, which are also common in higher S/N spectra of low redshift galaxies.
The 3.3-\um band, relatively strong in ISO SWS spectra of many local ULIRGs, seems absent;
but the IRS spectra of the ELAIS-IRS starbursts are very noisy in this range.
  
Some ionic lines frequently observed in local starbursts also appear in the average spectrum,
most significantly \arii 6.99 \uu, \ariii 8.99 \uu, \neii 12.81 \um 
(which overlaps with the 12.7-\um PAH) and \neiii 15.56 \uu. It is noteworthy the absence
of rotational transitions of the H$_{\rm 2}$ molecule, with the possible exception of S(3) 9.67 \uu.
The weak continuum shortwards of 6 \um indicates that the relative contribution of
AGN or obscured star formation is small. If the four starburst-dominated sources with suspected or known 
AGN contribution are discarded, the average spectrum of the remaining starburst
sources (EIRS-2, 3, 9, 13, 14, 16 and 32) has a lower continuum emission at $\lambda$ $<$ 6 \uu, but 
remains unchanged at longer wavelengths while the PAH features get significantly boosted.

\section{Discussion}

\subsection{Comparison with other high-$z$ samples\label{otrasmuestras}}

Before Spitzer, the spectroscopic study of ULIRGs was restricted to a few
local sources due to the low sensitivity of previous instruments; but in the last 
four years, thanks to the unprecedented sensitivity of IRS, it has been possible to 
obtain MIR spectra for several hundred ULIRGs up to $z$ $\sim$ 3.  

Most of these sources were selected from wide-area MIR surveys,
and are usually heavily obscured, because only sources with high MIR-to-optical flux
ratios were selected for observation with IRS \citep[e.g.][]{Houck05,Yan07}. 
In some cases, additional constraints were imposed to favour the selection of
starburst or AGN-dominated sources \citep{Weedman06,Farrah08,Polletta08}.

By contrast, the selection criteria for the ELAIS-IRS sample put no constraints on the
MIR-to-optical flux ratio of the sources, but required optical spectroscopic or photometric
redshifts. This introduced a bias that is not easy to quantify, but certainly favored sources
with little or no obscuration at moderate redshifts. In addition, selection at 15 \um instead 
of 24 \um also favours lower-redshift sources, because the PAH features and the steep 
slope shortwards of 6 \um in obscured AGN enter the selection band at lower $z$.  
In this sense, the ELAIS-IRS sample can
be seen as complementary to other higher-$z$ spectroscopic samples.

\begin{figure} 
\begin{center}
\includegraphics[width=8.5cm]{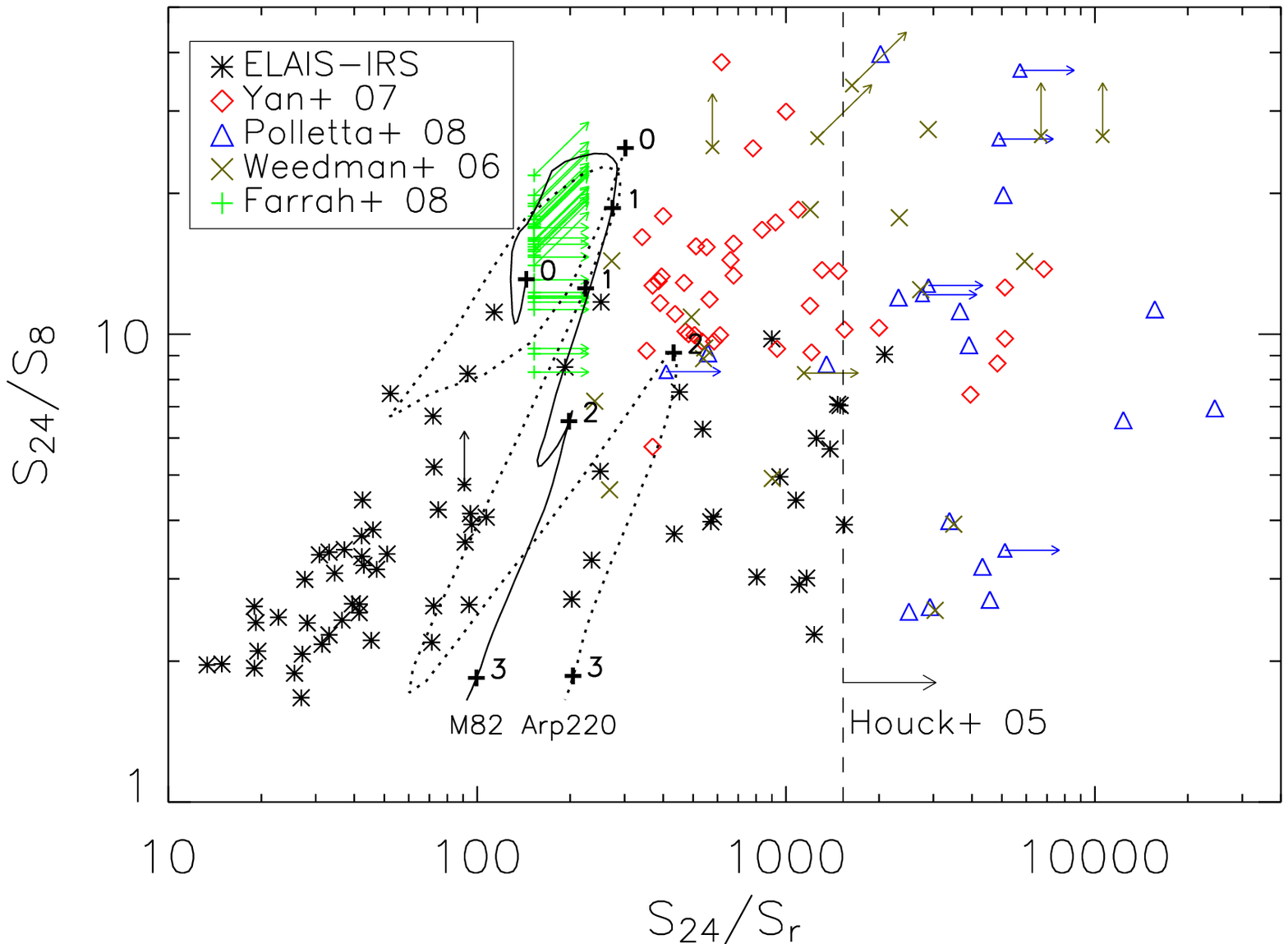}
\end{center}
\caption[Selection Criteria Other Samples]{Colour-colour diagram for the most representative high-$z$
ULIRG samples observed with the IRS: \citet{Weedman06} (x signs), \citet{Yan07} (diamonds),
\citet{Farrah08} (plus signs), \citet{Polletta08} (triangles) and ELAIS-IRS
(asterisks). The dashed vertical line indicates the lower limit in
$S\!_{\rm{24}}$/$S\!_r$ for the sample of \citet{Houck05}.
The continuous and dotted lines represent, respectively, tracks for M82 and Arp220 redshifted from $z$ = 0
to $z$ = 3.\label{S24_S8_Sr}}
\end{figure}

The selection criteria leave their fingerprint in a colour-colour diagram like the one in 
Fig. \ref{S24_S8_Sr}. Indeed, the ratio between the flux density at 24 \um and an optical band, 
used as selection criterion
for all samples except ELAIS-IRS, determines to a large extent the nature of the sources selected. 
The most extreme cases ($S\!_{\rm{24}}$/$S\!_r$ $>$ 2000) dominate the samples from Houck and 
Polletta. These sources are strongly obscured in the optical and show steep IR spectra,
with strong silicate absorption, and tend to be at high redshift ($z$ $>$ 1.7).

Ratios of 300 $<$ $S\!_{\rm{24}}$/$S\!_r$ $<$ 2000 correspond to the obscured AGN in ELAIS-IRS,
the starburst galaxies from Farrah and the composite sources from Yan.
The distribution of $z$ goes down to $\sim$0.5 with a gap around $z$ $\sim$ 1.4 because of the
silicate feature entering the MIPS24 band.

At $S\!_{\rm{24}}$/$S\!_r$ $\sim$ 100 we find the ELAIS-IRS starburst at 0.6 $<$ $z$ $<$ 1, which are
bluer in $S\!_{\rm{24}}$/$S\!_r$ and  $S\!_{\rm{24}}$/$S\!_{\rm 8}$ than starburst sources from Farrah or Yan; 
and at $S\!_{\rm{24}}$/$S\!_r$ $<$ 100 we find most of the ELAIS-IRS quasars, which are slightly or not obscured,
spanning a wide range in redshift up to $z$ $\sim$ 3.

Note that excluding the ELAIS-IRS sources, the rest are significantly 
redder in $S\!_{\rm{24}}$/$S\!_r$ than would be M82 or Arp220 at the same redshift, indicating that these
sources are more obscured than many local ULIRGs, and probably represent only the most extreme
cases of the ULIRG population at high redshift.

\begin{figure} 
\begin{center}
\includegraphics[width=8.5cm]{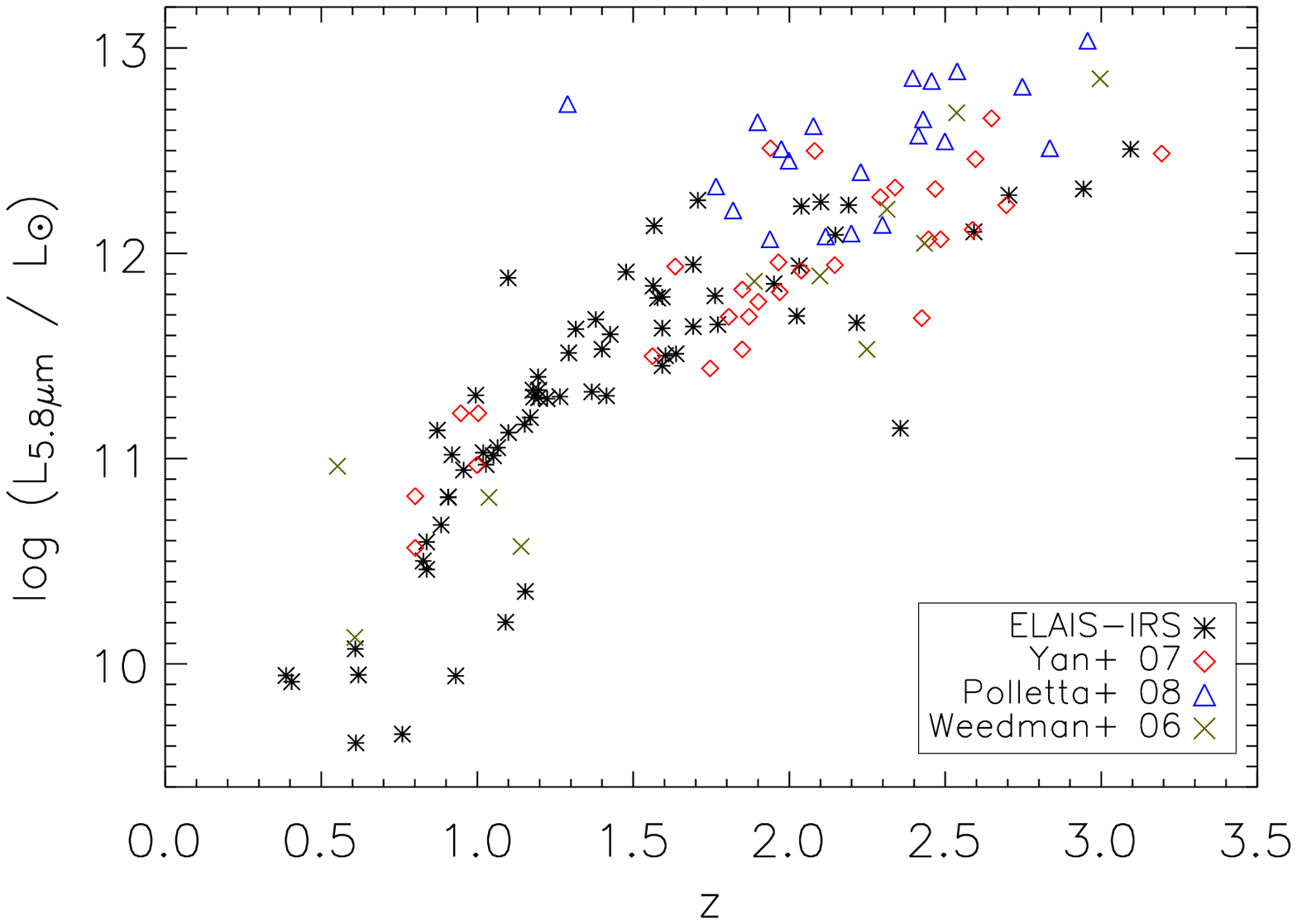}
\end{center}
\caption[$L$--$z$ plane]{MIR luminosity as a function of redshift for some high-$z$
ULIRG samples. Symbols are as in Fig. \ref{S24_S8_Sr}. The eight ELAIS-IRS
sources with lower 5.8-\um luminosity that depart from the main trend are starbursts,
while most of the other sources are AGN or AGN-dominated composite sources.\label{L-z_plane}}
\end{figure}
   
Since most of the sources in these samples have $S\!_{\rm{24}}$ $\sim$ 1 mJy, their distribution in the
$L$--$z$ plane covers a very narrow strip (Fig. \ref{L-z_plane}), indicating that, except for
a few sources with ultra-deep IRS spectra down to $\sim$0.1 mJy \citep{Teplitz07}, 
only the most luminous sources at each redshift have been probed so far. A deeper insight into the
dominant population of LIRGs at $z$ $>$ 1 will be possible once the next generation of MIR spectrographs
onboard \textit{JWST} and \textit{SPICA} become available.

\subsection{Starburst ULIRGs at $z$ $\sim$ 1}

We have classified 10 of the 27 optically-faint ELAIS-IRS sources as starburst-dominated galaxies,
and there is also an optical quasar (EIRS-41) dominated by starburst features in the MIR.

The selection criteria required $z_{\rm{phot}}$ $>$ 1 for galaxies, but $z_{\rm{IRS}}$ estimates
are in the 0.6--1.2 range for all starburst galaxies, with $\langle z \rangle$ = 0.8. 
This is caused by uncertainties in $z_{\rm{phot}}$ and by the selection at 15 \uu, 
which favours sources with strong PAHs at $z$ $\sim$ 1, while at higher redshift requires 
a strong continuum at $\lambda$ $<$ 6\uu, favoring AGN.

In spite of this bias towards PAH-strong sources at $z$ $\sim$ 1, only 1/3 of the 
$0.6<z<1.2$ ELAIS-IRS sources are starbursts, while 2/3 are AGN. Most significantly, if we exclude the
QSOs only half of the galaxies in this range are starbursts, the rest being
obscured AGN. This indicates that the population of 15-\um sources at $z$ $\sim$ 1 is comprised mostly
of AGN for $S\!_{\rm{15}}$ $>$ 1 mJy. The starbursts could still dominate the highly obscured
galaxy population, since they are excluded from the ELAIS-IRS sample because they are too faint in 
the optical; but this is probably not the case, because the AGN also dominate the 24-\um counts
of sources brighter than 1 mJy at $z$ $\sim$ 1.7 \citep{Weedman05}.   

\begin{figure} 
\begin{center}
\includegraphics[width=8.5cm]{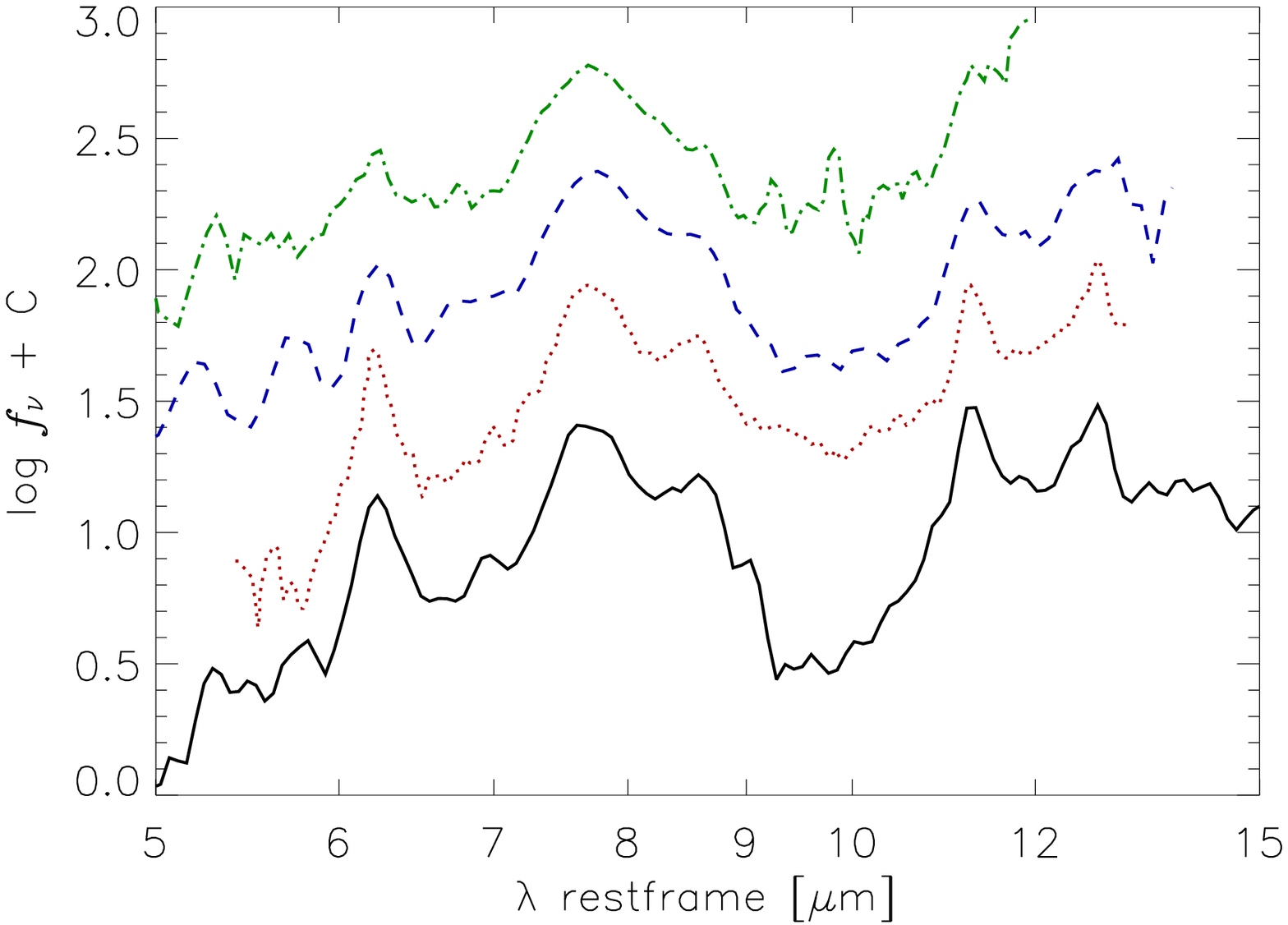}
\end{center}
\caption[High z Starbursts]{Composite spectra for sources with strong PAHs in four high-$z$ samples:
ELAIS-IRS starbursts (solid line), $z\sim1.8$ starbursts from \citet{Farrah08} (dotted line), PAH-strong sources 
from \citet{Sajina07} (dashed line) and $z\sim1.8$ starbursts from \citet{Weedman06} (dot-dashed line).\label{highz_SB}}
\end{figure}

The averaged spectrum of the ELAIS-IRS starbursts is similar to that found in higher redshift samples 
selected at 24 \um (Fig. \ref{highz_SB}). Differences between them in the depth of the silicate feature and 
PAH equivalent width can be interpreted as a combination of a `pure starburst' (HII+PDR) spectrum
and an AGN spectrum in variable proportions. In this case, starburst galaxies from Farrah and 
ELAIS-IRS would be the less contaminated by AGN dust continuum, while in those of Yan
the continuum would dominate de MIR spectrum. 

\begin{figure} 
\begin{center}
\includegraphics[width=8.5cm]{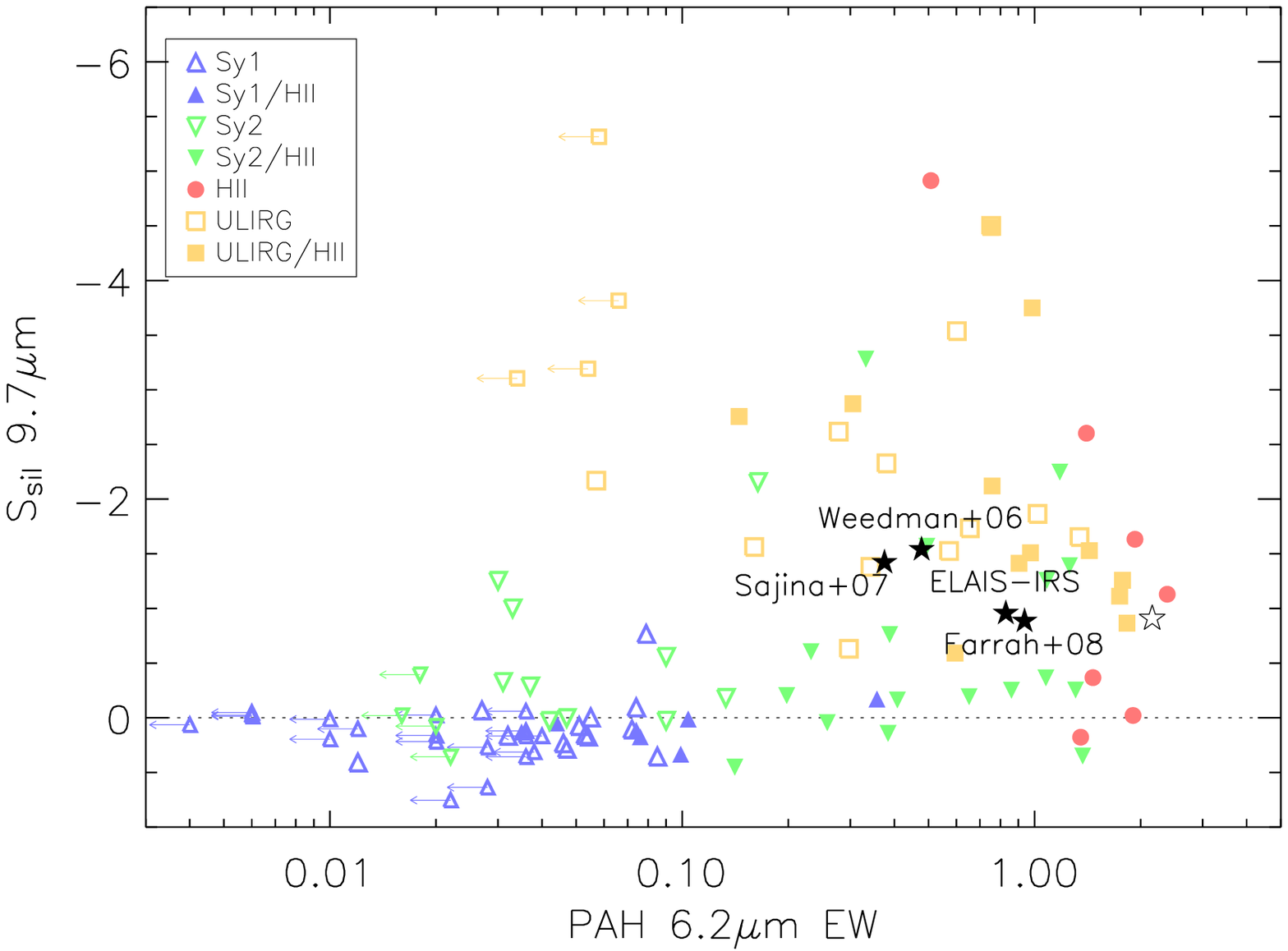}
\end{center}
\caption[Spoon diagram for averaged spectra]{Location in the Spoon diagram of the
composite spectra of several high-$z$ starburst-dominated samples (filled
stars). The open star represents the averaged spectrum of starburst-dominated ELAIS-IRS
sources after removing the four sources with strong MIR continuum. Positions for all 
sources in the Library are also shown as a reference.\label{spoon_promedio}}
\end{figure}

The Spoon diagram for averaged spectra (Fig. \ref{spoon_promedio}) indicates that the
composite spectrum of the ELAIS-IRS starbursts is in the range of local starburst ULIRGs.
If the starburst-dominated sources with strong continuum are removed from the composite,  
the resulting $EW_{\rm{62}}$ increases threefold and resembles that of lower luminosity local starbursts.

\begin{figure} 
\begin{center}
\includegraphics[width=8.5cm]{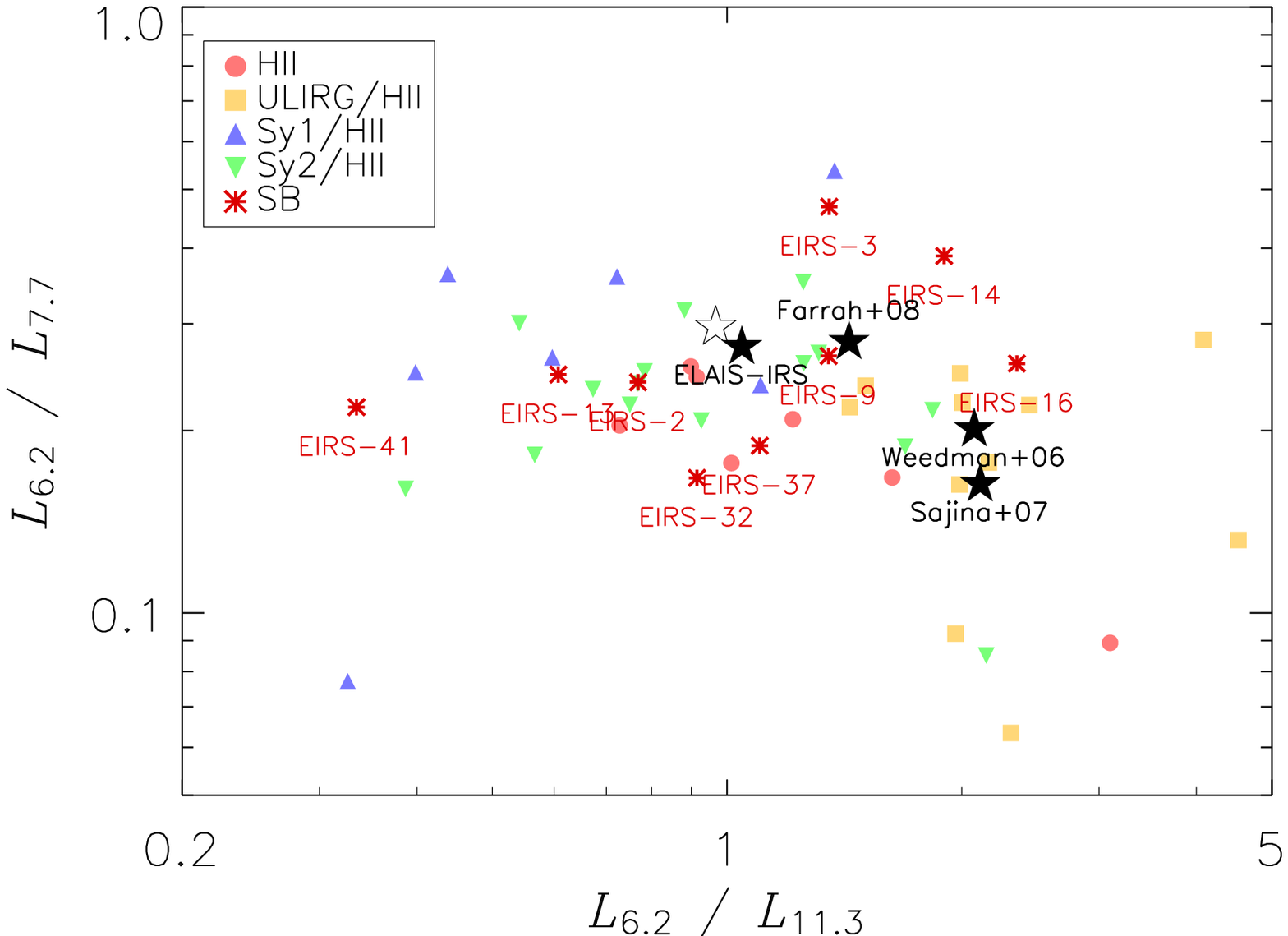}
\end{center}
\caption[PAH ratios]{Luminosity ratio for the 6.2, 7.7, and 11.3 \um PAH features
in the composite spectra of starburst-dominated sources in high-$z$ samples (big solid stars).
The open star represents the averaged spectrum of starburst-dominated ELAIS-IRS
sources after removing the four sources with strong MIR continuum, while individual starburst-dominated
sources from ELAIS-IRS are shown as asterisks. 
Sources from the Library with a starburst component are also shown as a reference: 
starburst galaxies (circles), 
starburst ULIRGs (squares), type 1 AGN with starburst component (triangles) and
type 2 AGN with starburst (inverted triangles).\label{PAH_ratios}}
\end{figure}

In the diagram of PAH luminosity ratios (Fig. \ref{PAH_ratios}) the composite spectrum of
ELAIS-IRS starburst-dominated sources also situates among local starburst
galaxies, with $L_{\rm{62}}$/$L_{\rm{113}}$ substantially higher than in typical starburst-dominated
ULIRGs. This ratio is related to the ionization state of the PAH molecules \citep{Draine01,Rapacioli05,
Brandl06,Galliano08}, a high ratio indicating abundance of cationic species, 
while a low ratio indicates most PAH carriers are neutral.

$L_{\rm{62}}$/$L_{\rm{77}}$ shows anticorrelation with the extinction \citep{Rigopoulou99},
and is usually lower in local ULIRGs than in less luminous starburst galaxies \citep{Lutz98}.
In the ELAIS-IRS average, it is somewhat higher than in most local ULIRGs, suggesting a lower
extinction level.

Lower ionization an extinction values than in local ULIRGs have also been found in higher-$z$
starburst-dominated galaxies \citep{Farrah08}, and could indicate 
that the star formation is extended over larger regions relative to 
their local counterparts, or, alternatively, that their gas-to-dust ratio is higher.

\subsection{Star formation in the AGN\label{SFR_in_AGN}}

It has been stated that the SFR of Seyfert galaxies correlates with the luminosity of the AGN
\citep[e.g.][]{Schweitzer06,Shi07,Maiolino07} and that the SFR is 
systematically higher in type 2 Seyferts relative to type 1 \citep{Maiolino95,Buchanan06}. 

For higher luminosity AGN there are contradictory results: most local QSOs are hosted
by elliptical galaxies, with very little star formation \citep{Dunlop03}, and only 30 per 
cent of them show signs of interaction with other galaxies \citep{Guyon06}. 
Measurements in the \oii 3727-$\AA$ line indicate that the SFR is very low 
(a few \msunyr) in optically selected QSOs \citep{Ho05,Kim06}. In the IR,
\citet{Schweitzer06} find in IRS spectra of Palomar-Green QSOs that more than 30 per cent of 
the IR luminosity is due to star formation, but a later work from \citet{Shi07} reduces the starburst
contribution to $\sim$25 per cent of the flux at 70 and 160 \um --which translates to $\sim$10 per cent
of \lir\ if we assume templates like Dale26 for the starburst and QSO-norm for the AGN component--
for a larger sample of optically selected QSOs.
Using MIR and FIR photometry from SWIRE, \citet{Hatziminaoglou08} estimate relative AGN and SB 
contributions to \lir\ in a sample of 70 SDSS QSOs with 70-\um detection, and find the AGN 
contribution varies from $\sim$100 to $\sim$20 per cent of \lir.

Optically selected QSOs, however, represent less than half of the total QSO 
population \citep{Martinez-Sansigre05,Stern05}, the remaining being obscured quasars (QSO2s) that show up in
the infrared \citep{Cutri01,Lacy04} radio \citep{White03} or X-rays \citep{Norman02}. 

Conversely to what is found in local QSOs, HST images of a sample of $0.3<z<0.8$ QSO2s show signs of
perturbations and intense star formation in the host galaxy \citep{Lacy07}, and in the MIR 
there are signs of recent or ongoing star formation \citep{Lacy05,Yan07}.

At higher redshift there are samples of sub-mm selected QSOs with very high SFR \citep[1000--3000 
\msunyr;][]{Bertoldi03,Beelen06,Lutz07,Lutz08}, but most
of the high-$z$ QSOs are undetected in the sub-mm range \citep[$\sim$70 per cent;][]{Omont03}, and
the observed trend indicates that the starburst-to-AGN luminosity ratio decreases with increasing 
AGN luminosity \citep{Haas03, Maiolino07}, thus suggesting that the observed 
correlation betwen star formation and AGN luminosity saturates at higher luminosities.

The same pattern seems to apply to QSO2s: \citet{Sturm06} find that IRS spectra of X-ray
selected obscured AGN show very weak or no PAH bands, and \citet{Polletta08} estimate that
the star formation contributes $<$20 per cent to the bolometric luminosity of a sample of obscured
AGN with $L_{\rm{6 \mu m}}$ $>$ 10$^{12}$ \lsun.

The ELAIS-IRS AGN comprise a very heterogeneous population, since it covers a wide range of 
PAH equivalent width and SFR. In some of the sources, the estimated starburst-induced IR luminosity 
($SFL_{\rm{IR}}$)
calculated from the PAH luminosity is higher than the total IR luminosity (\lir) obtained from 
the SED fitting. Since it is unlikely that $L_{\rm{PAH}}$/$SFL_{\rm{IR}}$ be lower for
these sources (indeed, PAH destruction by the AGN would lead to lower ratios) the most likely
interpretation is that the SED of these AGN has a cold dust component which is not accounted for
in the SED fitting --effectively underestimating \lir-- because of lack of deep enough photometry in the 
FIR. Incidentally, MIPS70 and MIPS160 upper limits for many ELAIS-IRS AGN cannot rule out a
significant cold dust component in the FIR (see fits in Fig. \ref{FullSED}).

\begin{figure} 
\begin{center}
\includegraphics[width=8.5cm]{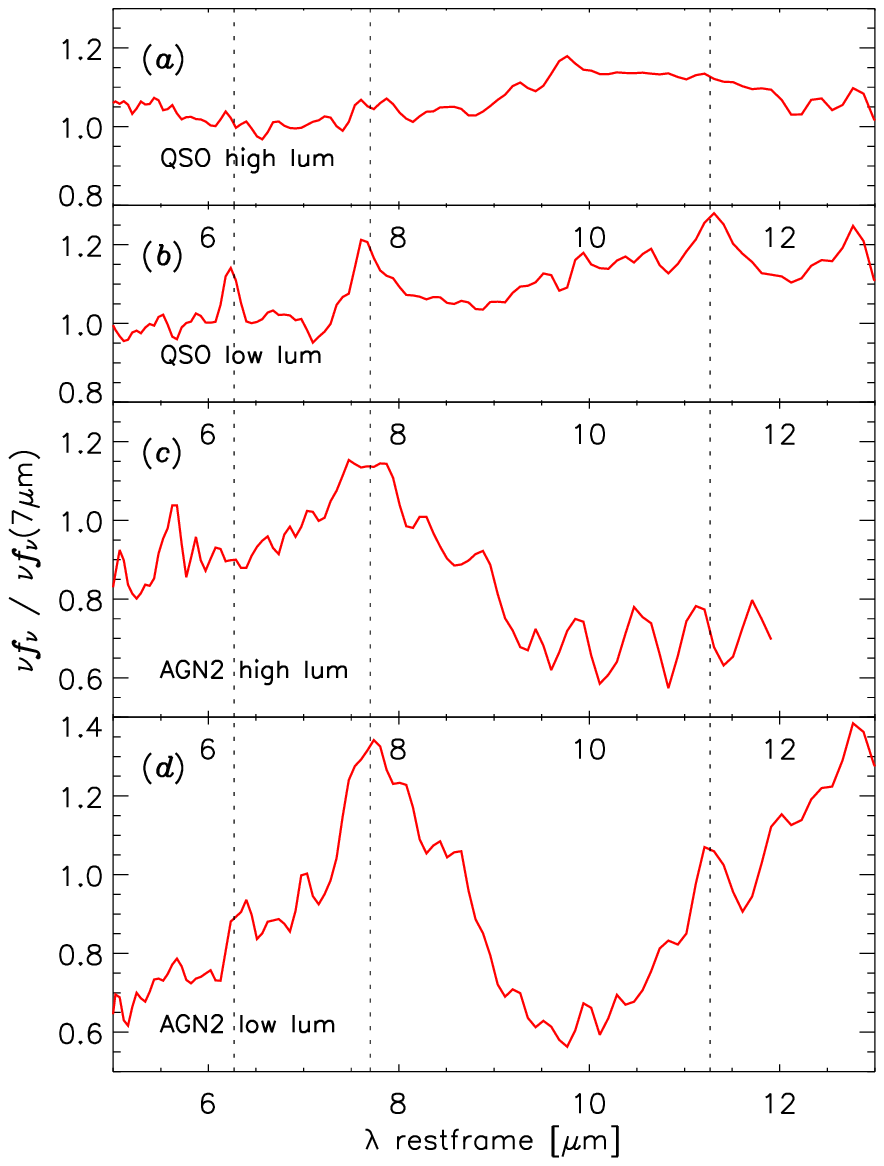}
\end{center}
\caption[AGN average templates by luminosity]{Composite spectra of ELAIS-IRS AGN split into four groups
based on their optical classification and 5.5-\um restframe continuum luminosity. (a) optical QSOs with 
$\nu L_\nu$ $>$ 4 $\times$ 10$^{11}$ \lsun. (b) optical QSOs with 
$\nu L_\nu$ $<$ 4 $\times$ 10$^{11}$ \lsun. (c) optical galaxies with
$\nu L_\nu$ $>$ 4 $\times$ 10$^{11}$ \lsun. (d) optical galaxies with
$\nu L_\nu$ $<$ 4 $\times$ 10$^{11}$ \lsun. The dashed lines mark the position of the
6.2, 7.7 and 11.3 \um PAH features. 
\label{espectros_AGN_lum}}
\end{figure}

The PAH features are comparatively very weak in most ELAIS-IRS AGN, and in many of them 
(35 sources, mostly unobscured AGN) 
only a loose upper limit of the (PAH-derived) $SFL_{\rm{IR}}$ is available.
To get a tighter estimate of its typical value, we measure the
strength of the PAH features in the averaged spectra of QSO and AGN2 sources.
Assuming that the equivalent width of a PAH feature in the averaged spectrum ($EW_{\rm a}$) 
is similar to those of the individual sources, we can roughly estimate the typical PAH 
luminosity of the sources by multiplying $EW_{\rm a}$ times the luminosity of the continuum underlying the
PAH feature. 
Since individual spectra are normalized at 7 \um before averaging them, the
mean PAH luminosity may be overestimated because the lower luminosity sources tend to show higher
PAH equivalent widths. To minimize this bias, we split the QSO and AGN2 populations in two subsamples
of high and low luminosity sources,
based on their restframe 5.5-\um luminosity. We place the limit between populations at 
$L_{\rm{55}}$ = 4 $\times$ 10$^{11}$ \lsun\ to include the same number of sources in both subsamples.
In the lower luminosity population the 6.2, 7.7 and 11.3 \um PAH features are detected
in both QSO and AGN2 averages, while in the higher luminosity ones they all but disappear 
(Fig. \ref{espectros_AGN_lum}). 

Table \ref{SFRtablepromedio} shows the estimated average PAH luminosity and associated $SFL_{\rm{IR}}$ 
and SFR for the averaged spectra of the ELAIS-IRS AGN. 
In the lower luminosity QSOs the average PAH luminosity corresponds to an IR luminosity in the LIRG
range, and accounts for 20--25 per cent of \lir\ as estimated from SED fitting, similar to the mean value
found by \citet{Shi07} in a sample of PG, 2MASS and 3CR QSOs at lower redshift.
The obscured AGN in the same luminosity range have PAH luminosities almost 3 times higher at 6.2 and
7.7\um (which suggests star formation is responsible for about half of \lir\ in these
sources), but they are similar to that of QSOs at 11.3 \uu. Estimated SFR is 
$\sim$50 \msunyr\ for the lower-luminosity QSOs and $\sim$100 \msunyr\ for the obscured AGN.

In the higher-luminosity AGN the PAH features are too weak to meassure them even in the composite
spectrum. Their equivalent width is at least 3 times lower in the high luminosity QSO composite than 
in the lower luminosity one, and the resulting upper limits in $L_{\rm{PAH}}$ constraint the 
mean SFR to $<$ 100 \msunyr. The implied IR luminosity from star formation
accounts for $<$10 per cent of the estimated \lir.
The high-luminosity obscured-AGN composite contains only five sources and has lower S/N than the others.
Upper limits for the 6.2 and 7.7 \um features suggest SFR $<$ 300 \msunyr, and a starburst
contribution to \lir\ of 10--15 per cent at most.

To better constrain the amount of FIR emission in the QSOs, we have performed stacking analysis of the
70 and 160 \um photometry. Fluxes for individual sources are estimated from the counts in the pixel
containing the coordinates of the source, and their uncertainties are obtained from a fit to the histogram
of (finite-valued) pixel values in the image. A mean sky level is computed for the image and subtracted from
the source flux in order to ensure that the mean value for random positions in the image is zero.
The mean flux $x$ and dispersion $\sigma_0$ for the population are estimated from the 
individual measurements ($x_i$, $\sigma_i$) as the maximum-likelihood solution for the distribution:
\begin{equation}
\rho(x,\sigma) = \frac{1}{\sqrt{2\pi(\sigma^2 + \sigma_0^2)}} exp \Big{(}-\frac{(x-\mu)^2}{2(\sigma^2 + \sigma_0^2)}\Big{)}
\end{equation} 

A detailed description of the whole stacking procedure can be found in \citet{Serjeant09}. 
The mean fluxes obtained at 70 and 160 \um are shown in Table \ref{stacking}. 
If we assume the FIR SED is dominated by a starburst component with an M82-like SED, the average IR luminosity
needed to reproduce the observed mean fluxes is 
$\sim$10$^{12}$ \lsun\ for the $z<1.5$ sources, and $\sim$5$\times$10$^{12}$ \lsun\ for those
at $z$ $>$ 1.5. Similar results are obtained if we assume an Arp220 SED instead of M82 for the starburst
component.

The $SFL_{\rm{IR}}$ extrapolated from the observed PAH luminosity is a factor 3 lower in the $z<1.5$ QSOs
(and a factor $\lesssim 10$ lower in the  $z>1.5$ QSOs) than that obtained
from the stacking analysis of 70 and 160 \um photometry, suggesting that the average PAH-to-FIR luminosity ratio is
significantly lower than in starburst galaxies.
A likely interpretation is reduced PAH emission in star-forming regions in the vicinity of an AGN due
to destruction of the PAH carriers by X-ray and far-UV photons from the AGN \citep{Voit92,LeFloch01,Siebenmorgen04}.

Incidentally, the factor $\gtrsim$3 reduction in $EW_{\rm a}$ and in the PAH-to-FIR luminosity ratio
between low- and high-luminosity QSOs suggests higher depletion rates of the PAH carriers in the most luminous AGN. 
Alternatively, it could indicate a saturation of the correlation between the luminosities of the starburst and AGN,
as found in other samples of $z<0.5$ QSOs \citep{Shi07} and higher-luminosity, high-$z$ QSOs \citep{Maiolino07,Lutz08},
but results from the stacking analysis suggest the FIR luminosity roughly scales with AGN luminosity.

Estimated average luminosity of the 6.2 and 7.7 \um PAH features in the low-luminosity composites 
are about 2--3 times higher in the obscured AGN than in the QSOs. 
This could indicate higher SFR in
the obscured AGN, or, alternatively, higher abundance of PAH carriers due to 
absorption by the obscuring material of the energetic photons from the AGN.
However, the statistical significance of this result is not strong due to the small number of sources contributing
to the obscured AGN average, and requires validation in larger samples. 

A selection bias favouring obscured sources with strong PAH emission could also be responsible for
the increased PAH luminosity, but we have not found any likely bias able to increase PAH luminosity
in obscured AGN but not in QSOs. In example, the PAH features increase the observed flux in the 
selection band at $z$ $\sim$ 1, where we find roughly half of the obscured AGN, but these sources are
bright enough to be selected even without contribution from PAH emission to the 15-\um flux.
Since the PAH features help to find a redshift estimate in many sources, this could play a role
in the observed asymmetry, because most redshifts for AGN2 sources rely on the IRS spectrum, while most
QSOs have optical spectroscopic redshifts. But in AGN2 sources a reliable redshift estimate
can usually be obtained from the (absorption) silicate feature alone, and indeed all sources with unreliable
redshifts are classified as AGN1. In addition, a lower accuracy in redshifts obtained from the silicate
feature relative to those from optical spectroscopy imply that PAH features may be blurred in the AGN2
composite, leading to an underestimation of their real strength.
 
Another possible interpretation could be lower MIR luminosity in obscured AGN than in QSOs of the
same intrinsic luminosity, as is expected if the putative torus is optically thick in the MIR; but this
is not supported by measurements of the luminosity ratio between X-rays and MIR \citep{Lutz04,Sturm06,
Polletta07,Horst08}.

\subsection{Implications for the dust distribution\label{geometry}}

A strong difference in SFR between optical type 1 and type 2 AGN has been observed in a sample 
of QSOs and QSO2s from the SDSS \citep{Kim06}. 
Using estimates from the \oii 3727 \AA\ line, they find a SFR 
an order of magnitude higher in type 2 sources, and this is probably a lower limit, due
to higher extinction in the optical. 

Such a large difference between type 1 and 2 sources can hardly be explained within the
AGN unification scheme, and might indicate instead that the AGN classification into type 1 or type 2
does not depend exclusively on the orientation of an AGN torus.
Recent papers suggest that a fraction of the obscuration in type 2 AGN occurs
in the host galaxy: \citet{Lacy07} find a correlation between silicate extinction and orientation
of the plane of the galaxy, while \citet{Polletta08} suggest that foreground extinction by cold
dust is required to reproduce both the silicate feature and NIR continuum for almost half of the sources
in a sample of MIR selected obscured AGN. 

If we assume that the extinction law for dust in the torus is similar to that found in local ULIRGs and in the
Galactic Centre, it reaches a minimum in opacity at $\sim$7 \um and increases towards shorter wavelengths; thus the
optical depth is higher in the NIR than at 7 \uu. If the silicate absorption feature originates in the torus, a decrease
in the viewing angle of the torus should produce a deeper silicate feature and weaker and steeper NIR continuum
peaking at longer wavelengths \citep[e.g.][]{Efstathiou95,Fritz06}. By contrast, if the silicate feature
originates in the host galaxy, a deep silicate profile could be observed in conjunction with an intense and flat NIR 
spectrum \citep{Polletta08}. 

\begin{figure} 
\begin{center}
\includegraphics[width=8.5cm]{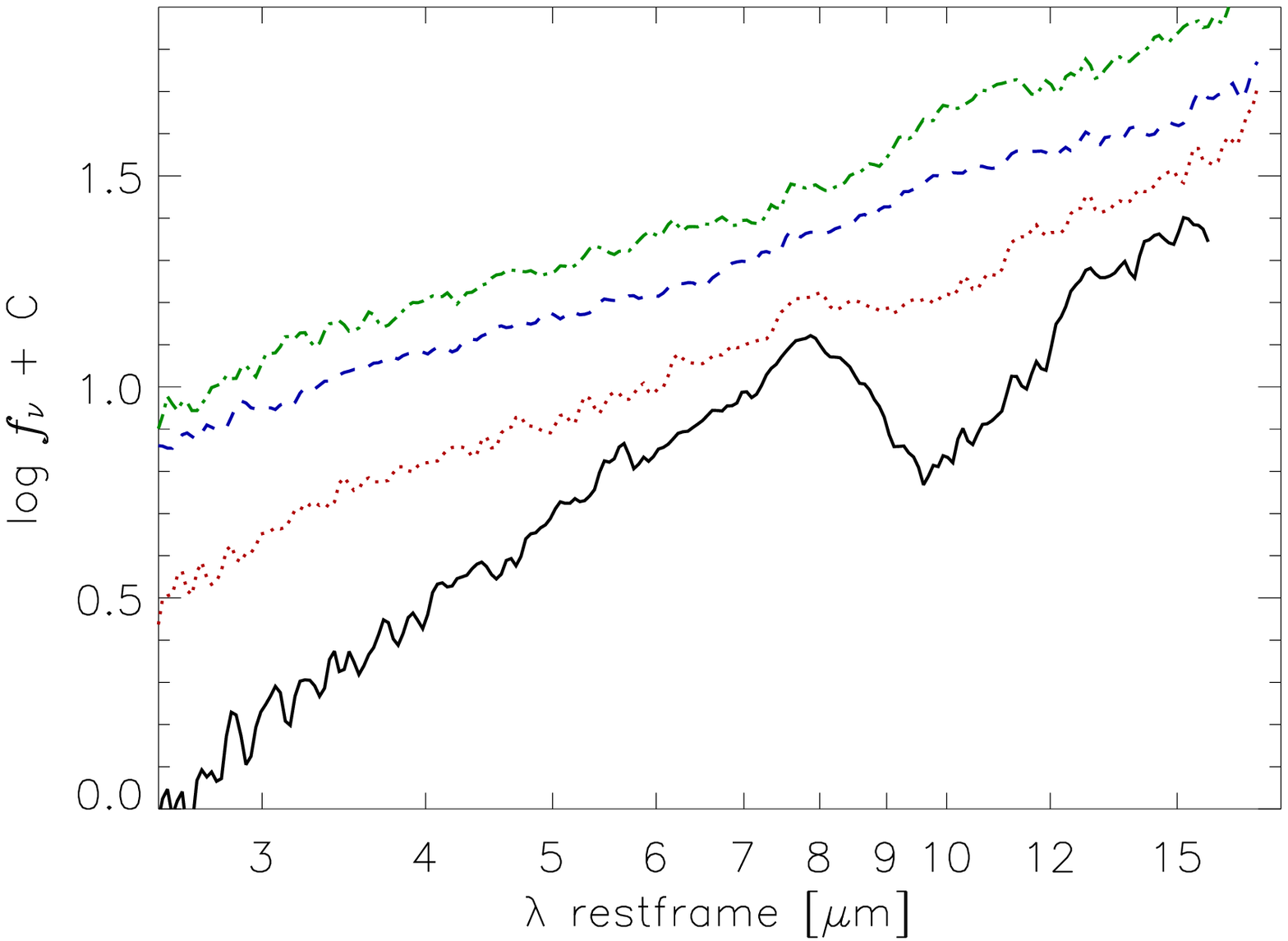}
\end{center}
\caption[Averaged spectra as a function of \ssil]{Composite spectra of ELAIS-IRS AGN separated into 4 groups
based on their measure of silicate strength: \ssil\ $<$ -0.5 (solid line), -0.5 $<$ \ssil\ $<$ 0 (dotted line),
0 $<$ \ssil\ $<$ 0.15 (dashed line), and \ssil\ $>$ 0.15 (dot-dashed line).\label{compara_Ssil}}
\end{figure}

\begin{figure} 
\begin{center}
\includegraphics[width=8.5cm]{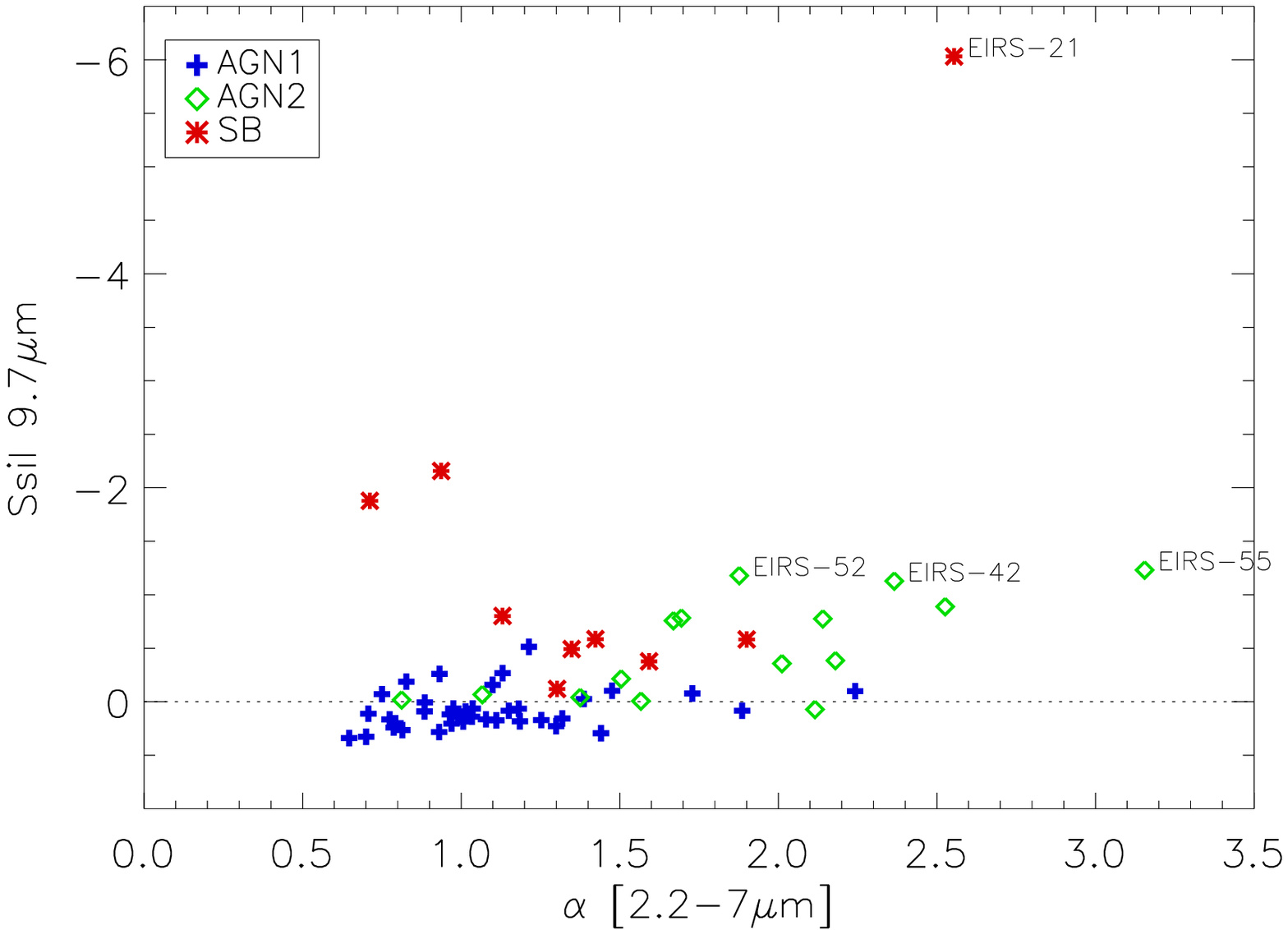}
\end{center}
\caption[NIR slope vs \ssil]{Relation between silicate strength and spectral index of the continuum between
2.2 and 7 \um restframe for ELAIS-IRS sources. Symbols represent: unobscured AGN (plus signs), 
obscured AGN (diamonds), and starbursts (asterisks).\label{alpha_vs_Ssil}}
\end{figure}

Fig. \ref{AGN2_highz} compares the average spectrum of ELAIS-IRS AGN2 sources with the average of sources
that fit the torus model (T) or torus+cold absorber (T+C) from \citet{Polletta08}. The NIR slope and strength
of the silicate feature are better reproduced by the T model. 

Figure \ref{compara_Ssil} shows the averaged spectrum of the ELAIS-IRS AGN separated in four groups 
attending to their measure of \ssil.
In sources with \ssil\ $>$ 0 (mostly QSOs) the NIR slope is small and independent of \ssil, but in 
sources with \ssil\ $<$ 0 the NIR slope is substantially higher, particularly for the subsample with strongest 
silicate absorption. This confirms the existence of a correlation between NIR slope and silicate absorption,
as expected if the obscuration originates mainly in the AGN torus. 

To search for sources that depart from this trend, we calculated the spectral index $\alpha$ of the 
continuum between 2.2 and 7 \um restframe for individual spectra (Fig. \ref{alpha_vs_Ssil}). Most QSOs 
yield $\alpha$ $<$ 1.5, while most AGN2 show $\alpha$ $>$ 1.5. There is a clear trend,
even if the dispersion is high, towards a simultaneous increase of $\alpha$ and \ssil\ in AGN, but not in
starbursts. 
Note that EIRS-21, which is suspected of harboring an obscured AGN due to its strong MIR continuum,
departs strongly from the AGN population. This suggests that the observed extinction is better reproduced 
by a screen of cold dust instead of a thick AGN torus. Incidentally, it is undetected 
down to $1\times 10^{-16}$ erg cm$^{-1}$ s$^{-1}$ in [2-10] keV band images from XMM 
(I. Valtchanov priv. comm.), so the presence of a dust enshrouded AGN is questionable.

The only AGN2 source that separates slightly from the general trend is EIRS-52, but its spectrum
is still better described by the torus model; therefore, a dusty torus is favored as the source
of the obscuration in the ELAIS-IRS AGN.

The lack of T+C sources in ELAIS-IRS is probably caused by the selection criteria, which prevent 
selection of highly obscured AGN as those from \citet{Polletta08}. 

\section{Conclusions} 

We have analyzed the IR properties of a sample of 70 MIR selected LIRGs and ULIRGs at 0.4 $<$ $z$ $<$ 3.2.
The MIR selection favours starburst-dominated sources in redshift intervals in which the strongest PAH bands overlap
with the passband of the selection filter, namely, $z$ $\sim$ 1 at 15 \um and $z$ $\sim$ 1.8 at 24 \uu.
For the AGN-dominated sources, the redshift distribution is wider because their IR SED has smoother features.
The relatively shallow depth of the optical data in the ELAIS-IRS sample and the requirement of optical 
photometric or spectroscopic redshifts for source candidates favours the selection of weakly obscured sources,
contrarily to other MIR high-$z$ ULIRG samples where highly obscured sources are preferred, and thus turns
the ELAIS-IRS sample into a valuable complement for the study of the high-$z$ ULIRG population.

Redshift estimates for the ELAIS-IRS sources from their MIR spectrum are reliable for obscured or star-forming
sources, and unreliable in many unobscured AGN. This behaviour is the opposite to that obtained through optical
spectroscopy, and thus guarantees that reliable redshift estimates can be obtained by one or other means for
most IR sources.

A combination of MIR diagnostic criteria efficiently separates starbursts, obscured and unobscured AGN,
with a good correlation to the optical classification. The selection criteria of the ELAIS-IRS sample 
favour the selection of starburst-dominated sources at 0.6 $<$ $z$ $<$ 1.2, yet
most galaxies selected in this range are AGN-dominated. This suggests that the population of $S\!_{15}$ $>$ 1 mJy 
sources at $z$ $\gtrsim$ 1 is dominated by AGN. 

The averaged spectrum of ELAIS-IRS starbursts is similar to that of local ULIRGs, but the observed depth of
the silicate feature, strength of PAH features and PAH flux ratios are more alike to those found in lower-luminosity 
starburst galaxies. This behaviour has already been observed in higher-redshift starburst samples
and could be related to star formation extending over larger regions or a higher gas-to-dust ratio in the galaxy.   

The correlation between MIR and FIR SEDs allows the estimation of IR luminosities 
from the MIR spectrum with a factor 3 uncertainty. If MIPS70 and/or MIPS160 photometry is available the
uncertainty reduces further and becomes dominated by uncertainties in the photometry and in the SED models.

The mean SFR in ELAIS-IRS QSOs estimated from PAH luminosity is 50-100 \msunyr, but its implied FIR
emission is 3--10 times lower than that obtained from stacking analysis in the 70 and 160 \um bands, 
thus suggesting the PAH emission in substantially depleted in most QSOs.

A lower PAH-to-continuum luminosity ratio is found in the most luminous QSOs relative to lower-luminosity ones,
suggesting a saturation of the correlation between star formation and AGN activity found in other samples of 
lower-luminosity sources or, alternatively, higher depletion rates of the PAH carriers in the galaxies hosting 
the most powerful AGN.

In the obscured AGN, the mean SFR is 2--3 times higher  
than in QSOs of the same luminosity. If confirmed in larger samples, this could indicate a 
connection between dust obscuration and star formation in the host galaxy. However, the observed 
correlation between silicate absorption and the slope of
the NIR-to-MIR continuum is consistent with the hypothesis of the obscuring dust being located 
in a dusty torus.

\section*{Acknowledgements}

This work is based on observations made with the \textit{Spitzer Space
Telescope}, which is operated by the Jet Propulsion Laboratory, Caltech
under NASA contract 1407.

\noindent This work was supported in part by the Spanish Plan Nacional del Espacio
(Grant ESP2007-65812-C02-02) and the UK PPARC.

\noindent We wish to thank the anonymous referee for their very useful comments and suggestions.

\clearpage
\onecolumn

\begin{deluxetable}{llcccrcrrrr} 
\tabletypesize{\scriptsize}
\tablecaption{IRS observations log\label{registro_IRS}}
\tablewidth{0pt}
\tablehead{
\colhead{ID} & \colhead{ELAIS name} & \colhead{R.A.\tablenotemark{a}} & \colhead{Dec.$^{a}$} & \colhead{$S\!_{\rm{15}}$} & \colhead{P.A.} & \colhead{Obs. date} & \multicolumn{4}{c}{Exposure time (s)} \\
\colhead{} & \colhead{} & \colhead{[h.m.s.]} & \colhead{[d.m.s]} &\colhead{[mJy]} & \colhead{[deg]} & \colhead{} & \colhead{SL2} &\colhead{SL1} &\colhead{LL2} &\colhead{LL1}}
\startdata
 EIRS-1 &         ELAISC15\_J161401.0+544733    &  16h14m30.9s  &   +54d52m15s   &  0.79 $\pm$  0.10 &    147.1&     2005-01-08  &    967      &       967      &      1950    &        1950   \\
 EIRS-2 &         ELAISC15\_J161350.0+542631    &  16h14m19.3s  &   +54d31m11s   &  0.80 $\pm$  0.16 &    146.9&     2005-01-08  &   1451      &       967      &      1950    &        1950   \\
 EIRS-3 &         ELAISC15\_J163636.2+411049    &  16h36m42.9s  &   +41d17m04s   &  0.83 $\pm$  0.14 &    116.2&     2005-02-14  &    967      &       967      &      1950    &        1950   \\
 EIRS-4 &         ELAISC15\_J003640-433925      &  00h36m11.8s  &   -43d35m46s   &   0.98 $\pm$  0.17 &    50.0&     2005-06-08  &    967      &       967      &      1950    &        1950   \\
 EIRS-5 &         ELAISC15\_J003408-431011      &  00h34m41.4s  &   -43d12m31s   &  0.99 $\pm$  0.13 &    216.2&     2004-11-16  &    967      &       967      &      1950    &        1950   \\
 EIRS-6 &         ELAISC15\_J163721.3+411503    &  16h37m29.2s  &   +41d21m13s   &  1.00 $\pm$  0.14 &    118.4&     2005-02-12  &    484      &       484      &       975    &         975   \\
 EIRS-7 &         ELAISC15\_J163655.8+405909    &  16h37m03.2s  &   +41d05m23s   &  1.02 $\pm$  0.14 &    117.4&     2005-02-13  &    484      &       484      &       975    &         975   \\
 EIRS-8 &         ELAISC15\_J163455.0+412211    &  16h35m04.3s  &   +41d28m16s   &  1.04 $\pm$  0.15 &    120.9&     2005-02-08  &    484      &       484      &       975    &         975   \\
 EIRS-9 &         ELAISC15\_J160733.7+534749    &  16h08m01.6s  &   +53d52m41s   &  1.04 $\pm$  0.18 &    144.5&     2005-01-09  &    484      &       484      &       975    &         975   \\
 EIRS-10 &        ELAISC15\_J161511.2+550627    &  16h15m41.3s  &   +55d11m06s   &   1.06 $\pm$  0.12 &   147.5&     2005-01-08  &    484      &       484      &       975    &         975   \\
 EIRS-11 &        ELAISC15\_J163739.3+414348    &  16h37m47.5s  &   +41d49m59s   &   1.06 $\pm$  0.11 &   118.4&     2005-02-12  &    484      &       484      &       975    &         975   \\
 EIRS-12 &        ELAISC15\_J161229.0+542832    &  16h12m57.7s  &   +54d33m19s   &   1.06 $\pm$  0.17 &   145.8&     2005-01-09  &    484      &       484      &       975    &         975   \\
 EIRS-13 &        ELAISC15\_J163954.5+411109    &  16h40m02.3s  &   +41d17m20s   &   1.07 $\pm$  0.16 &   118.1&     2005-02-13  &    484      &       484      &       975    &         975   \\
 EIRS-14 &        ELAISC15\_J163536.6+404754    &  16h35m44.4s  &   +40d54m05s   &   1.12 $\pm$  0.12 &   118.1&     2005-02-12  &    484      &       484      &       975    &         975   \\
 EIRS-15 &        ELAISC15\_J161332.4+544830    &  16h14m02.5s  &   +54d53m13s   &   1.12 $\pm$  0.19 &   146.9&     2005-01-08  &    484      &       484      &       975    &         975   \\
\enddata
\tablenotetext{a}{coordinates of the optical ID of the 15-\um ELAIS source \citep{Gonzalez-Solares05}.}
\end{deluxetable}

\begin{deluxetable}{rcccccccc} 
\tabletypesize{\scriptsize}
\tablecaption{Optical and Near infrared photometry\label{fotometria-opt}}
\tablewidth{0pt}
\tablehead{\colhead{ID} & \colhead{$U$} & \colhead{$g$} & \colhead{$r$} & \colhead{$i$} & \colhead{$Z$} &
\colhead{$J$} & \colhead{$H$} & \colhead{$Ks$}}
\startdata
 EIRS-1  &  19.39 $\pm$  0.02 &     20.28 $\pm$   0.03 &     19.64 $\pm$   0.04 &     19.41 $\pm$   0.11 &     18.95 $\pm$   0.02 &         18.17 $\pm$ 0.02 &              &     16.42 $\pm$ 0.01  \\ 
 EIRS-2  &  20.51 $\pm$  0.04 &     23.08 $\pm$   0.05 &     22.14 $\pm$   0.06 &     21.25 $\pm$   0.12 &     20.40 $\pm$   0.04 &         19.10 $\pm$ 0.07 &              &     17.11 $\pm$ 0.03  \\ 
 EIRS-3  &                    &     23.22 $\pm$   0.10 &     22.27 $\pm$   0.05 &     21.01 $\pm$   0.10 &     20.40 $\pm$   0.05 &                      &                  &                       \\ 
 EIRS-4  &                    &                        &                        &                        &                        &                      &                  &                       \\
 EIRS-5  &                    &                        &                        &                        &                        &                      &                  &                       \\
 EIRS-6  &  20.51 $\pm$  0.03 &     20.77 $\pm$   0.02 &     20.59 $\pm$   0.04 &     20.14 $\pm$   0.06 &     19.97 $\pm$   0.04 &                      &                  &                       \\ 
 EIRS-7  &  22.63 $\pm$  0.10 &     22.87 $\pm$   0.04 &     22.92 $\pm$   0.08 &     23.02 $\pm$   0.23 &                        &                      &                  &                       \\ 
 EIRS-8  &  21.94 $\pm$  0.05 &     22.39 $\pm$   0.03 &     21.80 $\pm$   0.06 &     20.90 $\pm$   0.10 &     20.20 $\pm$   0.05 &                      &                  &                       \\ 
 EIRS-9  &                    &                        &     22.71 $\pm$   0.07 &     21.88 $\pm$   0.06 &     21.11 $\pm$   0.12 &                      &                  &                       \\ 
 EIRS-10 &  19.07 $\pm$  0.02 &     19.58 $\pm$   0.02 &     19.07 $\pm$   0.02 &     19.00 $\pm$   0.02 &     18.72 $\pm$   0.02 &                      &                  &                       \\ 
 EIRS-11 &  18.22 $\pm$  0.02 &     19.06 $\pm$   0.02 &     18.79 $\pm$   0.02 &     18.51 $\pm$   0.02 &     18.55 $\pm$   0.03 &                      &                  &                       \\ 
 EIRS-12 &  20.63 $\pm$  0.05 &     21.47 $\pm$   0.02 &     21.14 $\pm$   0.02 &     20.79 $\pm$   0.11 &     20.25 $\pm$   0.04 &         20.23 $\pm$ 0.13 &              &     19.46 $\pm$ 0.15  \\ 
 EIRS-13 &                    &     23.57 $\pm$   0.07 &     22.76 $\pm$   0.06 &     22.10 $\pm$   0.11 &     20.49 $\pm$   0.06 &                      &                  &                       \\ 
 EIRS-14 &                    &                        &     23.46 $\pm$   0.14 &     21.78 $\pm$   0.12 &     21.18 $\pm$   0.11 &                      &                  &                       \\ 
 EIRS-15 &  22.49 $\pm$  0.10 &     22.72 $\pm$   0.04 &     21.74 $\pm$   0.05 &     20.49 $\pm$   0.11 &     19.68 $\pm$   0.02 &         18.08 $\pm$ 0.02 &              &     16.41 $\pm$ 0.01  \\ 
\enddata
\end{deluxetable}

\begin{deluxetable}{rcccccccc} 
\tabletypesize{\scriptsize}
\tablecaption{mid- and far-IR photometry\label{fotometria-IR}}
\tablewidth{0pt}
\tablehead{
 \colhead{ ID              } &
 \colhead{$S\!_{\rm{3.6}}$} & 
 \colhead{$S\!_{\rm{4.5}}$} &
 \colhead{$S\!_{\rm{5.8}}$} &
 \colhead{$S\!_{\rm{8.0}}$} &
 \colhead{$S\!_{\rm{15}}$} &
 \colhead{$S\!_{\rm{24}}$} &
 \colhead{$S\!_{\rm{70}}$} &
 \colhead{$S\!_{\rm{160}}$} \\
 \colhead{              } &
 \colhead{[\uuJy]} & 
 \colhead{[\uuJy]} &
 \colhead{[\uuJy]} &
 \colhead{[\uuJy]} &
 \colhead{[mJy]} &
 \colhead{[mJy]} &
 \colhead{[mJy]} &
 \colhead{[mJy]}
}
\startdata
 EIRS-1  &  297.0 $\pm$  1.9 &  366.0 $\pm$  2.4 &   471 $\pm$   6 &  789 $\pm$   5 &    0.79 $\pm$  0.10 &    2.48 $\pm$  0.02 &                     &                    \\ 
 EIRS-2  &  128.0 $\pm$  1.6 &  113.0 $\pm$  1.4 &    64 $\pm$   5 &   84 $\pm$   4 &    0.80 $\pm$  0.16 &    0.94 $\pm$  0.02 &    16.4 $\pm$   1.0 &   70.2 $\pm$   5.1 \\ 
 EIRS-3  &   63.4 $\pm$  1.1 &   40.7 $\pm$  1.2 &                &                 &    0.83 $\pm$  0.14 &                     &                     &                    \\ 
 EIRS-4  &  347.0 $\pm$  2.0 &  524.0 $\pm$  2.6 &   745 $\pm$   7 &  1090 $\pm$   7 &    0.98 $\pm$  0.17 &   4.33 $\pm$  0.04 &    33.1 $\pm$   1.7 &                    \\ 
 EIRS-5  &  276.0 $\pm$  2.2 &  359.0 $\pm$  2.7 &   458 $\pm$   8 &  707 $\pm$   8 &    0.99 $\pm$  0.13 &    2.18 $\pm$  0.03 &                     &                    \\ 
 EIRS-6  &   31.2 $\pm$  0.9 &   41.6 $\pm$  1.2 &    56 $\pm$   4 &  140 $\pm$   4 &    1.00 $\pm$  0.14 &    0.48 $\pm$  0.02 &                     &                    \\ 
 EIRS-7  &   27.6 $\pm$  0.5 &   52.6 $\pm$  1.0 &   121 $\pm$   3 &  399 $\pm$   4 &    1.02 $\pm$  0.14 &    2.39 $\pm$  0.02 &    13.0 $\pm$   1.2 &                    \\ 
 EIRS-8  &  205.0 $\pm$  1.8 &  247.0 $\pm$  1.6 &   345 $\pm$   6 &  516 $\pm$   4 &    1.04 $\pm$  0.15 &    1.40 $\pm$  0.02 &                     &                    \\ 
 EIRS-9  &   98.6 $\pm$  1.3 &   77.7 $\pm$  1.5 &    96 $\pm$   5 &  163 $\pm$   5 &    1.04 $\pm$  0.18 &    1.91 $\pm$  0.02 &    36.1 $\pm$   1.3 &                    \\ 
 EIRS-10 &  560.0 $\pm$  2.9 &  722.0 $\pm$  3.6 &   823 $\pm$   7 &  1100 $\pm$   6 &    1.06 $\pm$  0.12 &    2.73 $\pm$  0.02 &                    &                    \\ 
 EIRS-11 &  171.0 $\pm$  1.4 &  267.0 $\pm$  1.7 &   412 $\pm$   5 &   693 $\pm$   4 &    1.06 $\pm$  0.11 &    2.34 $\pm$  0.02 &                    &                    \\ 
 EIRS-12 &   27.1 $\pm$  0.7 &   45.0 $\pm$  1.1 &   102 $\pm$   4 &   248 $\pm$   4 &    1.06 $\pm$  0.17 &    2.11 $\pm$  0.02 &                    &                    \\ 
 EIRS-13 &   99.4 $\pm$  1.2 &   81.2 $\pm$  1.2 &    68 $\pm$   4 &                 &    1.07 $\pm$  0.16 &    0.48 $\pm$  0.02 &                    &                    \\ 
 EIRS-14 &  107.0 $\pm$  1.0 &   82.8 $\pm$  1.0 &   109 $\pm$   3 &   133 $\pm$   3 &    1.12 $\pm$  0.12 &    1.30 $\pm$  0.02 &   22.7 $\pm$   1.4 &   66.1 $\pm$   5.5 \\ 
 EIRS-15 &  264.0 $\pm$  1.8 &  230.0 $\pm$  1.7 &   336 $\pm$   6 &   347 $\pm$   4 &    1.12 $\pm$  0.19 &    1.41 $\pm$  0.02 &                    &   47.3 $\pm$   5.9 \\ 
\enddata
\end{deluxetable}

\begin{deluxetable}{l l c l l c l l} 
\tabletypesize{\scriptsize}
\tablecaption{Library of mid-IR spectra\label{info_biblioteca}}
\tablewidth{0pt}
\tablehead{
\colhead{ID} &
\colhead{$z$} &
\colhead{} &
\colhead{ID} &
\colhead{$z$} &
\colhead{} &
\colhead{ID} &
\colhead{$z$} }
\startdata
\multicolumn{8}{c}{ULIRGs; Armus \mbox{et al.} 2007}\\
\\
 IRAS 05189-2524  & 0.043 & &
 IRAS 08572+3915  & 0.058 & &
 IRAS 12112+0305  & 0.073 \\
 IRAS 14348-1445  & 0.083 & &
 IRAS 15250+3609  & 0.055 & &
 IRAS 22491-1808  & 0.078 \\
 Arp 220          & 0.018 & &
 Mrk 231          & 0.042 & &
 Mrk 273          & 0.038 \\
 UGC 5101         & 0.039 & & & & & \\
\\
\multicolumn{8}{c}{Starbursts; Brandl \mbox{et al.} 2006}\\
\\
 Mrk 52           & 0.007 & &    
 NGC 1222         & 0.008 & &    
 NGC 2146         & 0.003 \\   
 NGC 3628         & 0.003 & &    
 NGC 520          & 0.008 & &    
 NGC 7252         & 0.016 \\   
 NGC 7714         & 0.009 & & & & & \\   
\enddata
\end{deluxetable}

\begin{deluxetable}{l c c c c l} 
\tabletypesize{\scriptsize}
\tablecaption{Redshifts\label{origen_z}}
\tablewidth{0pt}
\tablehead{
\colhead{ID} & \colhead{$z$} &\colhead{opt. class\tablenotemark{a}} &
\colhead{$z$ type} &\colhead{$zQ$\tablenotemark{b}} & \colhead{Reference}}
\startdata
EIRS-1     &  0.387   & QSO     &   zspec       &  A   &   P\'erez-Fournon (priv. comm.) \\                                   
EIRS-2     &  1.154   & galaxy  &   zIRS        &  A   & \\                                                   
EIRS-3     &  0.676   & galaxy  &   zIRS        &  A   & \\                                                   
EIRS-4     &  1.181   & QSO     &   zspec       &  A   &   \citet{LaFranca04} \\                              
EIRS-5     &  1.065   & QSO     &   zspec       &  A   &   \citet{LaFranca04} \\                              
EIRS-6     &  2.356   & QSO     &   zspec       &  A   &   P\'erez-Fournon (priv. comm.) \\                                   
EIRS-7     &  2.592   & galaxy  &   zspec       &  A   &   \citet{Swinbank04}\\                               
EIRS-8     &  0.884   & galaxy  &   zIRS        &  B   & \\                                                   
EIRS-9     &  0.609   & galaxy  &   zIRS        &  A   & \\                                                   
EIRS-10    &  1.265   & QSO     &   zIRS        &  B   &        \\                                            
EIRS-11    &  1.414   & QSO     &   zspec       &  A   &   SDSS \\                                            
EIRS-12    &  2.024   & galaxy  &   zIRS        &  A   &                \\                                    
EIRS-13    &  1.091   & galaxy  &   zIRS        &  A   &                \\                                    
EIRS-14    &  0.619   & galaxy  &   zIRS        &  A   &                \\                                    
EIRS-15    &  0.827   & galaxy  &   zIRS        &  A   &        \\                                            
\enddata
\tablenotetext{a}{Classification of the source based on its optical SED or spectrum}
\tablenotetext{b}{Reliability of the redshift estimation. A = reliable, B = unreliable}
\end{deluxetable}

\begin{deluxetable}{l r@{ $\pm$ }lr@{ $\pm$ }lr@{ $\pm$ }lr@{ $\pm$ }lr@{ $\pm$ }l r@{ $\pm$ }lr@{ $\pm$ }lr@{ $\pm$ }lr@{ $\pm$ }lr@{ $\pm$ }l}             
\tabletypesize{\scriptsize}
\tablecaption{Monochromatic fluxes and luminosities\label{tabla_continuo}}
\tablewidth{0pt}
\tablehead{
\colhead{ID} &
\multicolumn{2}{c}{$f_\nu$ 2.2 \um} &
\multicolumn{2}{c}{$f_\nu$ 5.5 \um} &
\multicolumn{2}{c}{$f_\nu$ 7 \um} &
\multicolumn{2}{c}{$f_\nu$ 10 \um} &
\multicolumn{2}{c}{$f_\nu$ 15 \um} &

\multicolumn{2}{c}{$\nu L_\nu$ 2.2 \um} &
\multicolumn{2}{c}{$\nu L_\nu$ 5.5 \um} &
\multicolumn{2}{c}{$\nu L_\nu$ 7 \um} &
\multicolumn{2}{c}{$\nu L_\nu$ 10 \um} &
\multicolumn{2}{c}{$\nu L_\nu$ 15 \um} \\

\colhead{} &
\multicolumn{2}{c}{[mJy]} &
\multicolumn{2}{c}{[mJy]} &
\multicolumn{2}{c}{[mJy]} &
\multicolumn{2}{c}{[mJy]} &
\multicolumn{2}{c}{[mJy]} &
\multicolumn{2}{c}{[10$^{10}$ \lsun]} &
\multicolumn{2}{c}{[10$^{10}$ \lsun]} &
\multicolumn{2}{c}{[10$^{10}$ \lsun]} &
\multicolumn{2}{c}{[10$^{10}$ \lsun]} &
\multicolumn{2}{c}{[10$^{10}$ \lsun]} 
}
\startdata
 EIRS-1 & \multicolumn{2}{c}{---} &   0.67 &   0.10 &    1.0 &    0.1 &    1.4 &    0.1 &    2.4 &    0.1 & \multicolumn{2}{c}{---} &    3 &    0 &    4 &    0 &    3 &    0 &    4 &    0 \\
 EIRS-2 &  0.104 &  0.004 &   0.18 &   0.07 &    0.5 &    0.1 &    0.5 &    0.2 &    1.0 &    0.2 &   13.0 &    0.5 &    9 &    3 &   21 &    3 &   13 &    4 &   19 &    3 \\
 EIRS-3 &  0.061 &  0.002 & \multicolumn{2}{c}{$<$   0.20} & \multicolumn{2}{c}{$<$    0.2} & \multicolumn{2}{c}{$<$    0.2} & \multicolumn{2}{c}{$<$    0.2} &    2.6 &    0.1 & \multicolumn{2}{c}{$<$    3} & \multicolumn{2}{c}{$<$    2} & \multicolumn{2}{c}{$<$    1} & \multicolumn{2}{c}{$<$    1} \\
 EIRS-4 &  0.575 &  0.008 &   1.56 &   0.10 &    2.0 &    0.1 &    3.9 &    0.2 &    6.3 &    0.2 &   75.0 &    1.0 &   81 &    5 &   82 &    4 &  111 &    5 &  120 &    3 \\
 EIRS-5 &  0.362 &  0.006 &   1.08 &   0.09 &    1.2 &    0.1 &    1.6 &    0.2 &    3.0 &    0.2 &   38.5 &    0.6 &   45 &    3 &   41 &    3 &   37 &    4 &   47 &    2 \\
 EIRS-6 &  0.117 &  0.009 & \multicolumn{2}{c}{$<$   0.40} &    0.3 &    0.2 &    0.9 &    0.4 &    1.0 &    0.2 &   55.1 &    4.2 & \multicolumn{2}{c}{$<$   75} &   50 &   24 &   94 &   38 &   69 &   10 \\
 EIRS-7 &  0.387 &  0.009 &   2.33 &   0.25 &    2.7 &    0.2 & \multicolumn{2}{c}{---} &    8.3 &    1.3 &  215.2 &    5.0 &  518 &   54 &  466 &   26 & \multicolumn{2}{c}{---} &  680 &  102 \\
 EIRS-8 &  0.230 &  0.003 &   0.66 &   0.10 &    0.9 &    0.1 &    0.9 &    0.2 &    1.8 &    0.2 &   16.9 &    0.2 &   19 &    2 &   21 &    2 &   14 &    3 &   19 &    1 \\
 EIRS-9 &  0.099 &  0.001 &   0.35 &   0.10 &    0.9 &    0.1 &    0.4 &    0.1 &    1.8 &    0.1 &    3.4 &    0.0 &    4 &    1 &    9 &    0 &    3 &    1 &    8 &    0 \\
EIRS-10 &  0.760 &  0.010 &   1.37 &   0.11 &    1.9 &    0.1 &    2.8 &    0.2 &    4.4 &    0.4 &  113.4 &    1.5 &   81 &    6 &   89 &    5 &   90 &    6 &   96 &    7 \\
EIRS-11 &  0.357 &  0.009 &   1.12 &   0.17 &    1.5 &    0.1 &    2.9 &    0.2 &    4.7 &    0.7 &   66.0 &    1.7 &   82 &   12 &   88 &    8 &  119 &    6 &  126 &   19 \\
EIRS-12 &  0.159 &  0.009 &   1.40 &   0.16 &    1.9 &    0.2 &    1.8 &    0.2 &    4.1 &    0.6 &   57.2 &    3.2 &  202 &   22 &  214 &   23 &  143 &   14 &  216 &   32 \\
EIRS-13 &  0.080 &  0.003 & \multicolumn{2}{c}{$<$   0.21} &    0.3 &    0.1 & \multicolumn{2}{c}{$<$    0.5} & \multicolumn{2}{c}{$<$    0.4} &    8.9 &    0.3 & \multicolumn{2}{c}{$<$    9} &   10 &    4 & \multicolumn{2}{c}{$<$   11} & \multicolumn{2}{c}{$<$    6} \\
EIRS-14 &  0.107 &  0.001 & \multicolumn{2}{c}{$<$   0.29} &    0.7 &    0.1 &    0.3 &    0.1 &    1.1 &    0.1 &    3.8 &    0.0 & \multicolumn{2}{c}{$<$    4} &    7 &    1 &    2 &    1 &    5 &    0 \\
EIRS-15 &  0.248 &  0.004 &   0.51 &   0.10 &    0.6 &    0.1 &    0.8 &    0.2 &    1.6 &    0.1 &   15.9 &    0.3 &   12 &    2 &   12 &    2 &   11 &    2 &   14 &    1 \\
\enddata
\end{deluxetable}

\begin{deluxetable}{lr@{ $\pm$ }lr@{ $\pm$ }lr@{ $\pm$ }lr@{ $\pm$ }lr@{ $\pm$ }lr@{ $\pm$ }lr@{ $\pm$ }lr}             
\tabletypesize{\scriptsize}
\tablecaption{PAH fluxes and luminosities. Silicate absorption\label{tabla_PAHfits}}
\tablewidth{0pt}
\tablehead{
\colhead{ID} &
\multicolumn{2}{c}{10$^{15}$ $\times$ $f_{\rm{62}}$} &
\multicolumn{2}{c}{$EW_{\rm{62}}$} &
\multicolumn{2}{c}{$L_{\rm{62}}$} &
\multicolumn{2}{c}{10$^{15}$ $\times$ $f_{\rm{77}}$} &
\multicolumn{2}{c}{$EW_{\rm{77}}$} &
\multicolumn{2}{c}{$L_{\rm{77}}$} &
\multicolumn{2}{c}{\tausil} &
\colhead{\ssil} \\
\colhead{} &
\multicolumn{2}{c}{[erg cm$^{-2}$ s$^{-1}$]} &
\multicolumn{2}{c}{[\uu]} &
\multicolumn{2}{c}{[10$^{10}$ \lsun]} &
\multicolumn{2}{c}{[erg cm$^{-2}$ s$^{-1}$]} &
\multicolumn{2}{c}{[\uu]} &
\multicolumn{2}{c}{[10$^{10}$ \lsun]} &
\multicolumn{2}{c}{} &
\colhead{}
}
\startdata
 EIRS-1 & \multicolumn{2}{c}{$<$   4.6} & \multicolumn{2}{c}{$<$   0.11} & \multicolumn{2}{c}{$<$   0.06} &  14.0 &   2.0 &   0.38 &   0.06 &   0.19 &   0.03 &   0.00 &   0.11 &   0.05 \\
 EIRS-2 &   9.6 &   0.9 &   1.43 &   0.14 &   1.89 &   0.18 &  40.0 &   1.5 &   5.70 &   0.21 &   7.88 &   0.29 &   0.61 &   0.61 &  -0.58 \\
 EIRS-3 &   6.7 &   2.2 &   5.86 &   2.00 &   0.35 &   0.12 &  14.3 &   2.6 &  13.61 &   2.60 &   0.74 &   0.14 &  -1.82 &   2.09 &   1.93 \\
 EIRS-4 & \multicolumn{2}{c}{$<$   2.5} & \multicolumn{2}{c}{$<$   0.04} & \multicolumn{2}{c}{$<$   0.52} & \multicolumn{2}{c}{$<$   2.5} & \multicolumn{2}{c}{$<$   0.05} & \multicolumn{2}{c}{$<$   0.52} &  -0.23 &   0.08 &   0.17 \\
 EIRS-5 &   5.1 &   1.5 &   0.12 &   0.04 &   0.81 &   0.24 &   5.1 &   1.5 &   0.16 &   0.05 &   0.83 &   0.25 &   0.08 &   0.15 &  -0.07 \\
 EIRS-6 & \multicolumn{2}{c}{$<$   2.8} & \multicolumn{2}{c}{$<$   0.31} & \multicolumn{2}{c}{$<$   3.20} & \multicolumn{2}{c}{$<$   2.4} & \multicolumn{2}{c}{$<$   0.34} & \multicolumn{2}{c}{$<$   2.76} &  -0.22 &   0.50 &   0.28 \\
 EIRS-7 & \multicolumn{2}{c}{$<$   2.7} & \multicolumn{2}{c}{$<$   0.05} & \multicolumn{2}{c}{$<$   3.90} & \multicolumn{2}{c}{$<$   2.2} & \multicolumn{2}{c}{$<$   0.05} & \multicolumn{2}{c}{$<$   3.20} &   0.95 &   0.25 &  -0.76 \\
 EIRS-8 & \multicolumn{2}{c}{$<$   4.4} & \multicolumn{2}{c}{$<$   0.15} & \multicolumn{2}{c}{$<$   0.45} & \multicolumn{2}{c}{$<$   6.9} & \multicolumn{2}{c}{$<$   0.27} & \multicolumn{2}{c}{$<$   0.70} &   0.48 &   0.25 &  -0.51 \\
 EIRS-9 &  22.0 &   1.7 &   2.00 &   0.15 &   0.89 &   0.07 &  82.9 &   2.7 &   7.32 &   0.24 &   3.34 &   0.11 &   0.77 &   0.38 &  -0.58 \\
EIRS-10 & \multicolumn{2}{c}{$<$   3.5} & \multicolumn{2}{c}{$<$   0.07} & \multicolumn{2}{c}{$<$   0.86} & \multicolumn{2}{c}{$<$   3.0} & \multicolumn{2}{c}{$<$   0.07} & \multicolumn{2}{c}{$<$   0.73} &  -0.05 &   0.11 &   0.13 \\
EIRS-11 &   7.8 &   2.6 &   0.20 &   0.07 &   2.56 &   0.87 &   6.7 &   2.3 &   0.19 &   0.07 &   2.20 &   0.76 &  -0.31 &   0.19 &   0.29 \\
EIRS-12 &   5.9 &   2.4 &   0.17 &   0.07 &   4.74 &   1.88 &  13.4 &   3.1 &   0.42 &   0.10 &  10.70 &   2.48 &   0.95 &   0.21 &  -0.77 \\
EIRS-13 &   5.3 &   1.6 &   1.21 &   0.38 &   0.90 &   0.28 &  21.2 &   2.4 &   5.49 &   0.63 &   3.64 &   0.42 &   0.80 &   1.98 & -$\infty$ \\
EIRS-14 &  25.0 &   2.9 &   3.11 &   0.37 &   1.05 &   0.12 &  64.5 &   3.5 &   8.47 &   0.47 &   2.71 &   0.15 &   0.84 &   0.60 &  -0.38 \\
EIRS-15 &   2.3 &   1.1 &   0.10 &   0.05 &   0.20 &   0.09 &  10.6 &   2.2 &   0.59 &   0.12 &   0.91 &   0.19 &   0.13 &   0.28 &  -0.02 \\
\enddata
\end{deluxetable}

\begin{deluxetable}{c | c c c c} 
\tabletypesize{\scriptsize}
\tablecaption{Diagnostic criteria\label{tabla_criterios}}
\tablewidth{0pt}
\tablehead{
\colhead{Criterion} & \colhead{Expression} & \colhead{AGN1} & \colhead{AGN2} & \colhead{SB}}
\startdata
C1 & $EW_{\rm{62}}$    & $<$ 0.2   & $<$ 0.2                 & $>$ 0.2 \\  
C2 & L$_{55}$/\lir\    & $>$ 0.1   & $>$ 0.1                 & $<$ 0.1 \\
C3 & $r_{\rm{PDR}}$    & $<$ 15\%  & $<$ 15\%                & $>$ 15\% \\
C4 & $l/c_{77}$        & $<$ 0.15  & 0.15--1.0               & $>$ 1.0 \\
C5 & \ssil\            & $>$ 0     & $<$ 0                   & $<$ 0   \\   
C6 & L$_{15}$/L$_{10}$ & $<$ 2     & 2--4                    & $>$ 4 \\    
\enddata
\end{deluxetable}

\begin{deluxetable}{l | c c c} 
\tabletypesize{\scriptsize}
\tablecaption{IR vs optical classification for the Library\label{resumen_diagnostico_biblioteca}}
\tablewidth{0pt}
\tablehead{
\colhead{Optical class} & \colhead{N AGN1} & \colhead{N AGN2} & \colhead{N SB}}
\startdata
           Sy1  & 31 &   2  &  0  \\
       Sy1/HII  &  3 &   3  &  1  \\
           Sy2  &  2 &  11  &  1  \\
       Sy2/HII  &  1 &   1  & 15  \\
           HII  &  0 &   0  &  7  \\
   ULIRG/LINER  &  0 &   3  & 12  \\
     ULIRG/HII  &  0 &   0  & 12  \\
\enddata
\end{deluxetable}

\begin{deluxetable}{lcrcrcrcrcrcrcc} 
\tabletypesize{\scriptsize}
\tablecaption{Diagnostic results for ELAIS-IRS\label{tabla_diagnostico}}
\tablewidth{0pt}
\tablehead{
\colhead{ID} &
\colhead{z} &
\colhead{$EW_{\rm{62}}$ [\uu]} & \colhead{C1} &
\colhead{$L_{\rm{55}}$/\lir} & \colhead{C2} &
\colhead{$r_{\rm{PDR}}$ [\%]} & \colhead{C3} & 
\colhead{$l/c_{\rm{77}}$} & \colhead{C4} &
\colhead{\ssil} & \colhead{C5} &
\colhead{$L_{\rm{15}}$/$L_{\rm{10}}$} & \colhead{C6} &
\colhead{Cfinal}} 
\startdata 
   EIRS-1 &   0.387 & $<$  0.112 & 1,2 &     0.486 & 1,2 &       3.9 & 1,2 &     0.260 &   2 &      0.05 &   1 &     1.777 &   1 &     1 \\
   EIRS-2 &   1.154 &      1.434 &   S &     0.030 &   S &      57.1 &   S &     8.010 &   S &     -0.58 & 2,S &     2.068 &   2 &     S \\
   EIRS-3 &   0.676 &      5.861 &   S &$<$  0.268 &   S &      57.5 &   S &    13.880 &   S &      1.28 &   1 &     0.721 &   1 &     S \\
   EIRS-4 &   1.181 & $<$  0.040 & 1,2 &     0.225 & 1,2 &       0.8 & 1,2 &$<$  0.046 &   1 &      0.16 &   1 &     1.614 &   1 &     1 \\
   EIRS-5 &   1.065 &      0.125 & 1,2 &     1.024 & 1,2 &       3.7 & 1,2 &     0.310 &   2 &     -0.07 & 2,S &     1.875 &   1 &     2 \\
   EIRS-6 &   2.356 & $<$  0.306 & 1,2 &$<$  0.558 &   S &       0.0 & 1,2 &$<$  0.340 &   1 &     -0.26 & 2,S &     1.101 &   1 &     1 \\
   EIRS-7 &   2.592 & $<$  0.052 & 1,2 &     0.316 & 1,2 &       0.0 & 1,2 &$<$  0.052 &   1 &     -0.76 & 2,S &        -- &   - &     2$^*$ \\
   EIRS-8 &   0.884 & $<$  0.152 & 1,2 &     0.528 & 1,2 &       1.1 & 1,2 &$<$  0.274 &   1 &     -0.51 & 2,S &     2.011 &   2 &     1 \\
   EIRS-9 &   0.609 &      2.004 &   S &     0.046 &   S &      46.0 &   S &     6.560 &   S &     -0.58 & 2,S &     4.036 &   S &     S \\
  EIRS-10 &   1.265 & $<$  0.066 & 1,2 &     0.899 & 1,2 &       0.0 & 1,2 &$<$  0.066 &   1 &      0.20 &   1 &     1.588 &   1 &     1 \\
  EIRS-11 &   1.414 &      0.199 & 1,2 &     0.654 & 1,2 &       3.3 & 1,2 &     0.300 &   2 &      0.17 &   1 &     1.599 &   1 &     1 \\
  EIRS-12 &   2.024 &      0.165 & 1,2 &     0.117 & 1,2 &       4.9 & 1,2 &     0.260 &   2 &     -0.77 & 2,S &     2.264 &   2 &     2 \\
  EIRS-13 &   1.091 &      1.208 &   S &$<$  0.084 &   S &      51.5 &   S &     5.920 &   S &     -0.80 & 2,S &$>$  1.069 &   S &     S \\
  EIRS-14 &   0.619 &      3.113 &   S &$<$  0.048 &   S &      45.7 &   S &     7.690 &   S &     -0.38 & 2,S &     3.486 &   2 &     S \\
  EIRS-15 &   0.827 &      0.103 & 1,2 &     0.878 & 1,2 &      11.6 & 1,2 &     0.900 &   2 &     -0.02 & 2,S &     1.893 &   1 &     2 \\
\enddata
\tablenotetext{*}{Assuming the unknown diagnostic from $L_{\rm{15}}$/$L_{\rm{10}}$ agrees with that of \ssil.}
\tablenotetext{+}{Source classifies as both AGN1 and SB, but priority is given to the later.}
\end{deluxetable}

\begin{deluxetable}{l | c c c } 
\tablecaption{IR vs optical classification for ELAIS-IRS\label{resumen_diagnostico_ELAIS-IRS}}
\tabletypesize{\scriptsize}
\tablewidth{0pt}
\tablehead{
\colhead{Optical class} & \colhead{N AGN1} & \colhead{N AGN2} & \colhead{N SB}}
\startdata
QSOs    &  33 &   5  &   1 \\
Galaxy &   7 &  12  &  10 \\
\enddata
\end{deluxetable}

\begin{deluxetable}{c c c l} 
\tabletypesize{\scriptsize}
\tablecaption{Recent measurements of $L_{\rm{PAH}}$/\lir\ from the Literature\label{tabla_LPAH_LIR}}
\tablewidth{0pt}
\tablehead{
\colhead{Reference} &
\colhead{Galaxy type} &
\colhead{PAH feature(s)} &
\colhead{Result}}
\startdata
Rigopoulou \mbox{et al.} (1999)\tablenotemark{a}  &  starburst            & 7.7 \um      &  $L_{\rm{77}}$/\lir\ = 0.0081    \\
                  ''                       &  ULIRG starburst      & 7.7 \um     &  L$_{77}$/\lir\ = 0.0055       \\
Lutz \mbox{et al.} (2003)\tablenotemark{b}        &  starburst            & 7.7 \um      &  L$_{77}$/\lir\ = 0.033        \\
Spoon \mbox{et al.} (2004)\tablenotemark{a}       &  normal and starburst & 6.2 \um      &  L$_{62}$/\lir\ = 0.0034    \\
Smith \mbox{et al.} (2007)\tablenotemark{c}       &  normal and starburst & 6.2 \um      &  L$_{62}$/\lir\ = 0.011    \\
                   ''                      &  normal and starburst & 7.7 \um      &  L$_{77}$/\lir\ = 0.041    \\
Farrah \mbox{et al.} (2008)\tablenotemark{a}      &  starburst            & 6.2+11.3 \um  &  L$_{62+113}$/\lir\ = 0.038 \\
\hline
This work                 &  ULIRG starburst      & 6.2 \um      &  L$_{62}$/\lir\ = 0.012        \\
    ''                    &  ULIRG starburst      & 7.7 \um      &  L$_{77}$/\lir\ = 0.038        \\
    ''                    &  ULIRG starburst      & 11.3 \um     &  L$_{113}$/\lir\ = 0.011
\enddata
\tablenotetext{a}{Estimates underlying continuum by interpolation in PAH-free intervals.}
\tablenotetext{b}{Uses $f_\nu$ in the peak of the feature. Converted to luminosity by Shi \mbox{et al.} (2007) 
assuming a Drude profile with FWHM = 0.6 \uu.}
\tablenotetext{c}{Estimates continuum by simultaneous fitting of the spectrum to a Lorentzians+continuum model.}
\end{deluxetable}

\begin{deluxetable}{lr@{ $\pm$ }lr@{ $\pm$ }lr@{ $\pm$ }lccr@{ $\pm$ }lr@{ $\pm$ }l} 
\tabletypesize{\scriptsize}
\tablecaption{IR Luminosity and SFR\label{LumIR_SFR}}
\tablewidth{0pt}
\tablehead{
\colhead{ID} &
\multicolumn{2}{c}{$S\!_{\rm{70}}$} &
\multicolumn{2}{c}{$S\!_{\rm{160}}$} &
\multicolumn{2}{c}{log \lir/\lsun} &
\colhead{SED\tablenotemark{a}} &
\colhead{PAH\tablenotemark{b}} &
\multicolumn{2}{c}{$SFL_{\rm{IR}}$/\lir \tablenotemark{c}} &
\multicolumn{2}{c}{SFR\tablenotemark{d}} \\

\colhead{} &
\multicolumn{2}{c}{[mJy]} &
\multicolumn{2}{c}{[mJy]} &
\colhead{} &
\colhead{} &
\colhead{} &
\colhead{} &
\multicolumn{2}{c}{} &
\multicolumn{2}{c}{[\msunyr]}}
\startdata
 EIRS-1 &\multicolumn{2}{c}{}&\multicolumn{2}{c}{}&    10.8 &    0.3   & QSO high & 6.2 11.3 & \multicolumn{2}{c}{$<$  0.44} & \multicolumn{2}{c}{$<$    5} \\
 EIRS-2 &    16.4 & 1.0 &   70.2 & 5.1 &   12.4 &    0.1   & Dale26 & 6.2 7.7 11.3 &  0.69 &  0.03 &  332 &   12 \\
 EIRS-3 &\multicolumn{2}{c}{}&\multicolumn{2}{c}{}&    11.0 &    0.3   & Seyfert 2 & 6.2 7.7 11.3 &  2.03 &  0.34 &   40 &    6 \\
 EIRS-4 &    33.1 & 1.7 &\multicolumn{2}{c}{}&    12.5 &    0.5   & Strong FIR QSO & 6.2 7.7 & \multicolumn{2}{c}{$<$  0.08} & \multicolumn{2}{c}{$<$   47} \\
 EIRS-5 &\multicolumn{2}{c}{}&\multicolumn{2}{c}{}&    11.6 &    0.3   & QSO low & 6.2 7.7 11.3 &  1.47 &  0.19 &  115 &   15 \\
 EIRS-6 &\multicolumn{2}{c}{}&\multicolumn{2}{c}{}&    12.1 &    0.3   & QSO high & 6.2 7.7 & \multicolumn{2}{c}{$<$  1.20} & \multicolumn{2}{c}{$<$  287} \\
 EIRS-7 &    13.0 & 1.2 &\multicolumn{2}{c}{}&    13.0 &    0.5   & Mrk 231 & 6.2 7.7 & \multicolumn{2}{c}{$<$  0.20} & \multicolumn{2}{c}{$<$  346} \\
 EIRS-8 &\multicolumn{2}{c}{}&\multicolumn{2}{c}{}&    11.3 &    0.3   & QSO low & 6.2 7.7 11.3 & \multicolumn{2}{c}{$<$  1.25} & \multicolumn{2}{c}{$<$   43} \\
 EIRS-9 &    36.1 & 1.3 &\multicolumn{2}{c}{}&    11.9 &    0.2   & NGC 6240 & 6.2 7.7 11.3 &  0.88 &  0.03 &  124 &    4 \\
EIRS-10 &\multicolumn{2}{c}{}&\multicolumn{2}{c}{}&    12.1 &    0.3   & Weak FIR QSO & 6.2 7.7 & \multicolumn{2}{c}{$<$  0.31} & \multicolumn{2}{c}{$<$   77} \\
EIRS-11 &\multicolumn{2}{c}{}&\multicolumn{2}{c}{}&    12.0 &    0.3   & QSO norm & 6.2 7.7 11.3 &  1.06 &  0.22 &  223 &   45 \\
EIRS-12 &\multicolumn{2}{c}{}&\multicolumn{2}{c}{}&    13.1 &    0.3   & M82 & 6.2 7.7 &  0.25 &  0.06 &  576 &  143 \\
EIRS-13 &\multicolumn{2}{c}{}&\multicolumn{2}{c}{}&    12.0 &    0.3   & GL12 & 6.2 7.7 11.3 &  0.99 &  0.10 &  173 &   17 \\
EIRS-14 &    22.7 & 1.4 &   66.1 & 5.5 &   11.9 &    0.1   & Dale26 & 6.2 7.7 11.3 &  0.86 &  0.05 &  119 &    6 \\
EIRS-15 &\multicolumn{2}{c}{}&    47.3 & 5.9 &   11.3 &    0.3   & Weak FIR QSO & 6.2 7.7 11.3 &  1.12 &  0.26 &   42 &    9 \\
\enddata
\tablenotetext{a}{template SED that best fits the MIR and FIR photometry.}
\tablenotetext{b}{PAH feature(s) used for star formation rate estimation.}
\tablenotetext{c}{ratio of star formation induced IR luminosity to total IR luminosity.}
\tablenotetext{d}{star formation rate as estimated from PAH luminosity}
\end{deluxetable}                            

\begin{deluxetable}{c| r c r r} 
\tabletypesize{\scriptsize}
\tablecaption{70 and 160\um stacking analysis of unobscured AGNs\label{stacking}}
\tablewidth{0pt}
\tablehead{\colhead{redshift} & \colhead{filter} & \colhead{mean flux} & 
\colhead{\lir\ M82\tablenotemark{a}}  & \colhead{\lir\ Arp220\tablenotemark{a}}\\ 
\colhead{}  & \colhead{} & \colhead{[mJy]}  & \colhead{[\lsun]}  & \colhead{[\lsun]}}
\startdata
z $<$ 1.5      &   70 \um &  11 $\pm$ 2    &   8.5$\times$10$^{11}$  &   8.4$\times$10$^{11}$ \\
(22 sources) &  160 \um &  27 $\pm$ 9    &   1.3$\times$10$^{12}$  &   8.9$\times$10$^{11}$ \\
\\
z $>$ 1.5      &   70 \um &  5.3 $\pm$ 1.1 &   4.4$\times$10$^{12}$  &   8.3$\times$10$^{12}$ \\
(18 sources) &  160 \um &  28 $\pm$ 7    &   6.3$\times$10$^{12}$  &   5.3$\times$10$^{12}$ \\
\enddata
\tablenotetext{a}{Infrared luminosity of the starburst component required to reproduce observed FIR flux, assuming M82 or Arp220 SED.}
\end{deluxetable}

\begin{deluxetable}{c c c c c c}   
\tabletypesize{\scriptsize}
\tablecaption{$SFR$ and $SFL_{\rm{IR}}$ in averaged spectra\label{SFRtablepromedio}}
\tablewidth{0pt}
\tablehead{
\colhead{Group} &
\colhead{PAH band} &
\colhead{$EW_{\rm a}$} &  
\colhead{$\langle$ $L_{\rm{PAH}}$ $\rangle$} &
\colhead{$\langle$ $SFL_{\rm{IR}}$ $\rangle$} &
\colhead{$\langle$ $SFR$ $\rangle$}\\

\colhead{} &
\colhead{} &
\colhead{[\uu]} &
\colhead{[\lsun]} &
\colhead{[\lsun]} &
\colhead{[\msunyr]}}
\startdata
          QSOs       &  6.2 \um  &  $<$0.01 & $<$7.1 $\times$ 10$^{9}$   &   $<$5.9 $\times$ 10$^{11}$   &    $<$100 \\  
           high lum. &  7.7 \um  &  $<$0.03 & $<$1.7 $\times$ 10$^{10}$  &   $<$4.5 $\times$ 10$^{11}$   &     $<$80 \\  
        (18 sources) & 11.3 \um  &  $<$0.01 & $<$3.5 $\times$ 10$^{9}$   &   $<$3.1 $\times$ 10$^{11}$   &     $<$53 \\  
\hline
           QSOs      &  6.2 \um  &    0.03  &    3.3 $\times$ 10$^{9}$   &  2.7 $\times$ 10$^{11}$   &        46 \\  
           low lum.  &  7.7 \um  &    0.09  &    8.1 $\times$ 10$^{9}$   &  2.1 $\times$ 10$^{11}$   &        37 \\  
        (18 sources) & 11.3 \um  &    0.05  &    3.4 $\times$ 10$^{9}$   &  3.0 $\times$ 10$^{11}$   &        52 \\  
\hline
           Obsc. AGN &  6.2 \um  &  $<$0.04 & $<$2.5 $\times$ 10$^{10}$  &  $<$2.1 $\times$ 10$^{12}$   &    $<$352 \\  
           high lum. &  7.7 \um  &  $<$0.10 & $<$5.6 $\times$ 10$^{10}$  &  $<$1.5 $\times$ 10$^{12}$   &    $<$257 \\  
         (5 sources) & 11.3 \um  &  $<$0.20 & $<$3.8 $\times$ 10$^{10}$  &  $<$3.5 $\times$ 10$^{12}$   &    $<$589 \\  
\hline
          Obsc. AGN  &  6.2 \um  &     0.10 &    8.1 $\times$ 10$^{9}$   &  6.7 $\times$ 10$^{11}$    &       114 \\  
            low lum. &  7.7 \um  &     0.34 &    2.6 $\times$ 10$^{10}$  &  6.8 $\times$ 10$^{11}$    &       121 \\  
         (6 sources) & 11.3 \um  &     0.07 &    3.6 $\times$ 10$^{9}$   &  3.3 $\times$ 10$^{11}$    &        54 \\  
\enddata

\end{deluxetable}

\bsp
\label{lastpage}
\end{document}